\documentclass[aps,preprint,showpacs,showkeys,superscriptaddress,groupedaddress]{revtex4-2}  
\usepackage{graphicx}  
\usepackage{float}
\usepackage{dcolumn}   
\usepackage{bm}        
\usepackage{amssymb}   
\usepackage{slashed}   
\usepackage{amsmath}   
\usepackage{simplewick} 
\usepackage{verbatim}  
\usepackage{color} 
\usepackage{appendix} 

\usepackage[bookmarks=true,colorlinks=true,linkcolor=blue,unicode=true]{hyperref}

\begin{document}

\title{Higgs mode in   $2+1$ dimensional $O(2)$ model}

\author{Ji-Chong Yang}

\affiliation{Department of Physics  \&  State Key Laboratory of Surface Physics,   Fudan University,\\ Shanghai 200433, China}

\affiliation{ Department of Physics, Liaoning Normal University, Dalian 116029, China}

\author{Yu Shi}
\email[]{yushi@fudan.edu.cn}
\thanks{Corresponding author}

\affiliation{Department of Physics  \&  State Key Laboratory of Surface Physics,   Fudan University,\\ Shanghai 200433, China}

\vskip 0.25cm

\date{\today}

\begin{abstract}
We investigate the spectral functions of the Higgs  mode in $O(2)$ model, which can be experimentally realized in  a 2D Bose gas. Zero temperature limit is considered. Our  calculation fully includes the 2-loop contributions. Peaks show up in the spectral functions of both the longitudinal and the scalar susceptibilities.  Thus this  model cannot explain the disappearance of the response at the weak interaction limit. Neither it can explain the similarity between the longitudinal and the scalar susceptibilities in the visibility of the Higgs mode. A  possible lower peak  at about $2m_{\sigma}$ is also noted.
\end{abstract}



\maketitle

\section{\label{sec:1}Introduction}

Higgs mode in condensed matter physics,  also called amplitude mode, was first discussed in details in superconductivity in 1980s~\cite{littlewood}.   Recently, it was  observed in ultra-cold Bose atoms in  three and two dimensional optical lattices~\cite{3DOpticalLattice,2DOpticalLattice}. It was also  studied in various other condensed matter systems~\cite{OtherExperiments,Cavity}.

In superfluids, Higgs modes  are difficult to observe. The first two time-reversal and gauge-invariant terms allowed in the action-density can be written as~\cite{Varma1,Varma2}
\begin{equation}
\begin{split}
&iK_1\Phi ^*\frac{\partial}{\partial t}\Phi - K_2 \frac{\partial}{\partial t}\Phi ^*\frac{\partial}{\partial t}\Phi ,
\end{split}
\label{eq:1.1}
\end{equation}
where $\Phi$ is the order parameter, $K_1$ and $K_2$ are constants. When $K_1\approx 0$ while  $K_2\neq 0$, Lorentz invariance is present, and there exists the  particle-hole symmetry, which is necessary  for the observation of the Higgs mode. Unfortunately, in the Gross-Pitaevskii model, which is often used in describing the superfluidity,  $K_1\neq 0$ and $K_2=0$, hence the particle-hole symmetry is absent.

In recent years, the ultra-cold dilute gases became a platform for many problems because of the convenience of tuning the parameters~\cite{BECReview}. The particle-hole symmetry can be provided by the periodicity of the optical  lattice~\cite{Varma1}. In the vicinity of superfluid-Mott insulating transition, the system can be described approximately  by a $O(2)$ model~\cite{O2Model1,O2Model2,3DO2}, $N=2$ version of  $O(N)$ model, which is important in the study of quantum phase transitions~\cite{Sachdev,SachdevBook}. The visibility of the Higgs mode in two dimensional $O(N)$ model has been studied~\cite{Podolsky1}. Evidence of the Higgs mode in the two dimensional optical lattice was found in a  quantum Monte Carlo simulation of the Bose-Hubbard model in the vicinity of the superfluid-Mott insulator transition~\cite{Simulationstudy}, not long before it was observed in an experiment~\cite{2DOpticalLattice}.

Previously,  Higgs mode in 2D Bose gas have been studied in the $O(N)$ model by using the large-N expansion~\cite{Podolsky1,Podolsky2}. However, in the $O(2)$ model, which describes the system in the vicinity of the critical point, the large-N expansion might be inefficient in convergence because $1/N$ is not so small. It is proposed to observe behavior of the Higgs mode through the scalar susceptibility. Thus it is  an interesting question whether such phenomenon can be understood by using the $O(2)$ model, without employing large-N expansion.

In this paper,  without using large-N expansion, we study the linear response of the Higgs mode in $2+1$ dimensional $O(2)$ model at zero temperature limit. We calculate the spectral function with full 2-loop contributions, and obtain the analytical results of some 2-loop diagrams with arbitrary external momenta, which  have not been obtained previously. We also calculate the dominant contributions up to infinite loop orders using variance summation methods. We find that there are peaks in the spectral functions of both the longitudinal and scalar susceptibilities  in the $2+1$ dimensional $O(2)$ model. However, the $O(2)$ model cannot reproduce the phenomenon that the peak of the spectral function of the amplitude broadens and then vanishes when the system is tuned away from the critical point into the superfluid phase, as observed in Ref.~\cite{2DOpticalLattice,FermionSuperfluid}. This may be because the $O(2)$ model can only approximately describe the system at the vicinity of the critical point, and the disappearance of the peak cannot be explained as long as one  uses  a relativistic model in zero temperature limit. Besides, we  find another lower  peak in the spectral function at about $2m_{\sigma}$.

The rest  of the paper is organized as the following. In Sec.~\ref{sec:2}, we briefly review the $O(2)$ model in two dimensions. The susceptibilities of the Higgs mode are calculated in Sec.~\ref{sec:3}. In Sec.~\ref{sec:4}, we present the numerical calculation.  Sec.~\ref{sec:5} is a summary.

\section{\label{sec:2}The \texorpdfstring{$O(N)$}{O(N)} model}

In imaginary-time representation,   $D$ dimensional $O(N)$ model is~\cite{SachdevBook,TsvelikBook}
\begin{equation}
\begin{split}
&S=\int d^{D} x\left\{\frac{1}{2}\left(\partial _{\mu} \Phi\right)^2 -\frac{m_0^2}{2}\Phi ^2 + \frac{U_0}{4}\Phi ^2\Phi ^2\right\},
\end{split}
\label{eq:2.1}
\end{equation}
which is also known as the $\phi ^4$ model with a negative mass term~\cite{PeskinBook}.  It can  also be  written as~\cite{Podolsky1}
\begin{equation}
\begin{split}
&S=\frac{1}{2g}\int d^{D} x\left\{\left(\partial _{\mu} \Phi_g\right)^2 -\frac{{m'_0}^2}{2}\Phi_g ^2+\frac{{m'_0}^2}{4N}\Phi_g ^2\Phi_g ^2 + \frac{{m'_0}^2N}{4}\right\},
\end{split}
\label{eq:2.2}
\end{equation}
which reduces to Eq.~(\ref{eq:2.1}) with
$\Phi \equiv \Phi _g / \sqrt{g}$, $m_0\equiv m_0'/\sqrt{2}$ and $U_0\equiv  g{m_0'}^2/2N$.

\subsection{\label{sec:2.1}Renormalization and Spontaneous Symmetry Breaking~(SSB)}

The $\phi ^4$ model can be renormalized with a field strength.   The renormalized action  can be written  as~\cite{PeskinBook}
\begin{equation}
\begin{split}
&S=\int d^{D} x\left\{\frac{1}{2}\left(\partial _{\mu} \Phi _r\right)^2 -\frac{m^2}{2}\Phi _r ^2 + \frac{U}{4}\Phi _r ^2\Phi _r ^2\right.\\
&\left.+\frac{1}{2}\delta _Z\left(\partial _{\mu} \Phi _r\right)^2 -\frac{\delta _m}{2}\Phi _r ^2 + \frac{\delta _U}{4}\Phi _r ^2\Phi _r ^2\right\},
\end{split}
\label{eq:2.4}
\end{equation}
with $\Phi _r$, $\delta _Z$, $\delta _m$ and $\delta _{U}$ defined as
\begin{equation}
\begin{split}
&\Phi =Z^{\frac{1}{2}}\Phi _r,\;\;\delta _Z=Z-1,\;\;\delta _m=m_0^2Z-m^2,\;\;\delta _{U}=U_0Z^2-U,
\end{split}
\label{eq:2.5}
\end{equation}
where $m_0$ and $U_0$ are bare mass and coupling constant, while $m$ and $U$ are the physical mass and coupling constant. The vacuum expectation value (VEV)   is  $|\Phi _r |^2=v^2,v=m/\sqrt{U}$. The order parameter can be parameterized as~\cite{Podolsky1}
\begin{equation}
\begin{split}
&\Phi _r = (v+\sigma, \mathbf{\pi}),
\end{split}
\label{eq:2.6}
\end{equation}
where $\mathbf{\pi} = (\pi_1, \cdots, \pi_{N-1})$.
Then the action can be expanded  as $S=S_0+S_A+S_C$, where
\begin{equation}
\begin{split}
&S_{0}=\int d^{D}x\left( \frac{1}{2}(\partial _{\mu}\sigma)^2 + m^2 \sigma ^2 + \frac{1}{2}(\partial _{\mu}\pi_k)^2\right),\\
&S_{A}=\int d^{D}x\left( mU^{\frac{1}{2}}\sigma (\sigma ^2 + \pi _k^2)+\frac{U}{4}(\sigma ^2 + \pi _k^2)^2\right),\\
\end{split}
\label{eq:2.7}
\end{equation}
and
\begin{equation}
\begin{split}
&S_{C}=\int d^{D}x\left( \frac{1}{2}\delta _Z (\partial _{\mu}\pi _k)^2 + \frac{1}{2}\left(\delta _{U}v^2-\delta _m\right)\pi_k^2\right.\\
&\left.+\frac{1}{2}\delta _Z (\partial _{\mu}\sigma )^2 + \frac{1}{2}(3 \delta _U v^2 - \delta _m)\sigma ^2+v(\delta _{U}v^2-\delta _m)\sigma\right.\\
&\left. +\delta _U v \sigma (\sigma ^2 + \pi_k^2)+\frac{\delta _U}{4}(\sigma ^2 +\pi _k^2)^2\right)\\
\end{split}
\label{eq:2.8}
\end{equation}
are harmonic, anharmonic, and counter terms, respectively. In imaginary-time representation, the momenta are in $D$ dimensional  Euclidean space, and can be represented as $p=(\bf{p},\omega)$, where ${\bf p}$ is the $D-1$ dimensional momentum, and $\omega$ is the bosonic Matsubara frequency.

\begin{figure*}
\begin{center}
\includegraphics[scale=0.5]{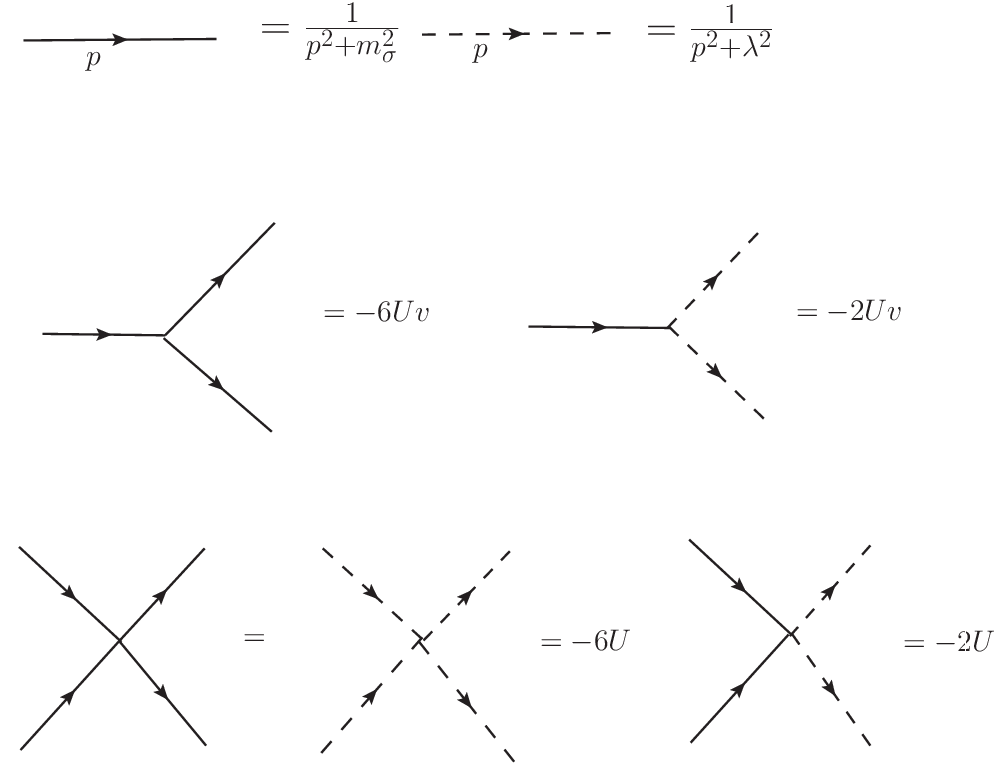}
\caption{Feynman rules of the propagators and the vertices of $O(2)$ model. }
\label{fig:vertex}
\end{center}
\end{figure*}

\begin{figure*}
\begin{center}
\includegraphics[scale=0.5]{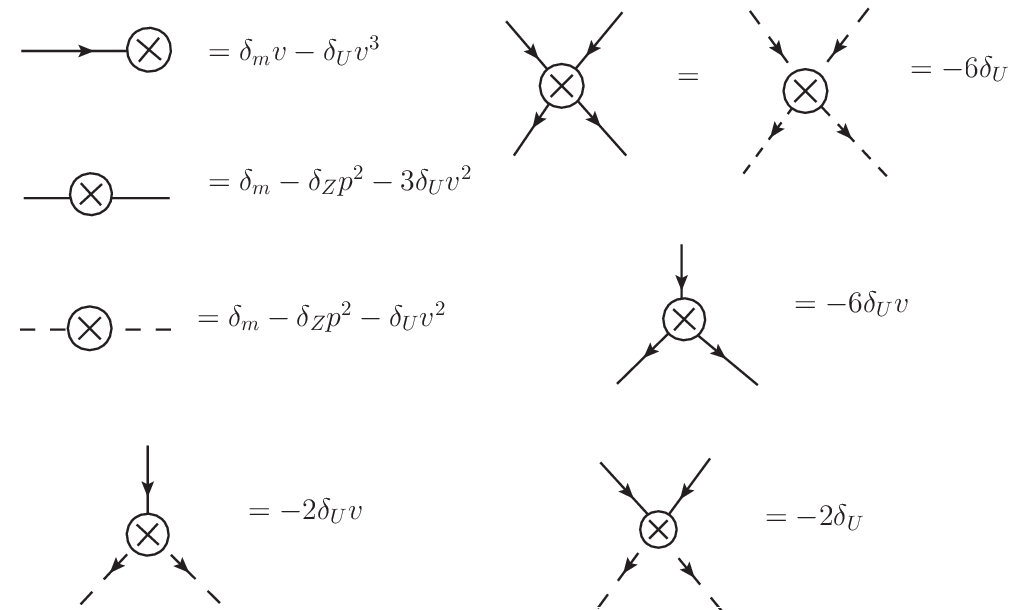}
\caption{Feynman rules of the counter terms of $O(2)$ model. }
\label{fig:counterterm}
\end{center}
\end{figure*}

We now specialize in the case of $N=2$. The corresponding propagators of $\sigma$ and $\pi$ are
\begin{equation}
\begin{split}
&G_{\sigma\sigma}=\frac{1}{p^2+2 m^2},\;\;\;\;\;G_{{\bf \pi}{\bf \pi}}=\frac{1}{p^2+\lambda ^2},
\end{split}
\label{eq:2.9}
\end{equation}
where a small mass $\lambda$ is assigned to the propagator of {\bf $\pi$} to regulate the possible infrared~(IR) divergences in the  loop integrals, so that the cancellation of the IR  divergence can be shown explicitly. $\lambda$ is set to be zero whenever possible. For simplicity, we define $m_{\sigma}\equiv \sqrt{2}m$. The Feynman rules of the propagators and the vertices  are shown in Fig.~\ref{fig:vertex}. The Feynman rules of the counter terms are shown in Fig.~\ref{fig:counterterm}.

\subsection{\label{sec:2.2}Susceptibilities}

We are interested in  spectral functions
\begin{equation}
\begin{split}
&\chi_{AB}''({\bf q}, \omega)=\lim _{\epsilon \to 0^+} {\rm Im}(\chi _{AB}({\bf q}, \omega +i \epsilon)),
\end{split}
\label{eq:2.10}
\end{equation}
where $\chi _{AB}({\bf q}, \omega)$ are dynamical  susceptibilities  defined in $D$ dimensions as~\cite{Podolsky1}
\begin{equation}
\begin{split}
&\chi _{AB}(q)=\int d^Dx e^{i q\cdot x}\langle A(x)B(0)\rangle _c ,
\end{split}
\label{eq:2.11}
\end{equation}
where $A,B=\sigma$, ${\bf \pi}$, ${\bf \pi}{\bf \pi}$, etc. The longitudinal susceptibility is $\chi _{\sigma\sigma}$.

The fluctuation can also be parameterized such that     the scalar fluctuation is~$ \left(\Phi ^2-v^2\right)/\sqrt{N}$~\cite{Podolsky1}. We define
\begin{equation}
\begin{split}
&\rho \equiv \frac{1}{v}\left(\Phi ^2-v^2\right)=2\sigma +\frac{\sigma ^2+\pi ^2}{v}
\end{split}
\label{eq:2.12}
\end{equation}
Similar to Eq.~(7) of Ref.~\cite{Podolsky1}, the scalar susceptibility can be written as a sum of longitudinal and cross susceptibilities,
\begin{equation}
\begin{split}
&\chi _{\rho\rho}=4\chi _{\sigma\sigma}+\frac{4}{v}\left(\chi _{\sigma ^2\sigma}+\chi _{\pi^2\sigma}\right)+\frac{1}{v^2}\left(\chi _{\sigma ^2\sigma^2}+\chi _{\pi^2\pi^2}+2\chi _{\sigma ^2\pi^2}\right)
\end{split}
\label{eq:2.13}
\end{equation}

\section{\label{sec:3}Calculation of susceptibilities}

Throughout this paper, we shall only consider zero temperature limit of  $O(2)$ model  in $D=2+1$ dimensions. We use dimensional regulation (DR)~\cite{DR} to regulate the ultraviolet (UV) divergence. In $D=3-\epsilon$ dimensions, we define $N_{\rm UV}$ as
\begin{equation}
\begin{split}
&N_{\rm UV}\equiv \frac{1}{\epsilon}-\gamma _E+\log (4\pi),
\end{split}
\label{eq:3.1}
\end{equation}
where $\gamma _E$ is the Euler constant.

The counter terms should be calculated order by order. In DR in $2+1$ dimensions, the UV divergences show up at 2-loop order,  and both  $\delta _Z$ and $\delta _U$ are UV finite. Since different renormalization conditions lead to different renormalization schemes, we choose to use $\delta _Z=\delta _U=0$ for simplicity. The only nonvanishing  counter term needs  one more renormalization condition. We also use $\langle \sigma \rangle =0$~\cite{Podolsky1,PeskinBook}, which requires that  the 1-particle-irreducible (1PI) tadpole diagrams of $\sigma$  vanish.

\subsection{\label{sec:3.1}1-loop level}

\subsubsection{\label{sec:3.1.1}Counter terms and Goldstone theorem at 1-loop order}

\begin{figure*}
\begin{center}
\includegraphics[scale=0.65]{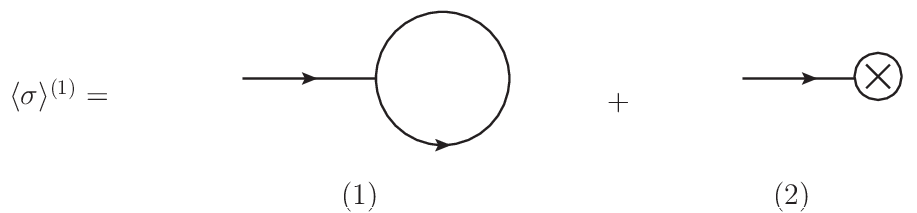}
\caption{The 1PI diagrams of $\langle \sigma \rangle$ at 1-loop order. The massless vacuum bubble diagram   vanishes in DR, and is thus not shown.}
\label{fig:1pitadpole1loop}
\end{center}
\end{figure*}

$\langle \sigma \rangle$ at 1-loop order is denoted as $\langle \sigma \rangle^{(1)}$. The non-vanishing  tadpole diagrams contributing  to $\langle \sigma \rangle^{(1)}$ are shown in Fig.~\ref{fig:1pitadpole1loop}, where the diagram  above label (i)  is denoted as $I_i^{\rm tad\;(1)}$, $i=1,2$. We find
\begin{equation}
\begin{split}
&\langle \sigma \rangle ^{(1)}= I_1^{\rm tad\;(1)}+I_2^{tad\;(1)},\\
&I_1^{\rm tad\;(1)}=3Uvf_a^{(1)},\;\;I_2^{\rm tad\;(1)}=\delta _m^{(1)}v,\\
\end{split}
\label{eq:3.1.1}
\end{equation}
where  $\delta _m^{(1)}$ denotes the counter term $\delta _m$ at 1-loop order, $f_a^{(1)}$ is given in Eq.~(\ref{eq:ap.3}). $\langle \sigma \rangle ^{(1)}=0$ leads to
\begin{equation}
\begin{split}
&\delta _m^{(1)}=-3U\frac{m_{\sigma}}{4\pi}.\\
\end{split}
\label{eq:3.1.2}
\end{equation}

\begin{figure*}
\begin{center}
\includegraphics[scale=0.65]{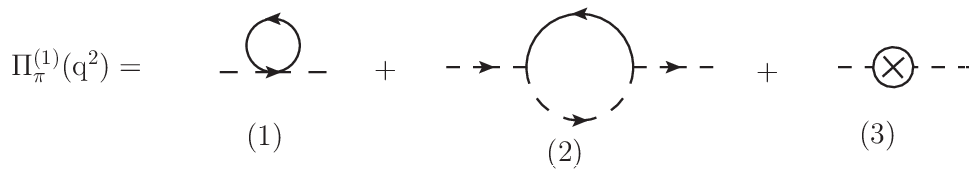}
\caption{The 1PI diagrams of $\Pi _{\pi}$ at the order of  $\mathcal{O}(U)$ , denoted as  $\Pi _{\pi}^{(1)}(q^2)$. The massless vacuum bubble    vanishes in DR, and is thus not shown.}
\label{fig:1pipi1loop}
\end{center}
\end{figure*}

We can use this result to confirm  the Goldstone theorem. The 1PI diagrams contributing  to self energy $\Pi_{\pi}$ at 1-loop are shown in Fig.~\ref{fig:1pipi1loop}. The diagram above  label (i) is denoted as $I_i^{\pi\;(1)}$, $i=1,2,3$. The Goldstone theorem requires $\Pi _{\pi}(q^2=0)=0$, consequently

\begin{equation}
\begin{split}
&\Pi _{\pi}^{(1)}(q^2)=\sum _{i=1,2,3} I_{i}^{\pi\;(1)}(q^2),\\
&I_1^{\pi\;(1)}(q^2)=Uf_a^{(1)},\;\;I_2^{\pi\;(1)}(q^2=0)=4U^2v^2\times f_e^{(1)}(q^2=0),\;\;I_3^{\pi\;(1)}(q^2)=\delta _m^{(1)},\\
\end{split}
\label{eq:3.1.3}
\end{equation}
where $\delta _m^{(1)}$ is given in Eq.~(\ref{eq:3.1.2}), $f_a^{(1)}$ and $f_e^{(1)}(q^2=0)$ are given in Eqs.~(\ref{eq:ap.3}) and (\ref{eq:ap.6}). At 1-loop order, $\Pi _{\pi}^{(1)}(q^2=0)=0$,  as expected.

\subsubsection{\label{sec:3.1.2} 1PI contribution to longitudinal susceptibility at \texorpdfstring{$\mathcal{O}(U)$}{O(U)} order}

\begin{figure*}
\begin{center}
\includegraphics[scale=0.65]{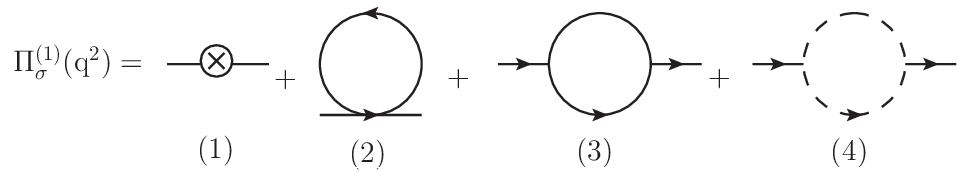}
\caption{The 1PI diagrams of $\Pi _{\pi}$ at $\mathcal{O}(U)$ order. The massless vacuum bubble vanishes  in DR, and is thus not shown.}
\label{fig:1pisigma1loop}
\end{center}
\end{figure*}

The 1PI self energy of $\sigma$ is denoted as $\Pi _{\sigma}$. The 1PI diagrams contributing  to $\Pi_{\sigma}$ at 1-loop are shown in Fig.~\ref{fig:1pisigma1loop}. The diagram above label  (i) is  denoted as $I_i^{\sigma\;(1)}$, $i=1,2,3,4$. We find
\begin{equation}
\begin{split}
&\Pi _{\sigma}^{(1)}(q^2)=\sum _{i=1}^{4} I_{i}^{\sigma\;(1)}(q^2),\\
&I_1^{\sigma\;(1)}(q^2)=\delta _m^{(1)},\;\;\;\;\;I_2^{\sigma\;(1)}(q^2)=3Uf_a^{(1)},\\
&I_3^{\sigma\;(1)}(q^2)=18U^2v^2f_c^{(1)}(q^2),\;\;I_4^{\sigma\;(1)}(q^2)=2U^2v^2f_d^{(1)}(q^2),\\
\end{split}
\label{eq:3.1.4}
\end{equation}
where $\delta _m^{(1)}$ is given in Eq.~(\ref{eq:3.1.2}), $f_a^{(1)}$, $f_c^{(1)}(q^2)$ and $f_d^{(1)}(q^2)$ are given in Eqs.~(\ref{eq:ap.3}) and (\ref{eq:ap.5}). As a result, we find both $\Pi _{\pi}^{(1)}$ and $\Pi _{\pi}^{(1)}$ are UV  and IR finite.
\subsubsection{\label{sec:3.1.3} 1PI contribution to cross-susceptibilities}

\begin{figure*}
\begin{center}
\includegraphics[scale=0.7]{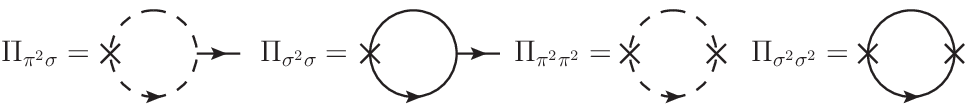}
\caption{The 1PI diagrams of $\Pi _{\pi^2\sigma}$, $\Pi _{\pi^2\pi^2}$, $\Pi _{\sigma^2\sigma}$ and $\Pi _{\sigma^2\sigma^2}$ at 1-loop order. There is no 1-loop contribution to $\chi _{\sigma^2\pi^2}$. The cross means that the propagators are disconnected. For $\Pi _{\pi^2\pi^2}^{(1)}$, there is another diagram obtained  by exchanging   $\pi$'s in the final states. Similarly there is another diagram for $\Pi _{\sigma^2\sigma^2}^{(1)}$.}
\label{fig:cross1loop}
\end{center}
\end{figure*}

The 1PI contributions to cross-susceptibilities $\chi _{\pi^2\sigma}$, $\chi _{\pi^2\pi^2}$, $\chi _{\sigma^2\sigma}$ and $\chi _{\sigma^2\sigma^2}$ are denoted as $\Pi _{\pi^2\sigma}^{(1)}$, $\Pi _{\pi^2\pi^2}^{(1)}$, $\Pi _{\sigma^2\sigma}^{(1)}$ and $\Pi _{\sigma^2\sigma^2}^{(1)}$,  respectively, as  shown in Fig.~\ref{fig:cross1loop}.

The results of $\Pi _{\pi^2\sigma}$, $\Pi _{\pi^2\pi^2}$, $\Pi _{\sigma^2\sigma}$ and $\Pi _{\sigma^2\sigma^2}$ can be written as
\begin{equation}
\begin{split}
&\Pi _{\pi^2\sigma}^{(1)}(q^2)=-2Uv f_c^{(1)}(q^2),\;\;\Pi _{\sigma^2\sigma}^{(1)}(q^2)=-6Uv f_d^{(1)}(q^2)\\
&\Pi _{\pi^2\pi^2}^{(1)}(q^2)=2f_c^{(1)}(q^2),\;\;\Pi _{\sigma^2\sigma^2}^{(1)}(q^2)=2f_d^{(1)}(q^2).\\
\end{split}
\label{eq:3.1.5}
\end{equation}

\subsection{\label{sec:3.2}2-loop level}

\subsubsection{\label{sec:3.2.1}Counter terms at 2-loop order}

\begin{figure*}
\begin{center}
\includegraphics[scale=0.7]{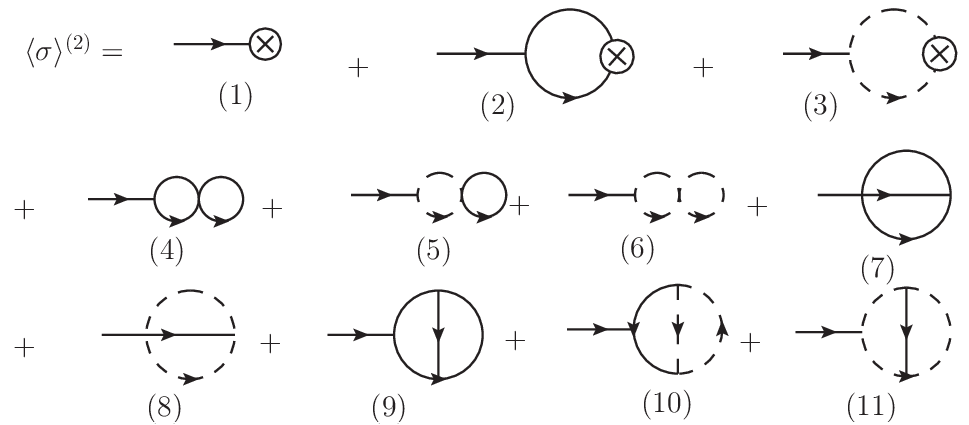}
\caption{The 1PI diagrams of $\langle \sigma \rangle$ at $\mathcal{O}(U^2)$ order.}
\label{fig:1pitadpole2loop}
\end{center}
\end{figure*}

The result of $\langle \sigma \rangle$ at 2-loop order is denoted as $\langle \sigma \rangle^{(2)}$. The nonvanishing  tadpole diagrams contributing  to $\langle \sigma \rangle^{(2)}$ are shown in Fig.~\ref{fig:1pitadpole2loop}. The diagram above label (i) is denoted as $I_i^{\rm tad\;(2)}$, $i=1,\cdots,11$. We find
\begin{equation}
\begin{split}
&\langle \sigma \rangle ^{(2)}=\sum _i I_i^{\rm tad \; (2)},\\
&I_1^{\rm tad\;(2)}=\delta _m^{(2)}v,\\
&I_2^{\rm tad\;(2)}=-3Uv\delta _m^{(1)}f_c^{(1)}(q^2=0),\\
&I_3^{\rm tad\;(2)}=-Uv\delta _m^{(1)}f_d^{(1)}(q^2=0),\\
&I_4^{\rm tad\;(2)}=-9U^2vf_a^{(1)}f_c^{(1)}(q^2=0),\\
&I_5^{\rm tad\;(2)}=-U^2vf_a^{(1)}f_d^{(1)}(q^2=0),\\
\end{split}
\quad
\begin{split}
&I_6^{\rm tad\;(2)}=-3U^2vf_b^{(1)}f_d^{(1)}(q^2=0),\\
&I_7^{\rm tad\;(2)}=6U^2vf_d^{(2)}(q^2=0),\\
&I_8^{\rm tad\;(2)}=2U^2vf_b^{(2)}(q^2=0),\\
&I_9^{\rm tad\;(2)}=-54U^3v^3f_j^{(2)}(q^2=0),\\
&I_{10}^{\rm tad\;(2)}=-6U^3v^3f_g^{(2)}(q^2=0),\\
&I_{11}^{\rm tad\;(2)}=-4U^3v^3f_e^{(2)}(q^2=0),\\
\end{split}
\label{eq:3.2.1}
\end{equation}
where $f_a^{(1)}$ and $f_b^{(1)}$ are given in Eq.~(\ref{eq:ap.3}). Here  $f_b^{(1)}=-\lambda/4\pi$. $f_c^{(1)}(q^2=0)$ and $f_d^{(1)}(q^2=0)$ are given in Eq.~(\ref{eq:ap.6}). $f_b^{(2)}(q^2=0)$ and $f_d^{(2)}(q^2=0)$ are given in Eq.~(\ref{eq:ap.18}). $f_e^{(2)}(q^2=0)$, $f_g^{(2)}(q^2=0)$ and $f_j^{(2)}(q^2=0)$ are given in Eq.~(\ref{eq:ap.33}).

$\langle \sigma \rangle=0$ requires  $\langle \sigma \rangle^{(1)}=0$ and $\langle \sigma \rangle^{(2)}=0$.  We find
\begin{equation}
\begin{split}
&\delta _m^{(2)}=\frac{3U^2}{2(4\pi)^2}-\frac{U^2}{2(4\pi)^2}\left(8N_{UV}+8\log \frac{\mu^2}{m_{\sigma}^2}-12\log(3)\right).\\
\end{split}
\label{eq:3.2.2}
\end{equation}
Although there exist IR divergences in $I_{3}^{\rm tad\;(2)}$, $I_{5}^{\rm tad\;(2)}$ and $I_{11}^{\rm tad\;(2)}$, those IR divergences are cancelled, consequently   $\delta _m^{(2)}$ is IR finite. However, $\delta _m^{(2)}$ is UV divergent. We will show that the UV  divergence in $\delta _m^{(2)}$ cancels the UV divergence in 1PI self energies  of $\pi$ and $\sigma$.

\subsubsection{\label{sec:3.2.2}Cancellation of divergences and Goldstone theorem at 2-loop level}

\begin{figure*}
\begin{center}
\includegraphics[scale=0.8]{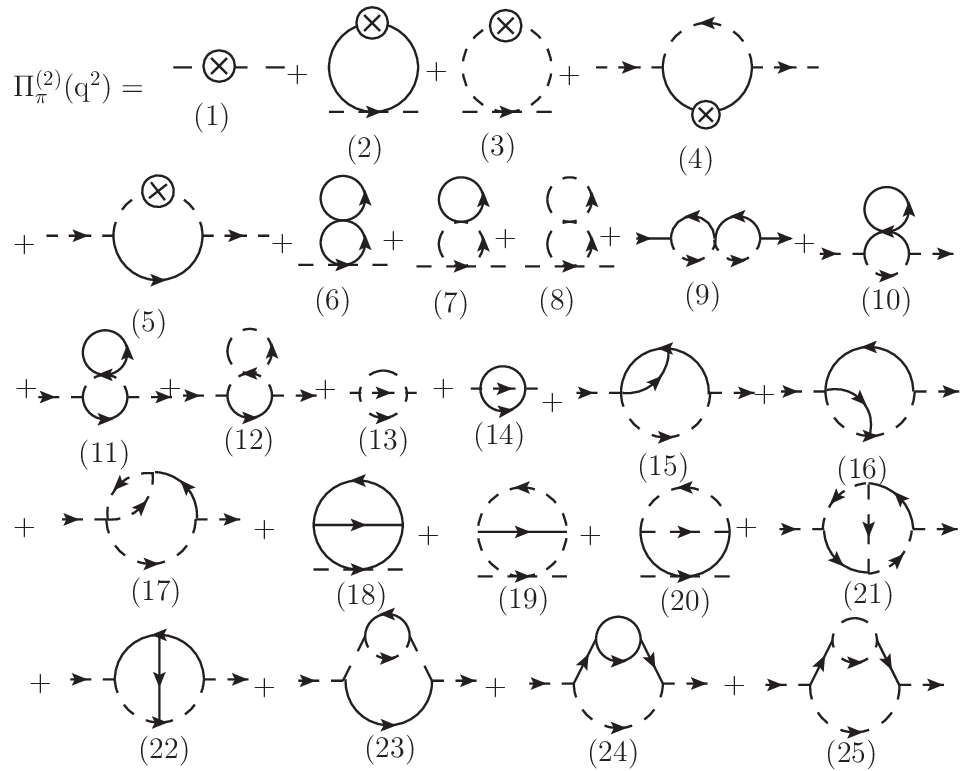}
\caption{The 1PI diagrams of $\Pi _{\pi}$ at 2-loop order. There are three other diagrams that are the horizontal mirrors of diagrams (15), (16) and (17), respectively. For brevity, they are not shown. They are included in $I_{15}^{\pi\;(2)}$, $I_{16}^{\pi\;(2)}$ and $I_{17}^{\pi\;(2)}$, respectively.}
\label{fig:1pipi2loop}
\end{center}
\end{figure*}

\begin{figure*}
\begin{center}
\includegraphics[scale=0.8]{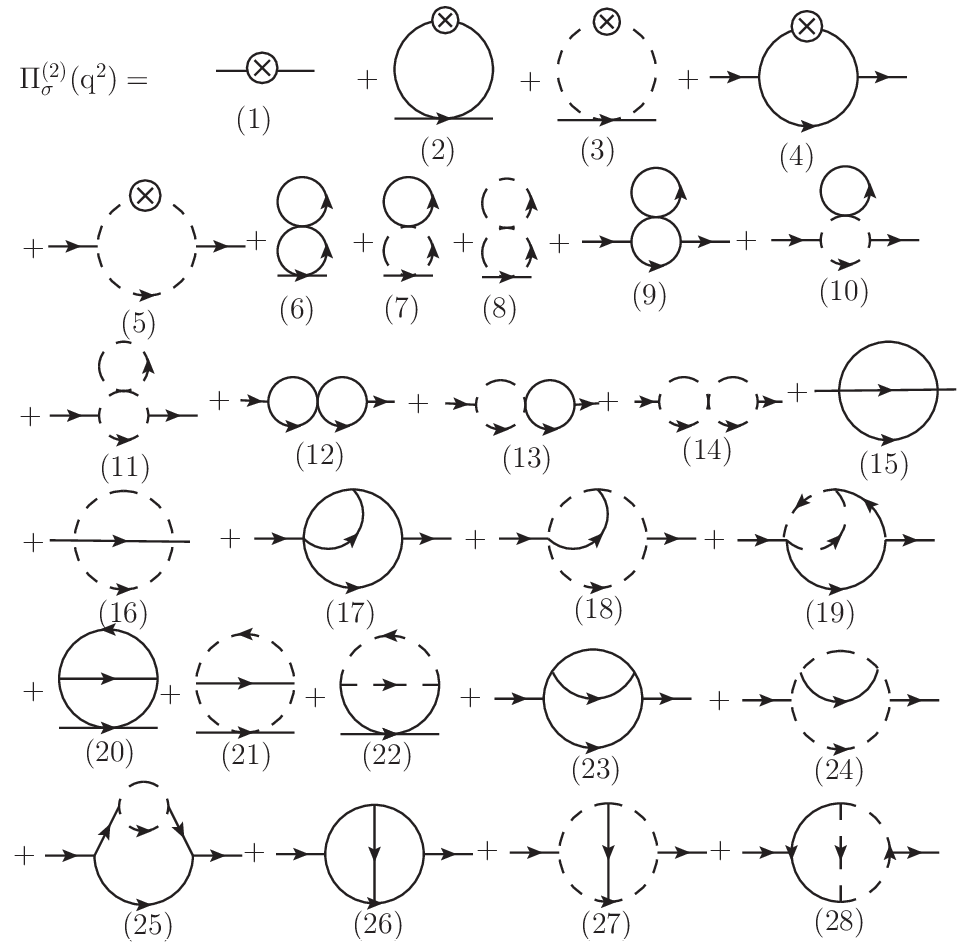}
\caption{The 1PI diagrams of $\Pi _{\sigma}$ at 2-loop order. There are five other diagrams that  are the horizontal mirrors of diagrams (13), (17), (18), (19) and (28), respectively. For brevity, they are not shown. They are   included in $I_{13}^{\sigma\;(2)}$, $I_{17}^{\sigma\;(2)}$, $I_{18}^{\sigma\;(2)}$, $I_{19}^{\sigma\;(2)}$ and $I_{28}^{\sigma\;(2)}$, respectively.}
\label{fig:1pisigma2loop}
\end{center}
\end{figure*}

The diagrams contributing  to $\Pi _{\pi}$ and $\Pi _{\sigma}$ at 2-loop level are shown in Fig.~\ref{fig:1pipi2loop} and Fig.~\ref{fig:1pisigma2loop}. The 1PI self-energies  of $\pi$ and $\sigma$ at 2-loop level are denoted as $\Pi _{\pi}^{(2)}$ and $\Pi _{\sigma}^{(2)}$. In these two figures, the diagrams above the label (i) is denoted as $I_i^{\pi\;(2)}$ and  $I_i^{\sigma\;(2)}$.   One obtains
\begin{equation}
\begin{split}
&\Pi _{\pi}^{(2)}(q^2)=\sum _{i=1}^{25}I_i^{\pi\;(2)},\\
&I_1^{\pi\;(2)}=\delta _m^{(2)},\\
&I_2^{\pi\;(2)}=-U \times f_c^{(1)}(q^2=0)\times \delta _m^{(1)},\\
&I_3^{\pi\;(2)}=-3U \times f_d^{(1)}(q^2=0)\times \delta _m^{(1)},\\
&I_4^{\pi\;(2)}(q^2)=4U^2v^2 \times f_i^{(1)}(q^2)\times \delta _m^{(1)},\\
&I_5^{\pi\;(2)}(q^2)=4U^2v^2 \times f_h^{(1)}(q^2)\times \delta _m^{(1)},\\
&I_6^{\pi\;(2)}=3U^2 \times f_c^{(1)}(q^2=0)\times f_a^{(1)},\\
&I_7^{\pi\;(2)}=3U^2 \times f_d^{(1)}(q^2=0)\times f_a^{(1)},\\
&I_8^{\pi\;(2)}=9U^2 \times f_d^{(1)}(q^2=0)\times f_b^{(1)},\\
&I_9^{\pi\;(2)}(q^2)=-8U^3v^2 \times f_e^{(1)}(q^2)\times f_e^{(1)}(q^2),\\
&I_{10}^{\pi\;(2)}(q^2)=-12U^3v^2 \times f_i^{(1)}(q^2)\times f_a^{(1)},\\
&I_{11}^{\pi\;(2)}(q^2)=-4U^3v^2 \times f_h^{(1)}(q^2)\times f_a^{(1)},\\
&I_{12}^{\pi\;(2)}(q^2)=-12U^3v^2 \times f_h^{(1)}(q^2)\times f_b^{(1)},\\
\end{split}
\quad
\begin{split}
&I_{13}^{\pi\;(2)}(q^2)=6U^2 \times f_a^{(2)}(q^2),\\
&I_{14}^{\pi\;(2)}(q^2)=2U^2 \times f_c^{(2)}(q^2),\\
&I_{15}^{\pi\;(2)}(q^2)=-24U^3v^2 \times f_i^{(2)}(q^2),\\
&I_{16}^{\pi\;(2)}(q^2)=-16U^3v^2 \times f_h^{(2)}(q^2),\\
&I_{17}^{\pi\;(2)}(q^2)=-24U^3v^2 \times f_f^{(2)}(q^2),\\
&I_{18}^{\pi\;(2)}=-18U^3v^2 \times f_j^{(2)}(q^2=0),\\
&I_{19}^{\pi\;(2)}=-12U^3v^2 \times f_e^{(2)}(q^2=0),\\
&I_{20}^{\pi\;(2)}=-2U^3v^2 \times f_g^{(2)}(q^2=0),\\
&I_{21}^{\pi\;(2)}(q^2)=16U^4v^4 \times f_v^{(2)}(q^2),\\
&I_{22}^{\pi\;(2)}(q^2)=48U^4v^4 \times f_u^{(2)}(q^2),\\
&I_{23}^{\pi\;(2)}(q^2)=16U^4v^4 \times f_k^{(2)}(q^2),\\
&I_{24}^{\pi\;(2)}(q^2)=72U^4v^4 \times f_l^{(2)}(q^2),\\
&I_{25}^{\pi\;(2)}(q^2)=8U^4v^4 \times f_m^{(2)}(q^2),\\
\end{split}
\label{eq:3.2.3}
\end{equation}
and
\begin{equation}
\begin{split}
&\Pi_{\sigma}^{(2)}(q^2)=\sum _{i=1}^{28}I_i^{\sigma\;(2)},\\
&I_1^{\sigma\;(2)}=\delta _m^{(2)},\\
&I_2^{\sigma\;(2)}=-3U\times f_c^{(1)}(q^2=0)\times \delta _m^{(1)},\\
&I_3^{\sigma\;(2)}=-U\times f_d^{(1)}(q^2=0)\times \delta _m^{(1)},\\
&I_4^{\sigma\;(2)}(q^2)=36U^2v^2\times f_f^{(1)}(q^2)\times \delta _m^{(1)},\\
&I_5^{\sigma\;(2)}(q^2)=4U^2v^2\times f_g^{(1)}(q^2)\times \delta _m^{(1)},\\
&I_6^{\sigma\;(2)}=9U^2\times f_c^{(1)}(q^2=0)\times f_a^{(1)},\\
&I_7^{\sigma\;(2)}=U^2\times f_d^{(1)}(q^2=0)\times f_a^{(1)},\\
&I_8^{\sigma\;(2)}=3U^2\times f_d^{(1)}(q^2=0)\times f_b^{(1)},\\
&I_9^{\sigma\;(2)}(q^2)=-108U^3v^2\times f_f^{(1)}(q^2)\times f_a^{(1)},\\
&I_{10}^{\sigma\;(2)}(q^2)=-4U^3v^2\times f_g^{(1)}(q^2)\times f_a^{(1)},\\
&I_{11}^{\sigma\;(2)}(q^2)=-12U^3v^2\times f_g^{(1)}(q^2)\times f_b^{(1)},\\
&I_{12}^{\sigma\;(2)}(q^2)=-54U^3v^2\times f_c^{(1)}(q^2)\times f_c^{(1)}(q^2),\\
&I_{13}^{\sigma\;(2)}(q^2)=-12U^3v^2\times f_c^{(1)}(q^2)\times f_d^{(1)}(q^2),\\
&I_{14}^{\sigma\;(2)}(q^2)=-6U^3v^2\times f_d^{(1)}(q^2)\times f_d^{(1)}(q^2),\\
\end{split}
\quad
\begin{split}
&I_{15}^{\sigma\;(2)}(q^2)=6U^2\times f_d^{(2)}(q^2),\\
&I_{16}^{\sigma\;(2)}(q^2)=2U^2\times f_b^{(2)}(q^2),\\
&I_{17}^{\sigma\;(2)}(q^2)=-216U^3v^2\times f_j^{(2)}(q^2),\\
&I_{18}^{\sigma\;(2)}(q^2)=-16U^3v^2\times f_e^{(2)}(q^2),\\
&I_{19}^{\sigma\;(2)}(q^2)=-24U^3v^2\times f_g^{(2)}(q^2),\\
&I_{20}^{\sigma\;(2)}=-54U^3v^2\times f_j^{(2)}(q^2=0),\\
&I_{21}^{\sigma\;(2)}=-4U^3v^2\times f_e^{(2)}(q^2=0),\\
&I_{22}^{\sigma\;(2)}=-6U^3v^2\times f_g^{(2)}(q^2=0),\\
&I_{23}^{\sigma\;(2)}(q^2)=648U^4v^4\times f_o^{(2)}(q^2),\\
&I_{24}^{\sigma\;(2)}(q^2)=16U^4v^4\times f_n^{(2)}(q^2),\\
&I_{25}^{\sigma\;(2)}(q^2)=72U^4v^4\times f_p^{(2)}(q^2),\\
&I_{26}^{\sigma\;(2)}(q^2)=648U^4v^4\times f_q^{(2)}(q^2),\\
&I_{27}^{\sigma\;(2)}(q^2)=8U^4v^4\times f_r^{(2)}(q^2),\\
&I_{28}^{\sigma\;(2)}(q^2)=48U^4v^4\times f_s^{(2)}(q^2),\\
\end{split}
\label{eq:3.2.4}
\end{equation}
where $\delta _m^{(1)}$ and $\delta _m^{(2)}$ are given in Eqs.~(\ref{eq:3.1.2}) and (\ref{eq:3.2.2}). The functions $f^{(1)}$ and $f^{(2)}$ are given in Appendix~\ref{sec:a}.

The UV divergence of $\Pi _{\pi}^{(2)}$ appears in $I_1^{\pi\;(2)}$, $I_{13}^{\pi\;(2)}$ and $I_{14}^{\pi\;(2)}$,  and are
\begin{equation}
\begin{split}
&I_{1\;\rm UV}^{\pi\;(2)}=-\frac{U^2}{2(4\pi)^2}\left(8N_{UV}\right),\;\;I_{13\;\rm UV}^{\pi\;(2)}=\frac{6U^2}{2(4\pi)^2}N_{UV},\;\;I_{14\;\rm UV}^{\pi\;(2)}=\frac{2U^2}{2(4\pi)^2}N_{UV}.\\
\end{split}
\label{eq:3.2.5}
\end{equation}
The UV divergences of $\Pi _{\sigma}^{(2)}$ appear in $I_1^{\sigma\;(2)}$, $I_{15}^{\sigma\;(2)}$ and $I_{16}^{\sigma\;(2)}$, and are
\begin{equation}
\begin{split}
&I_{1\;\rm UV}^{\sigma\;(2)}=-\frac{U^2}{2(4\pi)^2}\left(8N_{UV}\right),\;\;I_{15\;\rm UV}^{\sigma\;(2)}=\frac{6U^2}{2(4\pi)^2}N_{UV},\;\;I_{16\;\rm UV}^{\sigma\;(2)}=\frac{2U^2}{2(4\pi)^2}N_{UV}.\\
\end{split}
\label{eq:3.2.6}
\end{equation}

As a result, we find that all the UV divergences in $\Pi _{\pi}^{(2)}$ and $\Pi _{\sigma}^{(2)}$ are cancelled by the UV divergence in   $\delta _m^{(2)}$, and the term $\log (\mu^2)$ is cancelled as well.

We can also show the cancellation of IR divergences explicitly. The IR divergences   in $\Pi _{\pi}^{(2)}$ are
\begin{equation}
\begin{split}
&I_{3\; \rm IR}^{\pi\;(2)}= \frac{9U^2m_{\sigma}}{2(4\pi)^2 \lambda},\;\;I_{5\; \rm IR}^{\pi\;(2)}(q^2)=\frac{-6U^2m_{\sigma}}{2(4\pi)^2 \lambda (1+t)},\\
&I_{7\; \rm IR}^{\pi\;(2)}= \frac{-3U^2m_{\sigma}}{2(4\pi)^2 \lambda},\;\;I_{11\; \rm IR}^{\pi\;(2)}(q^2)=\frac{2U^2m_{\sigma}}{2(4\pi)^2 \lambda (1+t)},\\
&I_{19\; \rm IR}^{\pi\;(2)}=\frac{-6U^2m_{\sigma}}{2(4\pi)^2 \lambda},\;\;I_{23\; \rm IR}^{\pi\;(2)}=\frac{4U^2m_{\sigma}}{2(4\pi)^2 \lambda (1+t)},\\
\end{split}
\label{eq:3.2.7}
\end{equation}
where $t\equiv q^2/m_{\sigma}^2$. Those IR divergences cancel each other explicitly.

The IR divergences in $\Pi _{\sigma}^{(2)}$ are
\begin{equation}
\begin{split}
&I_{3\; \rm IR}^{\sigma\;(2)}=3U^2\frac{m_{\sigma}}{2(4\pi)^2 \lambda},\;\;I_{7\; \rm IR}^{\sigma\;(2)}=-U^2\frac{m_{\sigma}}{2(4\pi)^2 \lambda},\\
&I_{5\; \rm IR}^{\sigma\;(2)}(q^2)=\frac{-12U^3v^2m_{\sigma}}{2(4\pi)^2\lambda q^2},\;\;I_{10\; \rm IR}^{\sigma\;(2)}(q^2)=\frac{4U^3v^2m_{\sigma}}{2(4\pi)^2\lambda q^2},\\
&I_{19\; \rm IR}^{\sigma\;(2)}=-2U^2\frac{m_{\sigma}}{2(4\pi)^2\lambda},\;\;I_{24\; \rm IR}^{\sigma\;(2)}(q^2)=\frac{8U^3v^2m_{\sigma}}{2(4\pi)^2 \lambda q^2}.\\
\end{split}
\label{eq:3.2.8}
\end{equation}

Similarly, the IR divergences in $\Pi _{\sigma}^{(2)}$ cancel each other explicitly. As result, both $\Pi _{\pi}^{(2)}$ and $\Pi _{\sigma}^{(2)}$ are UV finite and IR finite.

The Goldstone theorem also requires $\Pi _{\pi}^{(2)}(q^2=0)=0$. When $q^2=0$, using Eqs.~(\ref{eq:ap.6}), (\ref{eq:ap.18}), (\ref{eq:ap.33}), (\ref{eq:ap.48}) and (\ref{eq:ap.69}), we find
\begin{equation}
\begin{split}
&I_1^{\pi\;(2)}=\delta _m^{(2)},\\
&I_2^{\pi\;(2)}=3U^2 \frac{1}{2(4\pi)^2},\\
&I_3^{\pi\;(2)}=9U^2 \frac{m_{\sigma}}{2(4\pi)^2\lambda},\\
&I_4^{\pi\;(2)}=-12U^3v^2 \frac{1}{2(4\pi)^2 m_{\sigma}^2},\\
&I_5^{\pi\;(2)}=-12U^3v^2\frac{1}{2(4\pi)^2 m_{\sigma}^2}\left(\frac{m_{\sigma}}{\lambda}-2\right),\\
&I_6^{\pi\;(2)}=-3U^2 \frac{1}{2(4\pi)^2},\\
&I_7^{\pi\;(2)}=-3U^2 \frac{m_{\sigma}}{2(4\pi)^2\lambda},\\
&I_8^{\pi\;(2)}=-9U^2 \frac{1}{2(4\pi)^2},\\
&I_9^{\pi\;(2)}=-16U^3v^2 \frac{1}{2(4\pi)^2m_{\sigma}^2},\\
&I_{10}^{\pi\;(2)}=12U^3v^2 \frac{1}{2(4\pi)^2 m_{\sigma}^2},\\
&I_{11}^{\pi\;(2)}=4U^3v^2\frac{1}{2(4\pi)^2 m_{\sigma}^2}\left(\frac{m_{\sigma}}{\lambda}-2\right),\\
&I_{12}^{\pi\;(2)}=12U^3v^2 \frac{1}{2(4\pi)^2 m_{\sigma}^2},\\
&I_{13}^{\pi\;(2)}=6U^2 \frac{1}{2(4\pi)^2}\left(N_{UV}+1+\log \frac{\mu^2}{9\lambda^2}\right),\\
\end{split}
\quad
\begin{split}
&I_{14}^{\pi\;(2)}=2U^2 \frac{1}{2(4\pi)^2}\left(N_{UV}+1+\log \frac{\mu^2}{4m_{\sigma}^2}\right),\\
&I_{15}^{\pi\;(2)}=-24U^3v^2 \frac{1}{2(4\pi)^2m_{\sigma}^2}\log\left(\frac{9}{4}\right),\\
&I_{16}^{\pi\;(2)}=-16U^3v^2 \frac{1}{2(4\pi)^2m_{\sigma}^2}\log(4),\\
&I_{17}^{\pi\;(2)}=24U^3v^2 \frac{1}{2(4\pi)^2m_{\sigma}^2}\log \left(9\frac{\lambda^2}{m_{\sigma}^2}\right),\\
&I_{18}^{\pi\;(2)}=-6U^3v^2 \frac{1}{2(4\pi)^2m_{\sigma}^2},\\
&I_{19}^{\pi\;(2)}=-12U^3v^2\frac{1}{2(4\pi)^2m_{\sigma}^2}\left(\frac{m_{\sigma}}{\lambda}-2\right),\\
&I_{20}^{\pi\;(2)}=-2U^3v^2 \frac{1}{2(4\pi)^2m_{\sigma}^2},\\
&I_{21}^{\pi\;(2)}=-16U^4v^4 \frac{1}{2(4\pi)^3m_{\sigma}^4}\log\left(\frac{36\lambda^2}{m_{\sigma}^2}\right),\\
&I_{22}^{\pi\;(2)}=96U^4v^4 \frac{1}{2(4\pi)^2m_{\sigma}^4}\log (\frac{4}{3}),\\
&I_{23}^{\pi\;(2)}=16U^4v^4 \frac{1}{2(4\pi)^2m_{\sigma}^4}\left(\frac{m_{\sigma}}{\lambda}-2-\log (4)\right),\\
&I_{24}^{\pi\;(2)}=72U^4v^4 \frac{1}{2(4\pi)^2m_{\sigma}^4}\left(2 \log \left(\frac{3}{2}\right)-\frac{1}{3}\right),\\
&I_{25}^{\pi\;(2)}=8U^4v^4 \frac{1}{2(4\pi)^2m_{\sigma}^4}\left(-1-\log\left(\frac{9\lambda^2}{m_{\sigma}^2}\right)\right).\\
\end{split}
\label{eq:3.2.9}
\end{equation}

As a result, we find $\Pi _{\pi}^{(2)}(q^2=0)=0$.

\subsubsection{\label{sec:3.2.3}1PI contributions to Cross-Susceptibilities}

\begin{figure*}
\begin{center}
\includegraphics[scale=0.8]{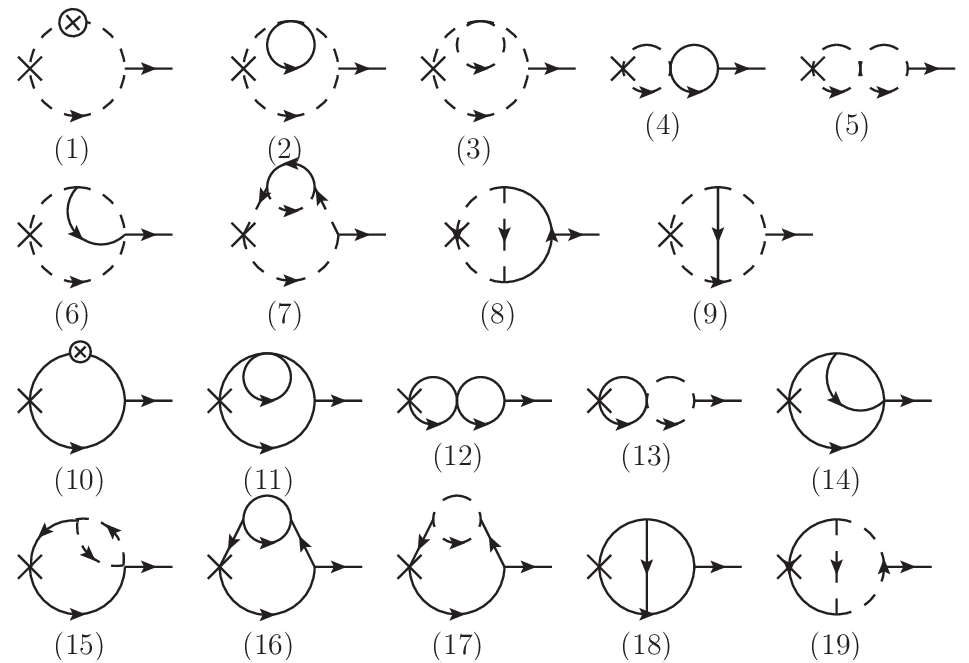}
\caption{Contributions to $\Pi _{\pi^2\sigma}^{(2)}$ and $\Pi _{\sigma^2\sigma}^{(2)}$.  For brevity,  the exchange of initial states   is not drawn. }
\label{fig:cross2loop1}
\end{center}
\end{figure*}

\begin{figure*}
\begin{center}
\includegraphics[scale=0.8]{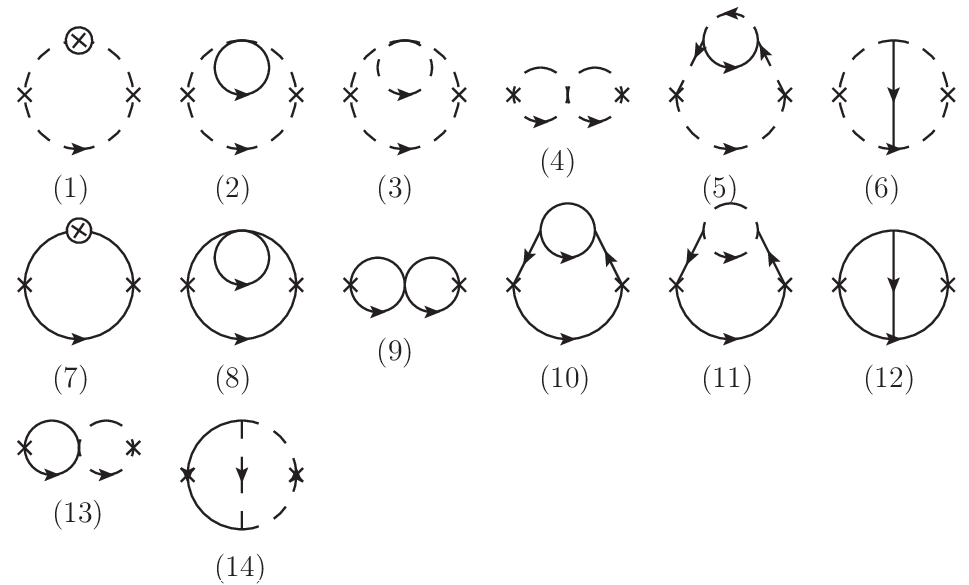}
\caption{Contributions to $\Pi _{\pi^2\pi^2}^{(2)}$, $\Pi _{\sigma^2\sigma^2}^{(2)}$ and $\Pi _{\sigma^2\pi^2}^{(2)}$.   For brevity, the exchanges of final states and initial states are not drawn.   }
\label{fig:cross2loop2}
\end{center}
\end{figure*}

The 1PI contributions to $\Pi _{\pi^2\sigma}$, $\Pi _{\sigma^2\sigma}$, $\Pi _{\pi^2\pi^2}$, $\Pi _{\sigma^2\sigma^2}$ and $\Pi _{\sigma^2\pi^2}$ at 2-loop level are shown in Fig.~\ref{fig:cross2loop1} and Fig.~\ref{fig:cross2loop2}. The diagrams above the label  (i) is denoted as $I_{i}^{A^2\sigma\;(2)}(q^2)$  and  $I_{i}^{A^2B^2\;(2)}(q^2)$, respectively.

One can see from Eq.~(\ref{eq:2.13}) that the cross-susceptibilities always appear in $\chi _{\rho\rho}$ as sums, so it is convenient to define
\begin{equation}
\begin{split}
&\Pi _{\rm A^2\sigma}^{(2)}=\Pi _{\pi^2\sigma}^{(2)}+\Pi _{\sigma^2\sigma}^{(2)},\;\;\Pi _{\rm A^2B^2}^{(2)}=\Pi _{\pi^2\pi^2}^{(2)}+\Pi _{\sigma^2\sigma^2}^{(2)}+2\Pi _{\sigma^2\pi^2}^{(2)}.\\
\end{split}
\label{eq:3.2.10}
\end{equation}
They can be written as
\begin{equation}
\begin{split}
&\Pi _{\pi^2\sigma}^{(2)}(q^2)=\sum _{i=1}^{9}I_{i}^{A^2\sigma\;(2)}(q^2),\\
&\Pi _{\sigma^2\sigma}^{(2)}(q^2)=\sum _{i=10}^{19}I_{i}^{A^2\sigma\;(2)}(q^2),\\
&I_{1}^{A^2\sigma\;(2)}(q^2)=-4Uv\delta _m^{(1)}f_g^{(1)}(q^2),\\
&I_{2}^{A^2\sigma\;(2)}(q^2)=4U^2vf_a^{(1)}f_g^{(1)}(q^2),\\
&I_{3}^{A^2\sigma\;(2)}(q^2)=12U^2vf_b^{(1)}f_g^{(1)}(q^2),\\
&I_{4}^{A^2\sigma\;(2)}(q^2)=6U^2vf_c^{(1)}(q^2)f_d^{(1)}(q^2)\\
&I_{5}^{A^2\sigma\;(2)}(q^2)=6U^2vf_d^{(1)}(q^2)f_d^{(1)}(q^2)\\
&I_{6}^{A^2\sigma\;(2)}(q^2)=8U^2vf_e^{(2)}(q^2)\\
&I_{7}^{A^2\sigma\;(2)}(q^2)=-16U^3v^3f_n^{(2)}(q^2)\\
&I_{8}^{A^2\sigma\;(2)}(q^2)=-24U^3v^3f_s^{(2)}(q^2)\\
\end{split}
\quad
\begin{split}
&I_{9}^{A^2\sigma\;(2)}(q^2)=-8U^3v^3f_r^{(2)}(q^2)\\
&I_{10}^{A^2\sigma\;(2)}(q^2)=-12Uv\delta _m^{(1)}f_f^{(1)}(q^2)\\
&I_{11}^{A^2\sigma\;(2)}(q^2)=36U^2vf_a^{(1)}f_f^{(1)}(q^2)\\
&I_{12}^{A^2\sigma\;(2)}(q^2)=18U^2vf_d^{(1)}(q^2)f_d^{(1)}(q^2)\\
&I_{13}^{A^2\sigma\;(2)}(q^2)=2U^2vf_d^{(1)}(q^2)f_c^{(1)}(q^2)\\
&I_{14}^{A^2\sigma\;(2)}(q^2)=36U^2vf_j^{(2)}(q^2)\\
&I_{15}^{A^2\sigma\;(2)}(q^2)=4U^2vf_g^{(2)}(q^2)\\
&I_{16}^{A^2\sigma\;(2)}(q^2)=-216U^3v^3f_o^{(2)}(q^2)\\
&I_{17}^{A^2\sigma\;(2)}(q^2)=-24U^3v^3f_p^{(2)}(q^2)\\
&I_{18}^{A^2\sigma\;(2)}(q^2)=-216U^3v^3f_q^{(2)}(q^2)\\
&I_{19}^{A^2\sigma\;(2)}(q^2)=-8U^3v^3f_s^{(2)}(q^2)\\
\end{split}
\label{eq:3.2.11}
\end{equation}
and
\begin{equation}
\begin{split}
&\Pi _{\pi^2\pi^2}(q^2)=\sum _{i=1}^{6}I_{i}^{A^2B^2\;(2)}(q^2),\\
&\Pi _{\sigma^2\sigma^2}(q^2)=\sum _{i=7}^{12}I_{i}^{A^2B^2\;(2)}(q^2),\\
&\Pi _{\sigma^2\pi^2}(q^2)=I_{13}^{A^2B^2\;(2)}(q^2)+I_{14}^{A^2B^2\;(2)}(q^2),\\
&I_{1}^{A^2B^2\;(2)}(q^2)=4\delta _m^{(1)}f_g^{(1)}(q^2),\\
&I_{2}^{A^2B^2\;(2)}(q^2)=-4Uf_a^{(1)}f_g^{(1)}(q^2),\\
&I_{3}^{A^2B^2\;(2)}(q^2)=-12Uf_b^{(1)}f_g^{(1)}(q^2),\\
&I_{4}^{A^2B^2\;(2)}(q^2)=-6Uf_d^{(1)}(q^2)f_d^{(1)(q^2)},\\
&I_{5}^{A^2B^2\;(2)}(q^2)=16U^2v^2f_n^{(2)}(q^2),\\
\end{split}
\quad
\begin{split}
&I_{6}^{A^2B^2\;(2)}(q^2)=8U^2v^2f_r^{(2)}(q^2),\\
&I_{7}^{A^2B^2\;(2)}(q^2)=4\delta _m^{(1)}f_f^{(1)}(q^2),\\
&I_{8}^{A^2B^2\;(2)}(q^2)=-12Uf_a^{(1)}f_f^{(1)}(q^2),\\
&I_{9}^{A^2B^2\;(2)}(q^2)=-6Uf_c^{(1)}(q^2)f_c^{(1)}(q^2),\\
&I_{10}^{A^2B^2\;(2)}(q^2)=72U^2v^2f_o^{(2)}(q^2),\\
&I_{11}^{A^2B^2\;(2)}(q^2)=8U^2v^2f_p^{(2)}(q^2),\\
&I_{12}^{A^2B^2\;(2)}(q^2)=72U^2v^2f_q^{(2)}(q^2),\\
&I_{13}^{A^2B^2\;(2)}(q^2)=-2Uf_c^{(1)}(q^2)f_d^{(1)}(q^2),\\
&I_{14}^{A^2B^2\;(2)}(q^2)=8U^2v^2f_s^{(2)}(q^2),\\
\end{split}
\label{eq:3.2.12}
\end{equation}
There are also IR divergent contributions in both $\Pi _{\rm cs\;1}^{(2)}$ and $\Pi _{\rm cs\;2}^{(2)}$,
\begin{equation}
\begin{split}
&I_{1\;\rm IR}^{A^2\sigma\;(2)}=12U^2\frac{m_{\sigma}}{2(4\pi)^2\lambda q^2},\;I_{2\;\rm IR}^{A^2\sigma\;(2)}=-4U^2v\frac{m_{\sigma}}{2(4\pi)^2\lambda q^2},\;I_{7\;\rm IR}^{A^2\sigma\;(2)}=-8U^2v\frac{m_{\sigma}}{2(4\pi)^2\lambda q^2}\\
&I_{1\;\rm IR}^{A^2B^2\;(2)}=-12U\frac{m_{\sigma}}{2(4\pi)^2\lambda q^2},\;I_{2\;\rm IR}^{A^2B^2\;(2)}=4U\frac{m_{\sigma}}{2(4\pi)^2\lambda q^2},\;I_{5\;\rm IR}^{A^2B^2\;(2)}=8U\frac{m_{\sigma}}{2(4\pi)^2\lambda q^2}\\
\end{split}
\label{eq:3.2.13}
\end{equation}
They are also cancelled, as expected.

\subsection{\label{sec:3.3}Higher order corrections}

\subsubsection{\label{sec:3.3.1}Contribution of higher orders: Random Phase Approximation~(RPA) like contributions.}

As shown in Fig.~\ref{fig:1pisigma1loop} and Fig.~\ref{fig:1pisigma2loop}, most of the contributions to $\Pi _{\sigma}$ can be classified to two classes.

The contributions $I_{1}^{\sigma\;(1)}$, $I_{2}^{\sigma\;(1)}$, $I_{1}^{\sigma\;(2)}$, $I_{2}^{\sigma\;(2)}$, $I_{3}^{\sigma\;(2)}$, $I_{6}^{\sigma\;(2)}$, $I_{7}^{\sigma\;(2)}$, $I_{8}^{\sigma\;(2)}$, $I_{18}^{\sigma\;(2)}$, $I_{19}^{\sigma\;(2)}$ and $I_{20}^{\sigma\;(2)}$ are of the first class. They are cancelled exactly by the counter terms because the same diagrams can be found in both  Fig.~\ref{fig:1pitadpole1loop} and Fig.~\ref{fig:1pitadpole2loop}. For higher orders, the counter terms also cancel such kind of contributions exactly.

\begin{figure*}
\begin{center}
\includegraphics[scale=0.6]{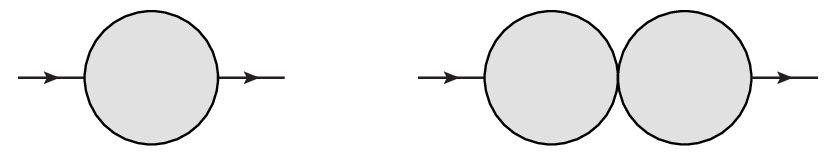}
\caption{The common structure of RPA-like contributions.}
\label{fig:rpaexample2}
\end{center}
\end{figure*}

\begin{figure*}
\begin{center}
\includegraphics[scale=0.6]{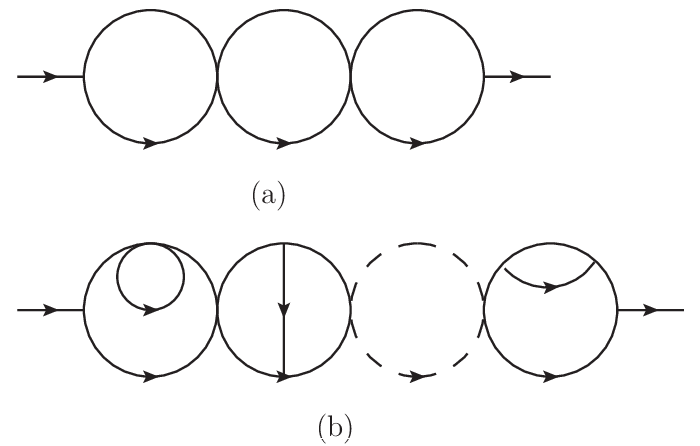}
\caption{Examples of RPA-like contributions at higher orders.}
\label{fig:rpaexample1}
\end{center}
\end{figure*}

The contributions $I_{3}^{\sigma\;(1)}$, $I_{4}^{\sigma\;(1)}$, $I_{4}^{\sigma\;(2)}$, $I_{5}^{\sigma\;(2)}$, $I_{9}^{\sigma\;(2)}$, $I_{10}^{\sigma\;(2)}$, $I_{11}^{\sigma\;(2)}$, $I_{12}^{\sigma\;(2)}$, $I_{13}^{\sigma\;(2)}$, $I_{14}^{\sigma\;(2)}$, $I_{17}^{\sigma\;(2)}$, $I_{18}^{\sigma\;(2)}$, $I_{19}^{\sigma\;(2)}$, $I_{23}^{\sigma\;(2)}$, $I_{24}^{\sigma\;(2)}$, $I_{25}^{\sigma\;(2)}$, $I_{26}^{\sigma\;(2)}$, $I_{27}^{\sigma\;(2)}$ and $I_{28}^{\sigma\;(2)}$ are of another class. They all have the structure  shown in Fig.~\ref{fig:rpaexample2}. For example,  each of $I_{3}^{\sigma\;(1)}$, $I_{4}^{\sigma\;(2)}$ and $I_{9}^{\sigma\;(2)}$ has one circle, while each of  $I_{12}^{\sigma\;(2)}$, $I_{13}^{\sigma\;(2)}$ and $I_{14}^{\sigma\;(2)}$    has  two circles. Other examples of higher order diagrams are shown in Fig.~\ref{fig:rpaexample1}. The diagrams in Fig.~\ref{fig:rpaexample1}.(a) is a 3-loop contribution with 3 circles, while the diagram in Fig.~\ref{fig:rpaexample1}.(b) is a 7-loop contribution with 4 circles.

The loop momenta in  circles are independent of each other. Therefore, similar to RPA, they are   the dominant contributions among all the higher orders contributions of 1PI diagrams~\cite{CondensedMatterBook}. We sum those RPA like contributions to infinite orders with each circle up to 2-loop orders.

Finally, when the RPA contributions of higher orders are included, the $\Pi _{\sigma}$ can be written as
\begin{equation}
\begin{split}
&\Pi _{\sigma}(q^2)=\Pi _{\sigma}^{\rm RPA}(q^2)+\Pi _{\sigma}^{\rm ct}+I_{15}^{\sigma\;(2)}(q^2)+I_{16}^{\sigma\;(2)}(q^2),\\
\end{split}
\label{eq:3.3.1}
\end{equation}
where $\Pi _{\sigma}^{\rm RPA}$ is given in Eq.~(\ref{eq:ap.79}), and $\Pi _{\sigma}^{\rm ct}$ is the sum of the first class of diagrams and can be written as
\begin{equation}
\begin{split}
&\Pi _{\sigma}^{\rm ct}=I_{1}^{\sigma\;(1)}+I_{2}^{\sigma\;(1)}+I_{1}^{\sigma\;(2)}+I_{2}^{\sigma\;(2)}+I_{3}^{\sigma\;(2)}+I_{6}^{\sigma\;(2)}\\
&+I_{7}^{\sigma\;(2)}+I_{8}^{\sigma\;(2)}+I_{18}^{\sigma\;(2)}+I_{19}^{\sigma\;(2)}+I_{20}^{\sigma\;(2)}\\
\end{split}
\label{eq:3.3.2}
\end{equation}

The 1PI diagrams with RPA like contributions to the cross-susceptibilities are denoted as $\Pi _{A^2\sigma}$ and $\Pi _{A^2B^2}$, and are given in Eqs.~(\ref{eq:ap.83}) and  (\ref{eq:ap.86}).

\subsubsection{\label{sec:3.3.2}IPI summation}

\begin{figure*}
\begin{center}
\includegraphics[scale=0.8]{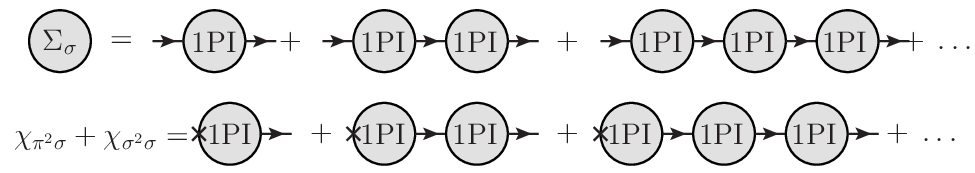}
\caption{Summation of 1PI diagrams.}
\label{fig:1pisum}
\end{center}
\end{figure*}

The self energy of $\sigma$ is denoted as $\Sigma _{\sigma}$, and can be obtained with the help of 1PI self energy $\Pi _{\sigma}$ as shown in Fig.~\ref{fig:1pisum}, and can be written as
\begin{equation}
\begin{split}
&\Sigma _{\sigma}(q^2)=\sum _{n=0}^{\infty}\frac{1}{q^2+m_{\sigma}^2}\left(\Pi _{\sigma}(q^2)\frac{1}{q^2+m_{\sigma}^2}\right)^n=\frac{1}{q^2+m_{\sigma}^2-\Pi _{\sigma}(q^2)},
\end{split}
\label{eq:3.3.3}
\end{equation}
where $\Pi _{\sigma}(q^2)$ is given in Eq.~(\ref{eq:3.3.1}).

Similarly, we can define
\begin{equation}
\begin{split}
&\chi_{A^2\sigma}=\chi_{\pi^2\sigma}+\chi_{\sigma^2\sigma},\;\;\chi_{A^2B^2}=\chi_{\pi^2\pi^2}+\chi_{\sigma^2\sigma^2}+2\chi_{\sigma^2\pi^2}.
\end{split}
\label{eq:3.3.4}
\end{equation}
The 1PI summation of $\chi_{A^2\sigma}$ is also shown in Fig.~\ref{fig:1pisum}, and can be written as
\begin{equation}
\begin{split}
&\chi _{A^2\sigma}(q^2)=\sum _{n=0}^{\infty}\frac{\Pi _{A^2\sigma}(q^2)}{q^2+m_{\sigma}}\times \left(\frac{1}{q^2+m_{\sigma}^2}\Pi _{\sigma}(q^2)\right)^n=\frac{\Pi _{A^2\sigma}(q^2)}{q^2+m_{\sigma}^2-\Pi _{\sigma}(q^2)},
\end{split}
\label{eq:3.3.5}
\end{equation}
where $\Pi _{A^2\sigma}(q^2)$ is given in Eq.~(\ref{eq:ap.83}) and the 1PI summation of $\chi _{A^2B^2}(q^2)$ is given in (\ref{eq:ap.88}).

\subsection{\label{sec:3.4}Summary of calculations of the susceptibilities.}

In $O(2)$ model, the coupling constant $U$ is not a dimensionless quantity in $D=2+1$ dimensions. We find from the calculations that the 1-loop order is at $\mathcal{O}(U/m_{\sigma})$ order, the 2-loop order is at $\mathcal{O}(U^2/m_{\sigma}^2)$ order. So the perturbation is expansion around $U/m_{\sigma}\sim 0$.

The longitudinal susceptibility and the scalar susceptibility, at 1-loop level,   can be written as
\begin{equation}
\begin{split}
&\chi _{\sigma\sigma}^{(1)}(q^2)=\frac{1}{q^2+m_{\sigma}^2}+\Pi _{\sigma}^{(1)}(q^2),\\
&\chi _{\rho\rho}^{(1)}(q^2)=4\chi _{\sigma\sigma}^{(1)}(q^2)+\frac{4}{v}\frac{\Pi _{\pi^2\sigma}^{(1)}(q^2)+\Pi _{\sigma^2\sigma}^{(1)}(q^2)}{q^2+m_{\sigma}^2}+\frac{1}{v^2}\left(\Pi _{\pi^2\pi^2}^{(1)}(q^2)+\Pi _{\sigma^2\sigma^2}^{(1)}(q^2)\right),\\
\end{split}
\label{eq:3.4.1}
\end{equation}
where $\Pi _{\sigma}^{(1)}(q^2)$ is given in Eq.~(\ref{eq:3.1.4}). $\Pi _{\pi^2\sigma}^{(1)}(q^2)$, $\Pi _{\sigma^2\sigma}^{(1)}(q^2)$, $\Pi _{\pi^2\pi^2}^{(1)}(q^2)$ and $\Pi _{\sigma^2\sigma^2}^{(1)}(q^2)$ are given in Eq.~(\ref{eq:3.1.5}).

When the 2-loop contributions, RPA like contributions and 1PI summation are included, and  the results are
\begin{equation}
\begin{split}
&\chi _{\sigma\sigma}(q^2)=\Sigma_{\sigma}(q^2),\\
&\chi _{\rho\rho}(q^2)=4\chi _{\sigma\sigma}(q^2)+\frac{4}{v}\chi_{A^2\sigma}(q^2)+\frac{1}{v^2}\chi_{A^2B^2}(q^2),\\
\end{split}
\label{eq:3.4.2}
\end{equation}
where $\Sigma_{\sigma}(q^2)$ is given in Eq.~(\ref{eq:3.3.3}), $\chi_{A^2\sigma}(q^2)$ is given in Eq.~(\ref{eq:3.3.5}), and $\chi_{A^2B^2}(q^2)$ is given in Eq.~(\ref{eq:ap.88}).

The spectral function can be obtained by using Eq.~(\ref{eq:2.10}). In a way similar to that in  Ref.~\cite{Podolsky1}, we use the analytical continuing $q=\sqrt{{\bf q}^2-(\omega + i\epsilon)^2}$ to obtain $\chi ''_{AB}({\bf q},\omega)$.

\section{\label{sec:4}Numerical results}

In Ref.~\cite{O2Model1}, the approximate $O(2)$ model can be written as
\begin{equation}
\begin{split}
&S=\frac{1}{8Jz\bar{n}^2}\int dt\int d^{d} x\left\{\left(\frac{\partial}{\partial t} \phi \right)^2-(2J\bar{n})^2z\left(\nabla\phi\right)^2 -(2J\bar{n}z)^2(u-1)\phi ^2 -(Jz)^2\bar{n}u\phi ^4\right\},
\end{split}
\label{eq:4.1}
\end{equation}
where $u=g/4Jz\bar{n}$, $z$ is the lattice coordinate number, $J$ is tunneling, $g$ is the coupling constant, $\bar{n}$ is boson occupation number. In experiments, typically,  $\bar{n}\approx 50$ and $2\pi \hbar /2Jz\bar{n}\approx 0.7{\rm ms}$.

We first rescale the coordinate as $x\to x/2J\bar{n}z$, then rescale the field as $\phi = \Phi /\sqrt{J}$, we find in $D=2+1$ dimensions, and in imaginary time representation, the model becomes
\begin{equation}
\begin{split}
&S=\int d^{2+1} x\left\{\frac{1}{2}\left(\partial_{\mu} \Phi\right)^2-\frac{(2J\bar{n}z)^2(1-u)}{2}\Phi ^2 +\frac{2Jz^2 \bar{n}u}{4}\Phi ^4\right\}.\\
\end{split}
\label{eq:4.2}
\end{equation}

Comparing Eq.~(\ref{eq:4.2}) with Eq.~(\ref{eq:2.1}), we find that when
\begin{equation}
\begin{split}
&m_{\sigma}=2\sqrt{2}Jz\bar{n}\sqrt{1-u},\;\;U=2Jz^2 \bar{n}u, \\
\end{split}
\label{eq:4.3}
\end{equation}
the model is as same as Eq.~(\ref{eq:2.1}). Note that $m_{\sigma}$ is as same as $\Delta$ in Ref.~\cite{O2Model1}, as expected.

Since in the experiment~\cite{2DOpticalLattice}, the result is given in terms of the  parameter  $j\equiv J/g$, we rewrite Eq.~(\ref{eq:4.3}) as
\begin{equation}
\begin{split}
&m_{\sigma}=2\sqrt{2}J\bar{n}z\sqrt{1-\frac{j_c}{j}},\;\;U=\frac{Jz}{2j},\;\;j_c=\frac{1}{4\bar{n}z}.\\
\end{split}
\label{eq:4.4}
\end{equation}
In the experiment, the $j_c$ is found to be $\approx 0.06$~\cite{2DOpticalLattice}.

Note that the corrections of perturbation at n-loop order is proportional to $\left(U/m_{\sigma}\right)^n$, which is independent of $J$ when $j$ is fixed. So $J$ only affects the amplitude of the spectral function. When the spectral function is normalized as in Ref.~\cite{2DOpticalLattice}, it is independent of $J$. For convenience, we use
\begin{equation}
\begin{split}
&J=1,\;\;\bar{n}=50,\;\;j_c=0.06.\\
\end{split}
\label{eq:4.5}
\end{equation}

\begin{figure*}
\begin{center}
\includegraphics[scale=0.6]{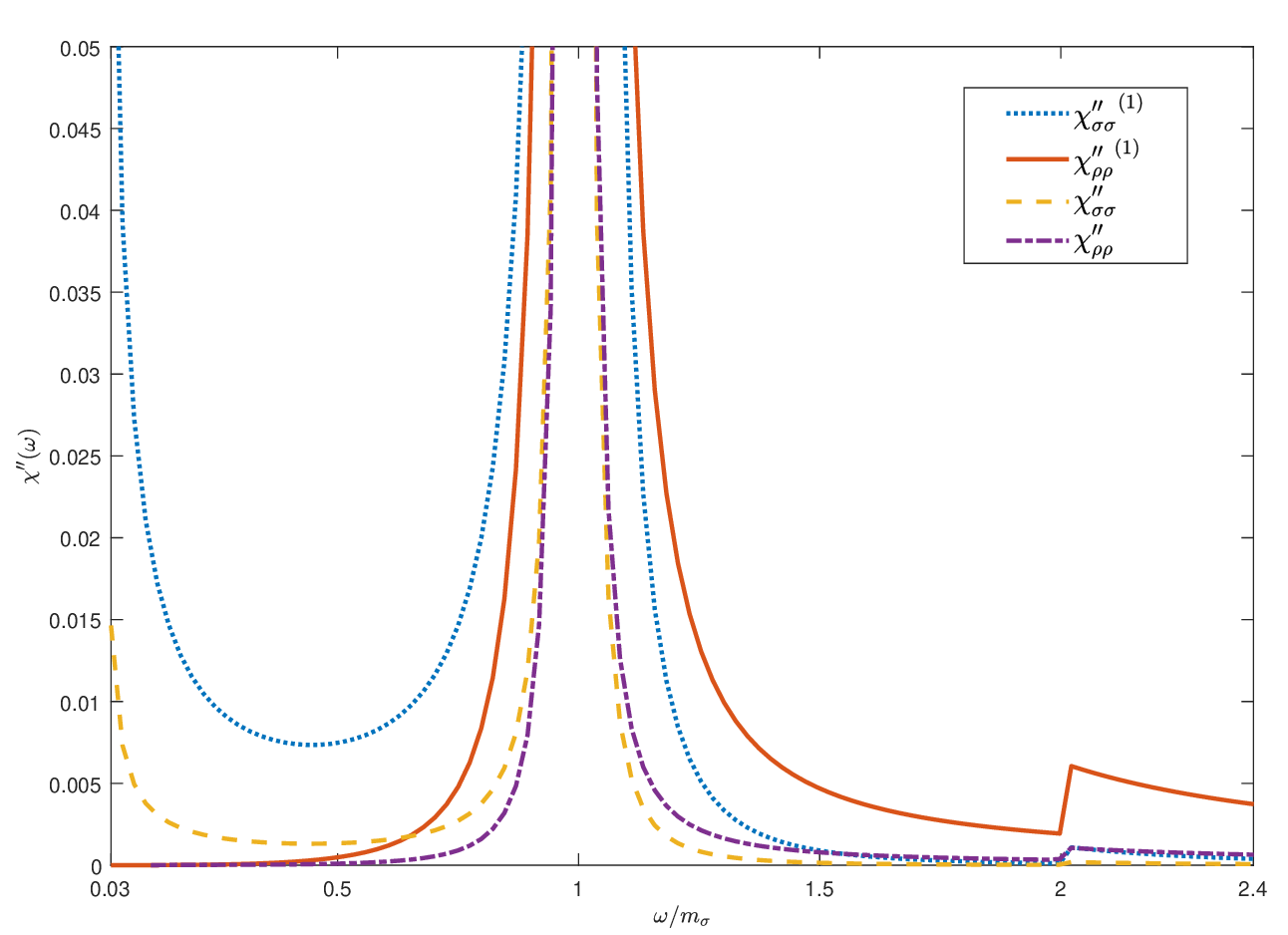}
\caption{The normalized spectral functions $\chi ''(\omega)$ when $j=2j_c$. The dotted line is ${\chi ''_{\sigma\sigma}}^{(1)}(\omega)$, the solid line is ${\chi ''_{\rho\rho}}^{(1)}(\omega)$, the dashed line is ${\chi ''_{\sigma\sigma}}(\omega)$, the dashed-dotted line is ${\chi ''_{\rho\rho}}(\omega)$.}
\label{fig:chiplot1}
\end{center}
\end{figure*}

The spectral functions  at ${\bf q}=0$, $\chi ''_{AB}(\omega)$,  can be numerically obtained.
We obtain $\chi ''_{\sigma\sigma}(\omega)$ and $\chi ''_{\rho\rho}(\omega)$ for    $j=2j_c$. The susceptibilities are calculated using Eqs.~(\ref{eq:2.10}), (\ref{eq:3.4.1}), (\ref{eq:3.4.2}), (\ref{eq:4.4}) and (\ref{eq:4.5}). The 1-loop level spectral functions of ${\chi ''_{\sigma\sigma}}^{(1)}(\omega)$ and ${\chi ''_{\rho\rho}}^{(1)}(\omega)$ are shown in Fig.~\ref{fig:chiplot1}.

\begin{figure*}
\begin{center}
\includegraphics[scale=0.6]{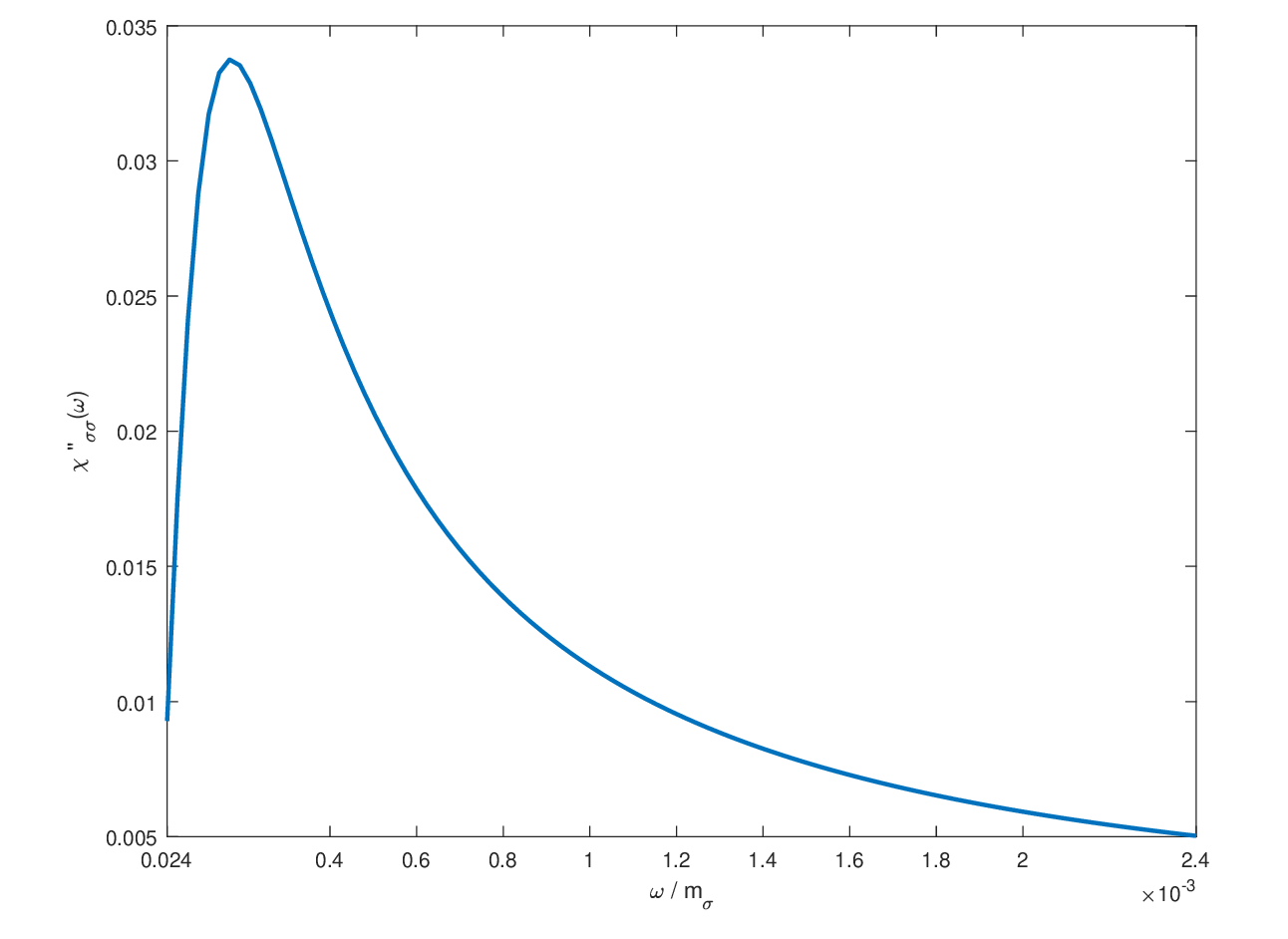}
\caption{The spectral functions $\chi ''_{\sigma\sigma}(\omega)$ when $j=2j_c$ and $\omega$ is very small. The spectral functions are not normalized.}
\label{fig:chiplot2}
\end{center}
\end{figure*}

We also find from Fig.~\ref{fig:chiplot1} that, in contrast with the conclusion of Ref.~\cite{Podolsky1}, the singularity of the spectral function of longitudinal susceptibility at $\omega \to 0$ does not damage the visibility of the peak at $\omega=m_{\sigma}$.  The plot at small values of $\omega$, as magnified   in Fig.~\ref{fig:chiplot2}, indicates that the spectral function is convergent.

The behaviour of the singularity at $\omega=0$ of longitudinal susceptibility is changed when 1PI summation is included, as can also be seen in  the Eq.~(C.2) of Ref.~\cite{Podolsky1}, which can be written as
\begin{equation}
\begin{split}
&\chi _{\sigma\sigma}^{N=\infty}(q^2)=\frac{1}{q^2+m_{\sigma}^2-\Pi _{\sigma}^{\rm 1PI}}.
\end{split}
\label{eq:4.6}
\end{equation}
Even at 1-loop level and with only the leading contribution of large-N limit is included, i.e., $\Pi _{\sigma}^{\rm 1PI}=2U^2v^2\times \left(1/8q\right)$, the spectral function can be written as
\begin{equation}
\begin{split}
&{\chi ''_{\sigma\sigma}}^{N=\infty,\;(1)}(\omega)=\frac{8 m_{\sigma}^2 \omega U}{m_{\sigma}^4 \left(64 \omega^2+U^2\right)-128 m_{\sigma}^2 \omega^4+64 \omega^6}=\frac{8 \omega}{m_{\sigma}^2 U}+\mathcal{O}(\omega^3).\\
\end{split}
\label{eq:4.7}
\end{equation}
Therefore,  although $\chi ''_{\sigma\sigma} \sim \omega ^{-1}$ at 1-loop level,  in consistency with Ref.~\cite{Podolsky1}, we find that $\chi ''_{\sigma\sigma} \sim \omega $ when 1PI summation is included.

We also find  another small peak at about $2m_{\sigma}$.

\begin{figure*}
\begin{center}
\includegraphics[scale=0.7]{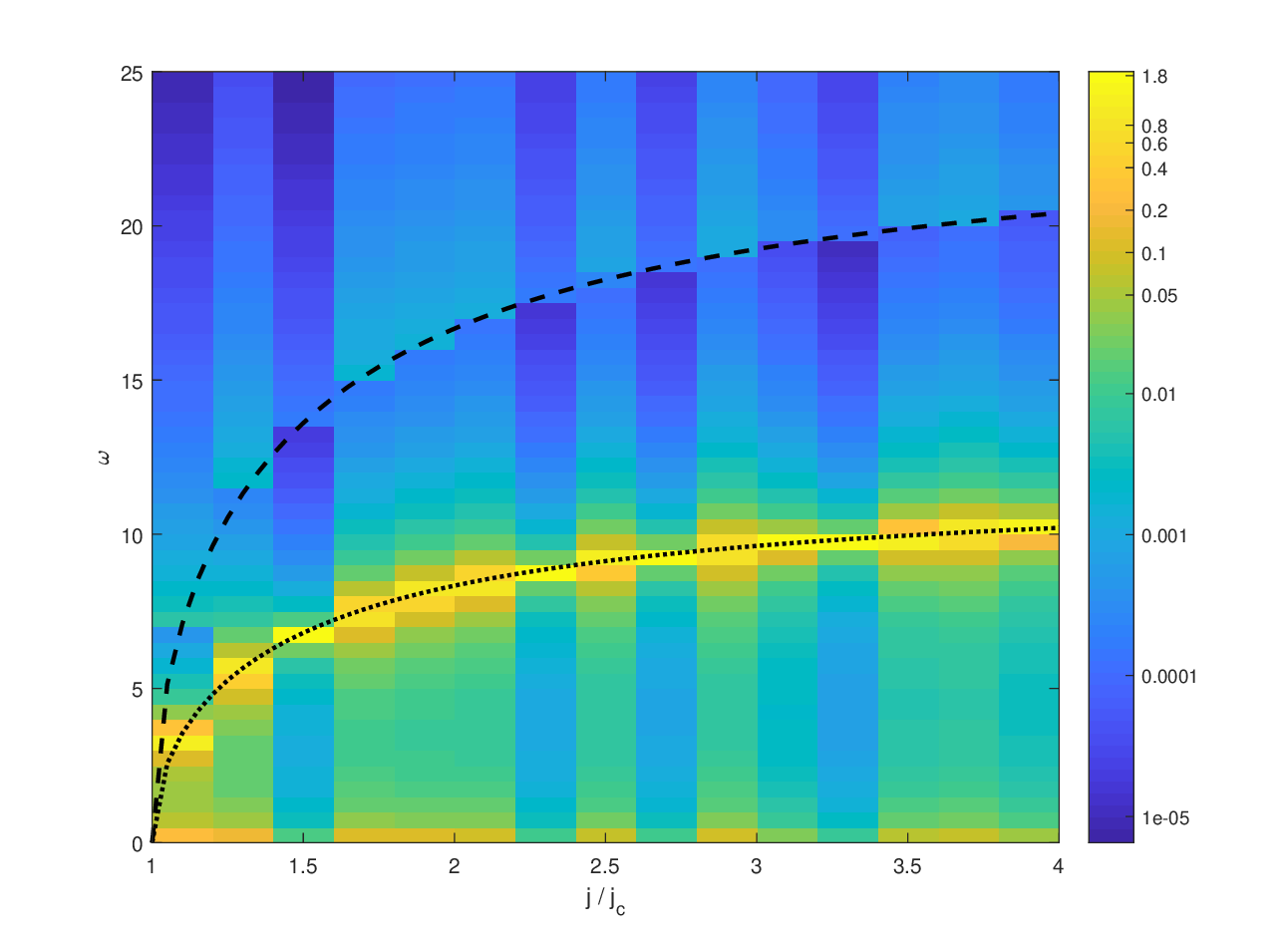}
\caption{The normalized spectral function $\chi ''_{\sigma\sigma}(\omega)$ when $j$ ranges from $1.1j_c$ to $3.9j_c$. The color map is logarithmically scaled. The dotted line is $m_{\sigma}(j/j_c)$ in Eq.~(\ref{eq:4.4}). The dashed line is $2m_{\sigma}(j/j_c)$.}
\label{fig:spectralsigma}
\end{center}
\end{figure*}

\begin{figure*}
\begin{center}
\includegraphics[scale=0.75]{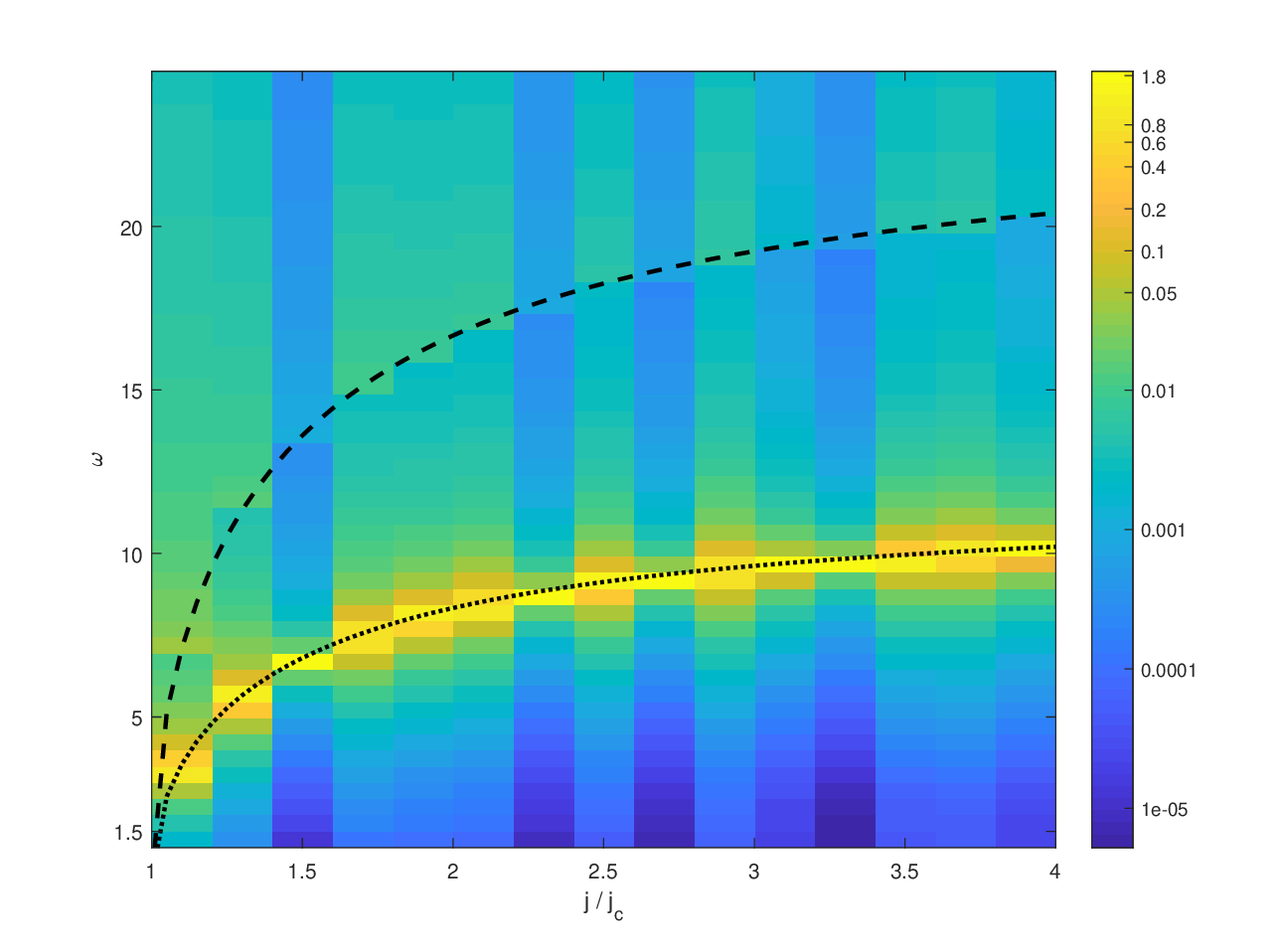}
\caption{The normalized spectral function $\chi ''_{\rho\rho}(\omega)$ when  $j$ ranges from $1.1j_c$ to $3.9j_c$.  The color map is logarithmically scaled. The dotted line is $m_{\sigma}(j/j_c)$ in Eq.~(\ref{eq:4.4}). The dashed line is $2m_{\sigma}(j/j_c)$.}
\label{fig:spectralrho}
\end{center}
\end{figure*}

Similar to Ref.~\cite{2DOpticalLattice}, we also show the spectral functions for  $j$ ranging from $1.1j_c$ to $3.9j_c$. The normalized spectral functions $\chi ''_{\sigma\sigma}(\omega)$ and  $\chi ''_{\rho\rho}(\omega)$ are shown in Fig.~\ref{fig:spectralsigma} and Fig.~\ref{fig:spectralrho}, respectively.
$\chi ''_{\rho\rho}(\omega)$ is almost as same as $\chi ''_{\sigma\sigma}(\omega)$ except that its peaks are a little wider.

We think that the disappearance of the peak of the spectral function  when the weak  interaction limit is approached, as observed in the  experiment~\cite{2DOpticalLattice},  cannot be explained by using the $O(2)$ model, because the $O(2)$ model is a relativistic model. As discussed in Ref.~\cite{Varma1}, the visibility of the Higgs mode depends on  whether the model is relativistic or not. So we cannot reproduce the result of the experiment as long as we use a relativistic model in zero temperature limit.

\begin{figure*}
\begin{center}
\includegraphics[scale=0.6]{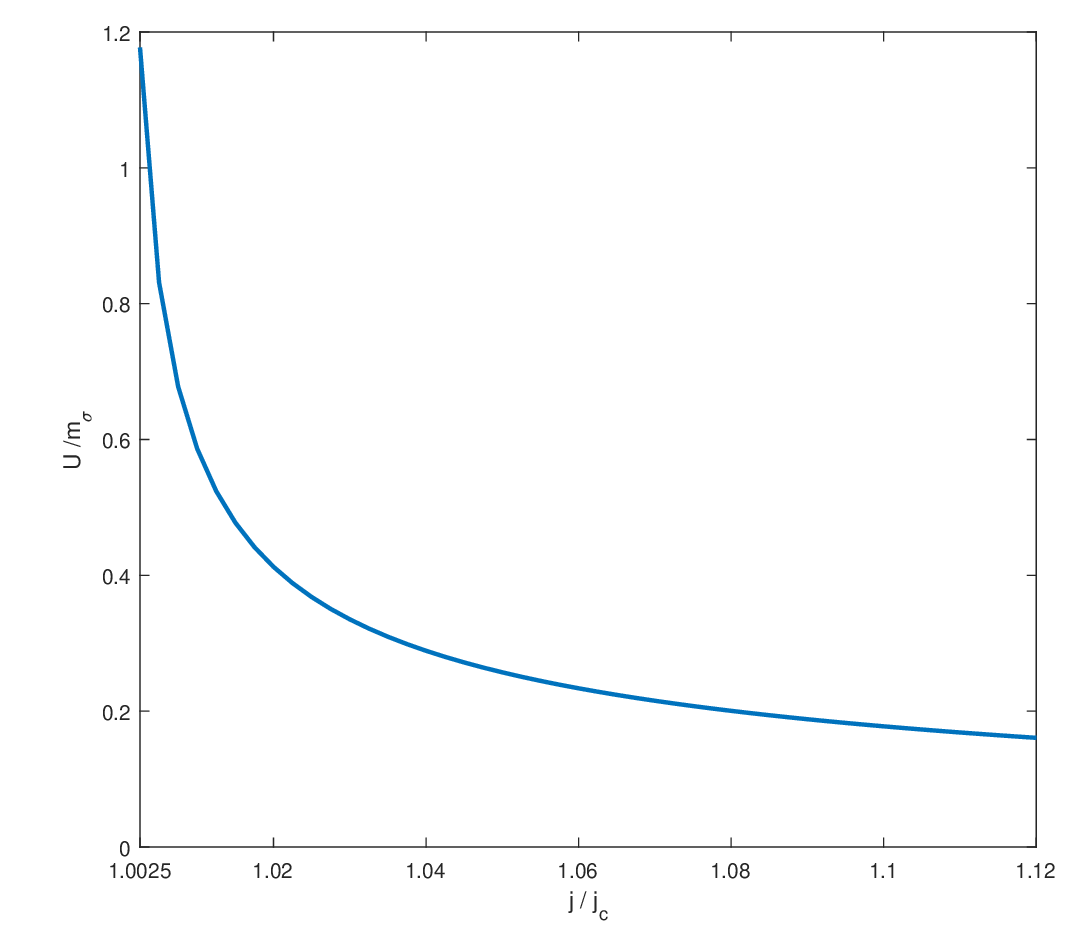}
\caption{$U/m_{\sigma}$ as a function of $j/j_c$.}
\label{fig:expa}
\end{center}
\end{figure*}

The perturbation  around $U/m_{\sigma}\sim 0$ is valid only  when $U/m_{\sigma}\ll 1$. $U/m_{\sigma}$ as a function of $j/j_c$ is shown in Fig.~\ref{fig:expa}, where it can be seen that  as $j$ decreases towards  $j_c$, $U/m_{\sigma}$ increases, invalidating the perturbation theory.

The monotonic decrease of $U/m_{\sigma}$ with the increase of $j/j_c$ is another evidence that the disappearance of the Higgs  peak  cannot be explained within the $O(2)$ model. With the increase of  $j/j_c$, $U/m_{\sigma}$ decreases, so the 1-loop or higher order contributions   become less important, hence the spectral function     approaches  the  delta function $\delta (\omega-m_{\sigma})$.

\section{\label{sec:5}Conclusion}

The Higgs mode observed in the two dimensional optical lattice~\cite{2DOpticalLattice} ended the debate  whether the Higgs mode can be observed in the two dimensional neutral superfluid system. However, it was argued that, to reproduce the disappearance of the response, a theory between strong  and weak interaction regimes is needed.

In this paper, we investigate the spectral function of the Higgs mode in  $O(2)$ model. Differing  from previous works, we calculate the spectral functions without using large-N expansion. The spectral function $\chi ''_{\sigma\sigma}(\omega)$ and $\chi ''_{\rho\rho}(\omega)$ are obtained in Eq.~(\ref{eq:3.4.2}),  as shown Fig.~\ref{fig:spectralsigma} and Fig.~\ref{fig:spectralrho}.

We argue that to use longitudinal susceptibility or scalar susceptibility is irrelevant to  the visibility of the Higgs mode. We also find that the disappearance of the response  with increase of $j/j_c$ cannot be explained within $O(2)$ model at zero temperature limit. We also find that  there is  a  small  peak in the spectral function at about $\omega \approx 2m_{\sigma}$.

\appendix

\section{\label{sec:a} Calculations of the Feynman diagrams}

Calculations of the Feynman diagrams needed are listed below.

\subsection{\label{sec:a1}1-loop diagrams}

\begin{figure*}
\begin{center}
\includegraphics[scale=0.8]{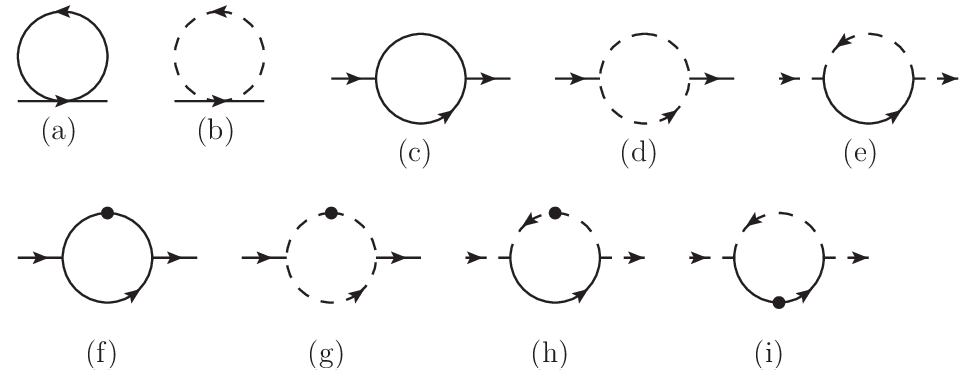}
\caption{All 1-loop diagrams.}
\label{fig:oneloopall}
\end{center}
\end{figure*}

\subsubsection{\label{sec:a11}Vacuum bubble diagrams}

The vacuum bubble diagrams are drawn in (a) and (b) in Fig.~\ref{fig:oneloopall},
\begin{equation}
\begin{split}
&f^{(1)}_a= \mu ^{\epsilon}\int \frac{d^Dk}{(2\pi)^D}\frac{1}{k^2 + m_{\sigma}^2},\;\;f^{(1)}_b= \mu ^{\epsilon}\int \frac{d^Dk}{(2\pi)^D}\frac{1}{k^2 + \lambda^2}.
\end{split}
\label{eq:ap.1}
\end{equation}

Using DR, for  $D\neq 2n$,
\begin{equation}
\begin{split}
&G_n(m)\equiv \mu ^{\epsilon}\int \frac{d^Dk}{(2\pi)^D}\frac{1}{(k^2 + m^2)^n}=\frac{\mu ^{\epsilon}}{(4\pi)^\frac{D}{2} }\frac{\Gamma (n-\frac{D}{2})}{\Gamma (n)}(m^2)^{\frac{D}{2}-n}.\\
\end{split}
\label{eq:ap.2}
\end{equation}

Unlike the cut-off regulator, in DR, the massless vacuum bubble diagrams vanish unless $D=2n$,  because the integral has   dimension $[p]^{D-2n}$ while there is no external momentum or mass, hence no dimensional variable,  the result can only be $0^{D-2n}$~\cite{GrozinMultiLoop}. Therefore for  $D=3-\epsilon$,
\begin{equation}
\begin{split}
&f^{(1)}_a=-\frac{m_{\sigma}}{4\pi},\;\;\;f^{(1)}_b=
\begin{cases}0,\;\;\text{no IR divergence},
\\
-\frac{\lambda}{4\pi},\;\;\text{with IR  divergence}\end{cases}\\.\\
\end{split}
\label{eq:ap.3}
\end{equation}

\begin{figure*}
\begin{center}
\includegraphics[scale=0.5]{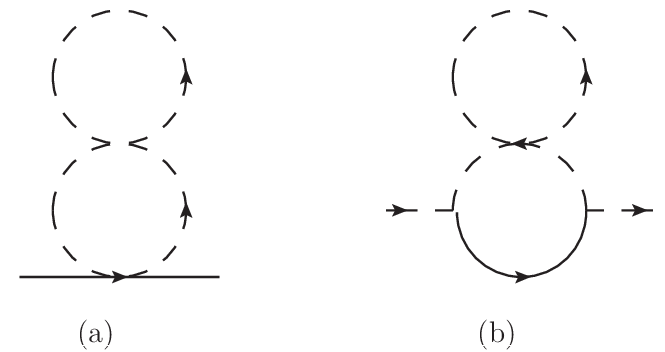}
\caption{The diagrams with a massless vacuum bubble acquires a finite result because of the IR divergence.}
\label{fig:2loopir}
\end{center}
\end{figure*}

When there is IR divergence, one should use $f_b^{(1)}=-\lambda /4\pi$, for example, as for  the diagrams in Fig.~\ref{fig:2loopir}, which is  nonzero  because of the IR divergence.

\subsubsection{\label{sec:a12}Other 1-loop diagrams}

Other 1-loop diagrams we need to calculate are drawn in (c) - (i) of Fig.~\ref{fig:oneloopall}, and are denoted as $f^{(1)}_{\alpha}(q^2)$,  where $\alpha =c,d,e,f,g,h,i$,
\begin{equation}
\begin{split}
&f^{(1)}_c(q^2)=\int \frac{d^3k}{(2\pi)^3}\frac{1}{(k-q)^2+m_{\sigma}^2}\frac{1}{k^2+m_{\sigma}^2},\\
&f^{(1)}_d(q^2)=\int \frac{d^3k}{(2\pi)^3}\frac{1}{(k-q)^2}\frac{1}{k^2},\\
&f^{(1)}_e(q^2)=\int \frac{d^3k}{(2\pi)^3}\frac{1}{(k-q)^2}\frac{1}{k^2+m_{\sigma}^2},\\
&f^{(1)}_f(q^2)=\int \frac{d^3k}{(2\pi)^3}\frac{1}{(k-q)^2+m_{\sigma}^2}\frac{1}{(k^2+m_{\sigma}^2)^2},\\
&f^{(1)}_g(q^2)=\int \frac{d^3k}{(2\pi)^3}\frac{1}{(k-q)^2+\lambda^2}\frac{1}{(k^2+\lambda^2)^2},\\
&f^{(1)}_h(q^2)=\int \frac{d^3k}{(2\pi)^3}\frac{1}{(k-q)^2+m_{\sigma}^2}\frac{1}{(k^2+\lambda^2)^2},\\
&f^{(1)}_i(q^2)=\int \frac{d^3k}{(2\pi)^3}\frac{1}{(k-q)^2}\frac{1}{
(k^2+m_{\sigma}^2)^2},\\
\end{split}
\label{eq:ap.4}
\end{equation}
which  are UV finite and are found to be
\begin{equation}
\begin{split}
&f^{(1)}_c(q^2)=\frac{1}{4\pi}\frac{\cot ^{-1}\left(\frac{2m_{\sigma}}{q}\right)}{q},\;f^{(1)}_d(q^2)=\frac{1}{8q},\;f^{(1)}_e(q^2)=\frac{1}{4\pi}\frac{\tan ^{-1}\left(\frac{q}{m_{\sigma}}\right)}{q},\\
&f^{(1)}_f(q^2)=\frac{1}{4\pi}\frac{1}{2m_{\sigma}^3(t+4)},\;\;f^{(1)}_g(q^2)=\frac{1}{4\pi}\frac{1}{2\lambda q^2},\\
&f^{(1)}_h(q^2)=\frac{1}{4\pi}\left(\frac{1}{2\lambda m_{\sigma}^2(1+t)}-\frac{1}{m_{\sigma}^3(1+t)^2}\right),\;\;f^{(1)}_i(q^2)=\frac{1}{4\pi}\frac{1}{2m_{\sigma}^3 (1+t)},\\
\end{split}
\label{eq:ap.5}
\end{equation}
where $t$ is defined as $t\equiv q^2/m_{\sigma}^2$. The result $f^{(1)}_d$ is as same as in~\cite{Podolsky1}.
When $q^2=0$, we have
\begin{equation}
\begin{split}
&f^{(1)}_c(q^2=0)=\frac{1}{8\pi m_{\sigma}},\;\;f^{(1)}_d(q^2=0)=\frac{1}{8\pi \lambda },\;\;f^{(1)}_e(q^2=0)=\frac{1}{4\pi m_{\sigma}},\\
&f^{(1)}_h(q^2=0)=\frac{1}{4\pi}\left(\frac{1}{2\lambda m_{\sigma}^2}-\frac{1}{m_{\sigma}^3}\right),\;\;f^{(1)}_i(q^2=0)=\frac{1}{4\pi}\frac{1}{2m_{\sigma}^3},\\
\end{split}
\label{eq:ap.6}
\end{equation}

\subsection{\label{sec:a2}2-loop diagrams}

\subsubsection{\label{sec:a21}Sunset diagrams}

\begin{figure*}
\begin{center}
\includegraphics[scale=0.7]{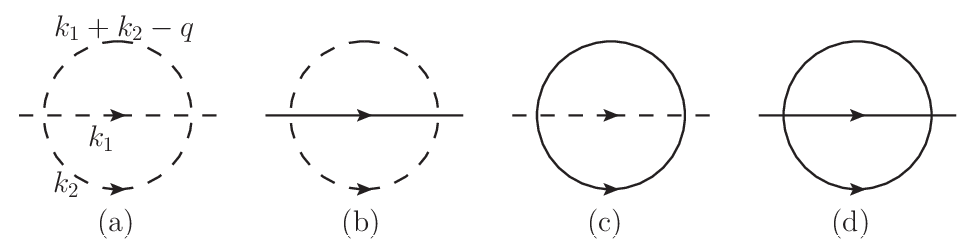}
\caption{The first type of 2-loop diagrams, i.e. the sunset diagrams.}
\label{fig:2loopclass1}
\end{center}
\end{figure*}

The sunset diagrams are shown in Fig.~\ref{fig:2loopclass1}. The diagrams in (a), (b), (c) and (d) of Fig.~\ref{fig:2loopclass1} are denoted as $f^{(2)}_{\alpha}(q^2)$, where $\alpha =a,b,c,d$,
\begin{equation}
\begin{split}
&f^{(2)}_a(q^2)=\mu ^{\epsilon} \int \frac{d^D k_1}{(2\pi)^D}\int \frac{d^D k_2}{(2\pi)^D}\frac{1}{k_1^2}\frac{1}{k_2^2}\frac{1}{(k_1+k_2-q)^2},\\
&f^{(2)}_b(q^2)=\mu ^{\epsilon} \int \frac{d^D k_1}{(2\pi)^D}\int \frac{d^D k_2}{(2\pi)^D}\frac{1}{k_1^2}\frac{1}{k_2^2}\frac{1}{(k_1+k_2-q)^2+m_{\sigma}^2},\\
&f^{(2)}_c(q^2)=\mu ^{\epsilon} \int \frac{d^D k_1}{(2\pi)^D}\int \frac{d^D k_2}{(2\pi)^D}\frac{1}{k_1^2+m_{\sigma}^2}\frac{1}{k_2^2+m_{\sigma}^2}\frac{1}{(k_1+k_2-q)^2},\\
&f^{(2)}_d(q^2)=\mu ^{\epsilon} \int \frac{d^D k_1}{(2\pi)^D}\int \frac{d^D k_2}{(2\pi)^D}\frac{1}{k_1^2+m_{\sigma}^2}\frac{1}{
k_2^2+m_{\sigma}^2}\frac{1}{(k_1+k_2-q)^2+m_{\sigma}^2},\\
\end{split}
\label{eq:ap.7}
\end{equation}
which are  UV divergent   in $D=3-\epsilon$.

The massless sunset diagrams with arbitrary powers of denominators are
\begin{equation}
\begin{split}
&G_{n_1,n_2,n_3}(q^2)=\mu ^{\epsilon} \int \frac{d^D k_1}{(2\pi)^D}\int \frac{d^D k_2}{(2\pi)^D}\frac{1}{(k_1^2)^{n_1}}\frac{1}{(k_2^2)^{n_2}}\frac{1}{((k_1+k_2-q)^2)^{n_3}}.\\
\end{split}
\label{eq:ap.8}
\end{equation}

An efficient way to calculate this integral is to use the Fourier transformation~\cite{GrozinMultiLoop}
\begin{equation}
\begin{split}
&G_{n_1,n_2,n_3}(q^2)=\frac{\mu ^{\epsilon}}{(4\pi)^D} \frac{\Gamma (\frac{D}{2}-n_1)\Gamma (\frac{D}{2}-n_2)\Gamma (\frac{D}{2}-n_3)}{\Gamma (n_1)\Gamma (n_2)\Gamma (n_3)} \\
&\times \frac{\Gamma (n_1+n_2+n_3-D)}{\Gamma (\frac{3D}{2}-n_1-n_2-n_3)}\frac{1}{(q^2)^{n_1+n_2+n_3-D}}.\\
\end{split}
\label{eq:ap.9}
\end{equation}

In $D=3-\epsilon$,   we need  $f^{(2)}_a(q^2)=G_{1,1,1}(q^2)$,  and obtain
\begin{equation}
\begin{split}
&f^{(2)}_a(q^2)=\frac{1}{2(4\pi)^2}\left(N_{\rm UV}+\log\frac{\mu^2}{q^2}+3\right).\\
\end{split}
\label{eq:ap.10}
\end{equation}

The sunset diagram with 1 internal mass can be calculated efficiently in Mellin-Barnes representation~\cite{MellinBarnes},
\begin{equation}
\begin{split}
&f^{(2)}_b(q^2)=\frac{\mu ^{\epsilon}}{2\pi i}\int _{-i\infty}^{i\infty}dz\Gamma (1+z)\Gamma (-z)(m^2)^z\\
&\times \int \frac{d^D k_1}{(2\pi)^D}\int \frac{d^D k_2}{(2\pi)^D}\frac{1}{k_1^2} \frac{1}{k_2^2}\frac{1}{((q-k_1-k_2)^2)^{1+z}}.\\
\end{split}
\label{eq:ap.11}
\end{equation}
By using  the massless sunset diagram Eq.~(\ref{eq:ap.8}),  we obtain
\begin{equation}
\begin{split}
&f^{(2)}_b(q^2)=\mu ^{\epsilon}\frac{(m_{\sigma}^2)^{D-3}\Gamma^2 (\frac{D}{2}-1)}{2\pi i (4\pi)^D}\\
&\times \int _{-i\infty}^{i\infty}dz\Gamma (-z) \frac{\Gamma (3-D+z)\Gamma (2-\frac{D}{2}+z)}{\Gamma (\frac{D}{2}+z)}t^{z}\\
&=\mu ^{\epsilon}\frac{(m_{\sigma}^2)^{D-3}\Gamma^2 (\frac{D}{2}-1)}{(4\pi)^D}\frac{\Gamma (3-D)\Gamma (2-\frac{D}{2})}{\Gamma (\frac{D}{2})}\;_2F_1\left(\left.\begin{array}{c}3-D,2-\frac{D}{2}\\ \frac{D}{2}\end{array}\right|-t\right),
\end{split}
\label{eq:ap.12}
\end{equation}
where   $t\equiv q^2/m_{\sigma}^2$.

In $D=3-\epsilon$, we use HypExp~\cite{HypExp} to expand it around small $\epsilon$, and  find
\begin{equation}
\begin{split}
&f^{(2)}_b(q^2)=\frac{1}{2(4\pi)^2}\left(N_{\rm UV}+3+\log \frac{\mu^2}{q^2+m_{\sigma}^2}-\frac{2\tan^{-1}\left(\frac{q}{m_{\sigma}}\right)}{\frac{q}{m_{\sigma}}}\right),\\
\end{split}
\label{eq:ap.13}
\end{equation}

Similar to the case with one internal mass, the sunset diagrams with two equal internal masses can be calculated in Mellin-Barnes representation. We obtain
\begin{equation}
\begin{split}
&f^{(2)}_c(q^2)\\
&=\mu ^{\epsilon}\frac{(m_{\sigma}^2)^{D-3}\Gamma (\frac{D}{2}-1)}{(4\pi)^D } \frac{\Gamma (3-D)\Gamma^2 (2-\frac{D}{2})}{\Gamma (4-D)\Gamma (\frac{D}{2})}\\
&\times \;_3F_2\left(\left.\begin{array}{c}3-D,2-\frac{D}{2},1\\ \frac{5-D}{2},\frac{D}{2}\end{array}\right|-\frac{t}{4}\right).\\
\end{split}
\label{eq:ap.14}
\end{equation}
When $D=3-\epsilon$,
\begin{equation}
\begin{split}
&f^{(2)}_c(q^2)=\frac{1}{2(4\pi)^2}\left(N_{\rm UV}+3+\log\frac{\mu^2}{ 4m_{\sigma}^2+q^2}-\frac{2\tan ^{-1}\left(\frac{q}{2m_{\sigma}}\right)}{\frac{q}{2m_{\sigma}}}\right).
\end{split}
\label{eq:ap.15}
\end{equation}
We also obtain
\begin{equation}
\begin{split}
&f^{(2)}_d(q^2)=\mu ^{\epsilon}\frac{\sqrt{\pi}(m_{\sigma}^2)^{D-3}}{2(2\pi i)^2(4\pi)^D} \int _{\frac{D-1}{2}-i\infty}^{\frac{D-1}{2}+i\infty}dz_1\int _{\frac{D+1}{2}-i\infty}^{\frac{D+1}{2}+i\infty}dz_2 t^{-z_1-z_2}\\
&\times \frac{4^{z_1}\Gamma (\frac{3-D}{2}-z_1)\Gamma (\frac{1}{2}-z_1)\Gamma (\frac{D-1}{2}-z_1)\Gamma(\frac{3-D}{2}-z_2)\Gamma (\frac{1}{2}-z_2)\Gamma (z_1+z_2)}{\Gamma (1-z_1)\Gamma (\frac{D}{2}-z_1-z_2)}.\\
\end{split}
\label{eq:ap:16}
\end{equation}
We expand it around small $\epsilon$ first, with the help of MB.m package~\cite{MB}. Then we use MBSums.m package~\cite{MBSum}, which   depends on AMBRE.m package~\cite{AMBRE},  to turn the integral into a summation. The result is
\begin{equation}
\begin{split}
&f^{(2)}_d(q^2)=\frac{1}{2(4\pi)^2}\left(N_{\rm UV}+3+\log \frac{\mu^2}{3m_{\sigma}^2}\right.\\
&\left.-\frac{3\tan ^{-1}\left(\frac{1}{2} \left(\sqrt{t}-i\right)\right)}{\sqrt{t}}+\frac{i \left(\sqrt{t}+3 i\right) \tan ^{-1}\left(\frac{1}{2} \left(\sqrt{t}+i\right)\right)}{\sqrt{t}}+\frac{3\tan ^{-1}\left(\frac{2 \sqrt{t}}{t+3}\right)}{\sqrt{t}}\right.\\
&\left.-\frac{1}{2} \left(\log \left(2 i \sqrt{t}+t+3\right)+\log \left(\frac{1}{9} \left(-2 i \sqrt{t}+t+3\right)\right)+2 \tanh ^{-1}\left(\frac{1}{2}+\frac{i \sqrt{t}}{2}\right)\right)\right) ,
\end{split}
\label{eq:ap.17}
\end{equation}
with  $t\equiv q^2\left./m_{\sigma}^2\right.$.

In absence of external momenta,
\begin{equation}
\begin{split}
&f^{(2)}_a(q^2=0)=\frac{1}{2(4\pi)^2}\left(N_{\rm UV}+1+\log \frac{\mu^2}{9\lambda^2}\right),\\
&f^{(2)}_b(q^2=0)=\frac{1}{2(4\pi)^2}\left(N_{\rm UV}+1+\log \frac{\mu^2}{m_{\sigma}^2}\right),\\
&f^{(2)}_c(q^2=0)=\frac{1}{2(4\pi)^2}\left(N_{\rm UV}+1+\log \frac{\mu^2}{4m_{\sigma}^2}\right),\\
&f^{(2)}_d(q^2=0)=\frac{1}{2(4\pi)^2}\left(N_{\rm UV}+1+\log \frac{\mu^2}{9m_{\sigma}^2}\right).\\
\end{split}
\label{eq:ap.18}
\end{equation}

\subsubsection{\label{sec:a22}The second type of 2-loop diagrams.}

\begin{figure*}
\begin{center}
\includegraphics[scale=0.7]{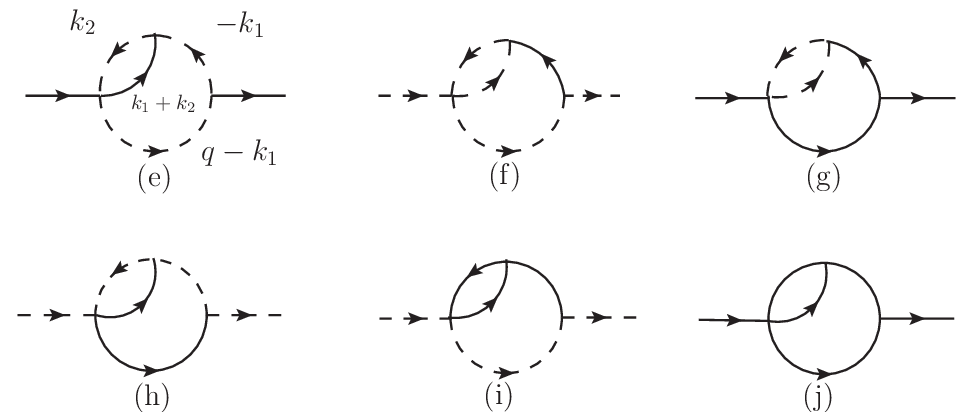}
\caption{The second type of 2-loop diagrams.}
\label{fig:2loopclass3}
\end{center}
\end{figure*}
The second type of 2-loop diagrams are shown in (e), (f), (g), (h), (i) and (j) of  Fig.~\ref{fig:2loopclass3}, and are denoted as $f^{(2)}_{\alpha}(q^2)$, with $\alpha = e,f,g,h,i,j$,
\begin{equation}
\begin{split}
&f_e^{(2)}(q^2)=\int \frac{d^3k_1}{(2\pi)^3}\int \frac{d^3k_2}{(2\pi)^3}\frac{1}{k_1^2}\frac{1}{(k_1-q)^2}\frac{1}{(k_1+k_2)^2}\frac{1}{k_2^2+m_{\sigma}^2},\\
&f_f^{(2)}(q^2)=\int \frac{d^3k_1}{(2\pi)^3}\int \frac{d^3k_2}{(2\pi)^3}\frac{1}{k_1^2+m_{\sigma}^2}\frac{1}{(k_1-q)^2}\frac{1}{(k_1+k_2)^2}\frac{1}{k_2^2},\\
&f_g^{(2)}(q^2)=\int \frac{d^3k_1}{(2\pi)^3}\int \frac{d^3k_2}{(2\pi)^3}\frac{1}{k_1^2+m_{\sigma}^2}\frac{1}{(k_1-q)^2+m_{\sigma}^2}\frac{1}{(k_1+k_2)^2}\frac{1}{k_2^2},\\
&f_h^{(2)}(q^2)=\int \frac{d^3k_1}{(2\pi)^3}\int \frac{d^3k_2}{(2\pi)^3}\frac{1}{k_1^2}\frac{1}{(k_1-q)^2+m_{\sigma}^2}\frac{1}{(k_1+k_2)^2+m_{\sigma}^2}\frac{1}{k_2^2},\\
&f_i^{(2)}(q^2)=\int \frac{d^3k_1}{(2\pi)^3}\int \frac{d^3k_2}{(2\pi)^3}\frac{1}{k_1^2+m_{\sigma}^2}\frac{1}{(k_1-q)^2}\frac{1}{(k_1+k_2)^2+m_{\sigma}^2}\frac{1}{k_2^2+m_{\sigma}^2},\\
&f_j^{(2)}(q^2)=\int \frac{d^3k_1}{(2\pi)^3}\int \frac{d^3k_2}{(2\pi)^3}\frac{1}{k_1^2+
m_{\sigma}^2}\frac{1}{(k_1-q)^2+m_{\sigma}^2}\frac{1}{
(k_1+k_2)^2+m_{\sigma}^2}\frac{1}{k_2^2+m_{\sigma}^2}, \\
\end{split}
\label{eq:ap.19}
\end{equation}
which are UV finite.

$f_e^{(2)}(q^2)$ can be efficiently calculated in Mellin-Barnes representation, by using MB.m package and MBSums.m package,
\begin{equation}
\begin{split}
&f_e^{(2)}(q^2)=\frac{1}{2(4\pi)^2m_{\sigma}^2}\frac{1}{t}\left(2\sqrt{t}\cot ^{-1}(\sqrt{t})+\log (1+t)\right),\\
\end{split}
\label{eq:ap.20}
\end{equation}
where $t\equiv q^2/m_{\sigma}^2$.

We integrate $f_f^{(2)}(q^2)$, $f_g^{(2)}(q^2)$ and $f_h^{(2)}(q^2)$ by  using the method in  calculating the $C^0$ Passinal-Veltman function~\cite{C0}. We take $f_f^{(2)}(q^2)$ for example. In terms of the  Feynman parameter, $f_f^{(2)}(q^2)$ can be written as
\begin{equation}
\begin{split}
&f_f^{(2)}(q^2)=\frac{\Gamma ^2(\frac{1}{2})}{(4\pi)^3}\int _0^1 dx\int _0^{1-x}dy\frac{x^{-\frac{1}{2}}}{(1-x-y)m_{\sigma}^2+y(1-y)q^2}.\\
\end{split}
\label{eq:ap.21}
\end{equation}
By making variable substitution $y\to (1-x)y$, $x\to x^2$, then $x\to x/y$, and then $x\to x+y$, the integral can be rewritten as
\begin{equation}
\begin{split}
&f_f^{(2)}(q^2)=\frac{2\Gamma ^2(\frac{1}{2})}{(4\pi)^3m_{\sigma}^2}\int _0^1dy\int _{-y}^0 dx\frac{1}{y(1-y+yt+x^2t+2xyt)},\\
\end{split}
\label{eq:ap.22}
\end{equation}
where $t\equiv q^2/m_{\sigma}^2$. Then by changing the order of integration over $x$ and $y$,  the integral can be written as
\begin{equation}
\begin{split}
&f_f^{(2)}(q^2)=\frac{2\Gamma ^2(\frac{1}{2})}{(4\pi)^3m_{\sigma}^2}\int _0^1 dx\int _x^1dy\frac{1}{y(1-y+yt+x^2t-2xyt)}.\\
\end{split}
\label{eq:ap.23}
\end{equation}
Then we find
\begin{equation}
\begin{split}
&f_f^{(2)}(q^2)=\frac{1}{2(4\pi)^2m_{\sigma}^2}\frac{i}{2 \sqrt{t}}\left\{ \text{Li}_2\left(\frac{1}{1+\frac{i}{\sqrt{t}}}\right)-\text{Li}_2\left(\frac{1}{1-\frac{i}{\sqrt{t}}}\right)+\text{Li}_2\left(\frac{1}{i \sqrt{t}+1}\right)\right.\\
&\left.-\text{Li}_2\left(\frac{1}{1-i \sqrt{t}}\right)-\frac{i \pi}{2} \log (t+1)+i \log (t) \tan ^{-1}\left(\sqrt{t}\right)\right\}. \\
\end{split}
\label{eq:ap.24}
\end{equation}
Similarly we find
\begin{equation}
\begin{split}
&f_g^{(2)}(q^2)=\frac{1}{2(4\pi)^2m_{\sigma}^2}
\frac{1}{2\sqrt{t}}\left\{\frac{1}{2} \left[-2 i \log \left(1-\frac{i}{\sqrt{t}}\right) \log \left(\frac{i}{\sqrt{t}}\right)+2 i \log \left(1+\frac{i}{\sqrt{t}}\right) \log \left(-\frac{i}{\sqrt{t}}\right)\right.\right.\\
&\left.\left.+2 i \log \left(\frac{\sqrt{t}-i}{\sqrt{t}-2 i}\right) \log \left(-\frac{i}{\sqrt{t}-2 i}\right)-2 i \log \left(\frac{\sqrt{t}+i}{\sqrt{t}+2 i}\right) \log \left(\frac{i}{\sqrt{t}+2 i}\right)\right.\right.\\
&\left.\left.+2 \pi  \log \left(2-i \sqrt{t}\right)+\pi  \left(\log \left(16 i \left(\sqrt{t}+i\right) t\right)-2 \log \left(\sqrt{t}-\sqrt{t+4}+2 i\right)\right)\right.\right.\\
&\left.\left.+i \left(\pi ^2+\log \left(-1+i \sqrt{t}\right) \left(\log (t)+2 \log \left(-8+4 i \sqrt{t}\right)-2 \log \left(\sqrt{t}-\sqrt{t+4}+2 i\right)\right)\right)\right.\right.\\
&\left.\left.-\log \left(1+i \sqrt{t}\right) \left(2 i \log \left(2+i \sqrt{t}\right)+2 i \log \left(-\frac{4 \sqrt{t}}{-\sqrt{t}+\sqrt{t+4}+2 i}\right)\right.\right.\right.\\
&\left.\left.\left.-2 i \log \left(\sqrt{t}+\sqrt{t+4}-2 i\right)+3 \pi \right)+2 \left(\pi +i \log \left(-1+i \sqrt{t}\right)\right) \log \left(-\frac{1}{\sqrt{t}+\sqrt{t+4}+2 i}\right)\right]\right.\\
&\left.-i \left[\text{Li}_2\left(-i \sqrt{t}\right)-\text{Li}_2\left(i \sqrt{t}\right)-2 \text{Li}_2\left(\frac{-i+\sqrt{t}}{-2 i+\sqrt{t}}\right)+2 \text{Li}_2\left(\frac{i+\sqrt{t}}{2 i+\sqrt{t}}\right)\right.\right.\\
&\left.\left.+\text{Li}_2\left(\frac{2 i}{2 i+\sqrt{t}-\sqrt{t+4}}\right)-\text{Li}_2\left(\frac{2 \left(i+\sqrt{t}\right)}{2 i+\sqrt{t}-\sqrt{t+4}}\right)-\text{Li}_2\left(\frac{2 i}{2 i-\sqrt{t}+\sqrt{t+4}}\right)\right.\right.\\
&\left.\left.+\text{Li}_2\left(\frac{2 i-2 \sqrt{t}}{2 i-\sqrt{t}+\sqrt{t+4}}\right)+\text{Li}_2\left(\frac{2 \left(-i+\sqrt{t}\right)}{-2 i+\sqrt{t}+\sqrt{t+4}}\right)-\text{Li}_2\left(-\frac{2 i}{-2 i+\sqrt{t}+\sqrt{t+4}}\right)\right.\right.\\
&\left.\left.+\text{Li}_2\left(\frac{2 i}{2 i+\sqrt{t}+\sqrt{t+4}}\right)-\text{Li}_2\left(\frac{2 \left(i+\sqrt{t}\right)}{2 i+\sqrt{t}+\sqrt{t+4}}\right)\right]\right\},
\end{split}
\label{eq:ap.25}
\end{equation}
\begin{equation}
\begin{split}
&f_h^{(2)}(q^2)=\frac{1}{2(4\pi)^2m_{\sigma}^2}\frac{1}{\sqrt{t} (t+1)}\left\{-2 (t+1) \cot ^{-1}\left(\frac{2}{\sqrt{t}}\right)-2 \tan ^{-1}\left(\sqrt{t}\right)\right.\\
&\left.+i \left[(t+1) \log \left(-\frac{\left(\sqrt{t}+i\right)^2}{t+1}\right)-i \sqrt{t} \log (t+4)+\log \left(\frac{t-i \sqrt{t}+2}{t+i \sqrt{t}+2}\right)\right]\right\}
\end{split}
\label{eq:ap.26}
\end{equation}

$f_i^{(2)}(q^2)$ and $f_j^{(2)}(q^2)$ are more difficult to calculate, we take $f_j^{(2)}(q^2)$ for example, it can written as
\begin{equation}
\begin{split}
&f_j^{(2)}(q^2)=f_g^{(2)}(q^2)-\frac{1}{2(4\pi)^2m_{\sigma}^2}2\left(-\frac{\log (3) \cot ^{-1}\left(\frac{2}{\sqrt{t}}\right)}{2 \sqrt{t}}+f_{j_1}(t)\right),\\
&f_{j_1}(t)=\int _0^1dx\frac{\coth ^{-1}\left(2 \sqrt{\frac{t (x-1) x-1}{x (t (x-1)+3)-4}}\right)}{2 \sqrt{1-t (x-1) x}},\\
\end{split}
\label{eq:ap.27}
\end{equation}
where
\begin{equation}
\begin{split}
&f_{j_1}(t)=\frac{1}{\sqrt{t}}f_{j_2}(u),\;\;\;f_{j_2}(u)=\int _0^1dx\frac{\coth ^{-1}\left(2 \sqrt{\frac{ (x-1) x-u}{x ((x-1)+3u)-4u}}\right)}{2 \sqrt{u-(x-1) x}},\\
\end{split}
\label{eq:ap.28}
\end{equation}
with  $u\equiv 1/t$.

It can be written that
\begin{equation}
\begin{split}
&f_{j_2}(u)=\int _0^u \left(\int _0^1dx \frac{\partial }{\partial u}\frac{\coth ^{-1}\left(2 \sqrt{\frac{ (x-1) x-u}{x ((x-1)+3u)-4u}}\right)}{2 \sqrt{u-(x-1) x}}\right)+C,\\
\end{split}
\label{eq:ap.29}
\end{equation}
where $C$ is a constant and can be obtained by comparing  the result with
\begin{equation}
\begin{split}
&f_{j_2}(u=0)=\int _0^1dx\frac{\coth ^{-1}\left(2 \sqrt{\frac{ (x-1) x}{x ((x-1))}}\right)}{2 \sqrt{(1-x) x}}.\\
\end{split}
\label{eq:ap.30}
\end{equation}
Eq.~(\ref{eq:ap.29}) is easier to integrate, one can integrate over $x$ and then over $u$ to obtain $f_{j_2}$.

Finally
\begin{equation}
\begin{split}
&f_i^{(2)}(q^2)=f_f^{(2)}(q^2)-\frac{1}{2(4\pi)^2m_{\sigma}^2}\frac{i}{2 \sqrt{t}}\left\{-\log \left(\frac{1}{\sqrt{t}}-\frac{i}{2}\right) \log \left(\frac{3 \left(\sqrt{t}+i\right)}{\sqrt{t}-i}\right)\right.\\
&\left.+\text{Li}_2\left(-1+\frac{2 i}{\sqrt{t}}\right)-\text{Li}_2\left(-1-\frac{2 i}{\sqrt{t}}\right)+\text{Li}_2\left(\frac{1}{3}+\frac{2 i}{3 \sqrt{t}}\right)-\text{Li}_2\left(\frac{1}{3}-\frac{2 i}{3 \sqrt{t}}\right)\right.\\
&\left.+\log (3) \left[\log (t+1)+\log \left(\frac{1}{\sqrt{t}}+\frac{i}{2}\right)-2 \log \left(1+i \sqrt{t}\right)\right]\right.\\
&\left.-2 i \log \left(\frac{2}{3} \left(\frac{2}{\sqrt{t}}+i\right)\right) \cot ^{-1}\left(\sqrt{t}\right)\right\},\\
\end{split}
\label{eq:ap.31}
\end{equation}
\begin{equation}
\begin{split}
&f_j^{(2)}(q^2)=f_g^{(2)}(q^2)-\frac{1}{2(4\pi)^2m_{\sigma}^2}\frac{1}{\sqrt{t}}\left\{\frac{1}{8} i \left[4 {\rm Re}(\text{Li}_2(3))-4 \text{Li}_2(3)-4 i \log \left(\frac{t}{t+4}\right) \tan ^{-1}\left(\frac{2 \sqrt{t}}{t+3}\right)\right.\right.\\
&\left.\left.-2 \log \left(\frac{1}{\sqrt{t}}+\frac{i}{2}\right) \log \left(-\frac{3 \left(t+2 i \sqrt{t}+3\right)}{5 \left(t-2 i \sqrt{t}+3\right)}\right)+\log (9) \log \left(\frac{1}{3}+\frac{2 i}{3 \left(\sqrt{t}-i\right)}\right)\right.\right.\\
&\left.\left.+4 i \log \left(\frac{1}{\sqrt{t}+2 i}+\frac{i}{2}\right) \tan ^{-1}\left(\frac{3}{\sqrt{t}}\right)-4 i \log \left(3 \left(\frac{1}{\sqrt{t}+2 i}+\frac{i}{2}\right)\right) \cot ^{-1}\left(\sqrt{t}\right)\right.\right.\\
&\left.\left.+2 \log \left(\frac{1}{\sqrt{t}}-\frac{i}{2}\right) \left(\log \left(\frac{3}{5} \left(-1+\frac{6 i}{\sqrt{t}+3 i}\right)\right)+2 i \tan ^{-1}\left(\frac{3}{\sqrt{t}}\right)\right)+8 i \log (3) \cot ^{-1}\left(\frac{2}{\sqrt{t}}\right)\right.\right.\\
&\left.\left.+2 \left(-\log (t+4)+2 \log \left(2+i \sqrt{t}\right)\right) \left(\log \left(-\frac{3}{5}+\frac{6 i}{5 \left(\sqrt{t}+3 i\right)}\right)+2 \coth ^{-1}\left(2-i \sqrt{t}\right)\right)\right.\right.\\
&\left.\left.+\pi ^2+\log ^2(3)+\log ^2(5)+2 i \pi  \log \left(\frac{20}{9}\right)\right]+\frac{1}{4} \left[-i \text{Li}_2\left(2 i \sqrt{\frac{1}{t}}-1\right)-i \text{Li}_2\left(6 i \sqrt{\frac{1}{t}}+3\right)\right.\right.\\
&\left.\left.-i \text{Li}_2\left(\frac{2}{3} i \sqrt{\frac{1}{t}}+\frac{1}{3}\right)+i \text{Li}_2\left(\frac{6}{5} i \sqrt{\frac{1}{t}}+\frac{3}{5}\right)+i \text{Li}_2\left(-2 i \sqrt{\frac{1}{t}}-1\right)+i \text{Li}_2\left(3-6 i \sqrt{\frac{1}{t}}\right)\right.\right.\\
&\left.\left.+i \text{Li}_2\left(\frac{1}{3}-\frac{2}{3} i \sqrt{\frac{1}{t}}\right)-i \text{Li}_2\left(\frac{3}{5}-\frac{6}{5} i \sqrt{\frac{1}{t}}\right)\right]-\frac{1}{12} i \left[3 \text{Li}_2\left(e^{2 i \tan ^{-1}\left(\frac{1}{2 \sqrt{\frac{1}{t}}}\right)+2 \tanh ^{-1}\left(\frac{1}{2}\right)}\right)\right.\right.\\
&\left.\left.+3 \text{Li}_2\left(e^{2 i \tan ^{-1}\left(\frac{1}{2 \sqrt{\frac{1}{t}}}\right)-2 \tanh ^{-1}\left(\frac{1}{2}\right)}\right)-3 \text{Li}_2\left(e^{2 i \tan ^{-1}\left(\frac{1}{2 \sqrt{\frac{1}{t}}}\right)+2 \tanh ^{-1}\left(\frac{3}{2}\right)}\right)\right.\right.\\
&\left.\left.-3 \text{Li}_2\left(e^{2 i \tan ^{-1}\left(\frac{1}{2 \sqrt{\frac{1}{t}}}\right)-2 \tanh ^{-1}\left(\frac{3}{2}\right)}\right)\right]\right\}.\\
\end{split}
\label{eq:ap.32}
\end{equation}

When $q^2=0$,
\begin{equation}
\begin{split}
&f_e^{(2)}{(q^2=0)}=\frac{1}{2(4\pi)^2m_{\sigma}^2}\left(\frac{m_{\sigma}}{\lambda}-2\right),\;\;\;f_f^{(2)}{(q^2=0)}=-\frac{1}{2(4\pi)^2m_{\sigma}^2}\log \left(9\frac{\lambda^2}{m_{\sigma}^2}\right),\\
&f_g^{(2)}{(q^2=0)}=\frac{1}{2(4\pi)^2m_{\sigma}^2},\;\;\;f_h^{(2)}{(q^2=0)}=\frac{1}{2(4\pi)^2m_{\sigma}^2}\log(4),\\
&f_i^{(2)}{(q^2=0)}=\frac{1}{2(4\pi)^2m_{\sigma}^2}\log\left(\frac{9}{4}\right),\;\;\;f_j^{(2)}{(q^2=0)}=\frac{1}{2(4\pi)^2m_{\sigma}^2}\frac{1}{3}.\\
\end{split}
\label{eq:ap.33}
\end{equation}

\begin{figure*}
\begin{center}
\includegraphics[scale=0.7]{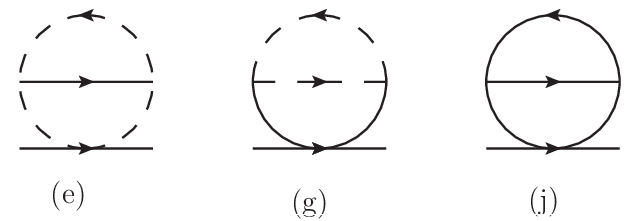}
\caption{The 2-loop diagrams without transition momenta,  in another form.}
\label{fig:class3zeroq}
\end{center}
\end{figure*}

Note that  the diagrams (e), (g), (j) of  Fig.~\ref{fig:class3zeroq} are  $f_e^{(2)}(q^2=0)$, $f_g^{(2)}(q^2=0)$ and $f_j^{(2)}(q^2=0)$.

Compared with the result of numerical integration, note  that in  making analytical  continuation $|q|\to \sqrt{-(\omega +i\epsilon)^2}$,  one  should use $\left(f_h((q^2)^*)\right)^*$ instead of $f_h(q^2)$.

\subsubsection{\label{sec:a23}Integral by part recursive relations}

The remaining diagrams are more difficult to calculate. We use the integral-by-part~(IBP) recursive relations~\cite{IBP}. For convenience,  we establish some definitions. The 2-loop diagrams can be written as
\begin{equation}
\begin{split}
&f(n_1,n_2,n_3,n_4,n_5)=\int \frac{d^Dk_1}{(2\pi)^D}\int \frac{d^Dk_2}{(2\pi)^D}\frac{1}{(k_1^2+m_3^2)^{n_3}}\\
&\times \frac{1}{((k_1+q)^2+m_1^2)^{n_1}}\frac{1}{((k_2-k_1)^2+m_5^2)^{n_5}}\frac{1}{(k_2^2+m_4^2)^{n_4}}\frac{1}{((k_2+q)^2+m_2^2)^{n_2}}.
\end{split}
\label{eq:ap.34}
\end{equation}

We define the ${\bf 1}^{\pm}$ operator as
\begin{equation}
\begin{split}
&{\bf 1}^{\pm}f(n_1,n_2,n_3,n_4,n_5)=f(n_1\pm 1,n_2,n_3,n_4,n_5),\\
\end{split}
\label{eq:ap.35}
\end{equation}
and ${\bf 2}^{\pm}$, ${\bf 3}^{\pm}$, ${\bf 4}^{\pm}$ and ${\bf 5}^{\pm}$ similarly.

With $k_2\cdot \frac{\partial }{\partial k_2}$, $k_1\cdot \frac{\partial }{\partial k_2}$ and $q\cdot \frac{\partial }{\partial k_2}$ acting on $f$, we  obtain the IBP relations
\begin{equation}
\begin{split}
&0=\left(D-n_5-2n_4-n_2-n_5({\bf 4}^--{\bf 3}^-+m_3^2-m_4^2-m_5^2){\bf 5}^+\right.\\
&\left.+2n_4m_4^2{\bf 4}^+-n_2({\bf 4}^--m_4^2-m_2^2-q^2){\bf 2}^+\right)f,\\
&0=\left(n_5-n_4-n_5({\bf 4}^--{\bf 3}^-+m_3^2-m_4^2+m_5^2){\bf 5}^+-n_4({\bf 3}^--{\bf 5}^-+m_5^2-m_3^2-m_4^2){\bf 4}^+\right.\\
&\left.-n_2({\bf 4}^-+{\bf 1}^--{\bf 5}^-+m_5^2-m_1^2-m_4^2-q^2){\bf 2}^+\right)f,\\
&0=\left(n_4-n_2-n_5({\bf 2}^-+{\bf 3}^--{\bf 1}^--{\bf 4}^-+m_1^2+m_4^2-m_2^2-m_3^2){\bf 5}^+\right.\\
&\left.-n_4({\bf 2}^-+m_4^2-m_2^2-q^2){\bf 4}^+-n_2(-{\bf 4}^-+m_4^2-m_2^2+q^2){\bf 2}^+\right)f.\\
\end{split}
\label{eq:ap.36}
\end{equation}

For example, the substraction of the first two equations in Eq.~(\ref{eq:ap.36}) is one of the triangle rules~\cite{GrozinMultiLoop}. When $n_i=1$ and $D=3$,
\begin{equation}
\begin{split}
&0=\left(-1+2m_5^2{\bf 5}^++({\bf 3}^--{\bf 5}^-+m_5^2-m_3^2+m_4^2){\bf 4}^+\right.\\
&\left.+({\bf 1}^--{\bf 5}^-+m_5^2-m_1^2+m_2^2){\bf 2}^+\right)f(1,1,1,1,1).\\
\end{split}
\label{eq:ap.37}
\end{equation}
Most of the remaining diagrams are calculated with the help of IBP relations.

\subsubsection{\label{sec:a24}The third type of 2-loop diagrams}

\begin{figure*}
\begin{center}
\includegraphics[scale=0.7]{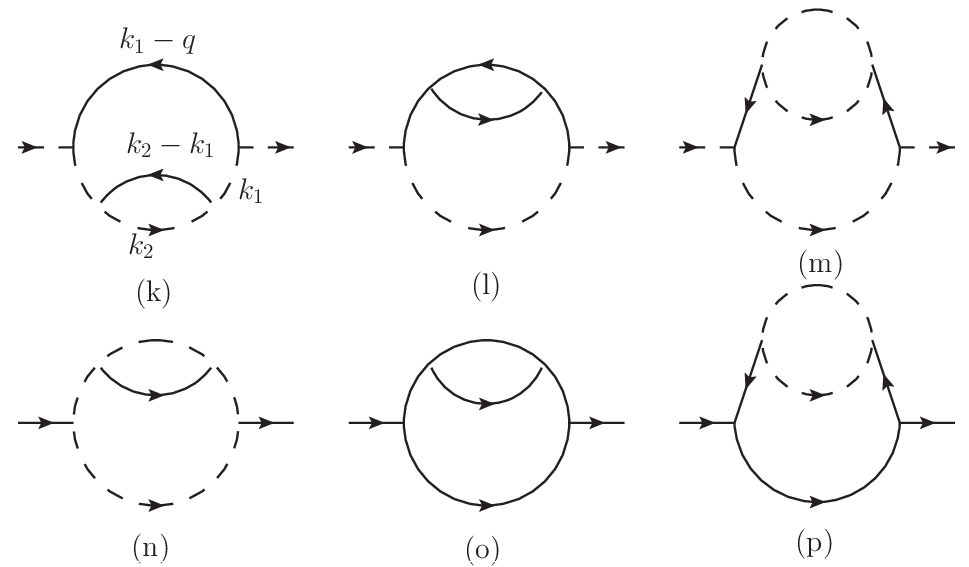}
\caption{The third type of 2-loop diagrams.}
\label{fig:2loopclass2}
\end{center}
\end{figure*}
The second type of 2-loop diagrams are shown in Fig.~\ref{fig:2loopclass2}. The diagrams in (k), (l), (m), (n), (o) and (p)  of Fig.~\ref{fig:2loopclass2} are denoted as $f^{(2)}_{\alpha }(q^2)$, where $\alpha = k,l,m,n,o,p$,
\begin{equation}
\begin{split}
&f^{(2)}_k(q^2)=\int \frac{d^3k_1}{(2\pi)^3}\int \frac{d^3k_2}{(2\pi)^3}\frac{1}{(k_1^2+\lambda^2)^2}\frac{1}{(k_1-q)^2+m_{\sigma}^2}\frac{1}{(k_1-k_2)^2+m_{\sigma}^2}\frac{1}{k_2^2+\lambda^2},\\
&f^{(2)}_l(q^2)=\int \frac{d^3k_1}{(2\pi)^3}\int \frac{d^3k_2}{(2\pi)^3}\frac{1}{(k_1^2+m_{\sigma}^2)^2}\frac{1}{(k_1-q)^2}\frac{1}{(k_1-k_2)^2+m_{\sigma}^2}\frac{1}{k_2^2+m_{\sigma}^2},\\
&f^{(2)}_m(q^2)=\int \frac{d^3k_1}{(2\pi)^3}\int \frac{d^3k_2}{(2\pi)^3}\frac{1}{(k_1^2+m_{\sigma}^2)^2}\frac{1}{(k_1-q)^2}\frac{1}{(k_1-k_2)^2}\frac{1}{k_2^2},\\
&f^{(2)}_n(q^2)=\int \frac{d^3k_1}{(2\pi)^3}\int \frac{d^3k_2}{(2\pi)^3}\frac{1}{(k_1^2+\lambda^2)^2}\frac{1}{(k_1-q)^2+\lambda^2}\frac{1}{(k_1-k_2)^2+m_{\sigma}^2}\frac{1}{k_2^2+\lambda^2},\\
&f^{(2)}_o(q^2)=\int \frac{d^3k_1}{(2\pi)^3}\int \frac{d^3k_2}{(2\pi)^3}\frac{1}{(k_1^2+m_{\sigma}^2)^2}\frac{1}{(k_1-q)^2+m_{\sigma}^2}\frac{1}{(k_1-k_2)^2+m_{\sigma}^2}\frac{1}{k_2^2+m_{\sigma}^2},\\
&f^{(2)}_p(q^2)=\int \frac{d^3k_1}{(2\pi)^3}\int \frac{d^3k_2}{(2\pi)^3}\frac{1}{(k_1^2+m_{\sigma}^2)^2}
\frac{1}{(k_1-q)^2+m_{\sigma}^2}\frac{1}{(k_1-k_2)^2}
\frac{1}{k_2^2},\\
\end{split}
\label{eq:ap.38}
\end{equation}
which  are UV finite.   $f_k^{(2)}$ in Eq.~(\ref{eq:ap.38}) and $f_h^{(2)}$ in Eq.~(\ref{eq:ap.19}) can be rewritten as
\begin{equation}
\begin{split}
&f^{(2)}_k(q^2)=\int \frac{d^3k_1}{(2\pi)^3}\int \frac{d^3k_2}{(2\pi)^3}\frac{1}{k_1^2+\lambda^2}\frac{1}{(k_1-k_2)^2+m_{\sigma}^2}\frac{1}{(k_2^2+\lambda^2)^2}\frac{1}{(k_2-q)^2+m_{\sigma}^2},\\
&f^{(2)}_h(q^2)=\int \frac{d^3k_1}{(2\pi)^3}\int \frac{d^3k_2}{(2\pi)^3}\frac{1}{k_1^2+\lambda^2}\frac{1}{(k_2-k_1)^2+m_{\sigma}^2}\frac{1}{k_2^2+\lambda^2}\frac{1}{(k_2+q)^2+m_{\sigma}^2}.\\
\end{split}
\label{eq:ap.39}
\end{equation}

We find, at order $\mathcal{O}(\lambda ^0)$,
\begin{equation}
\begin{split}
&\left(-1+2m_{\sigma}^2{\bf 5}^++({\bf 3}^--{\bf 5}^-+m_{\sigma}^2){\bf 4}^++(({\bf 1}^--m_1^2)-{\bf 5}^-+2m_{\sigma}^2){\bf 2}^+\right)f_h^{(2)}=0,\\
\end{split}
\label{eq:ap.40}
\end{equation}
where
\begin{equation}
\begin{split}
&{\bf 4}^+f_h^{(2)}=f_k^{(2)},\\
&({\bf 1}^--m_1^2){\bf 2}^+f_h^{(2)}\\
&=\int \frac{d^3k_1}{(2\pi)^3}\int \frac{d^3k_2}{(2\pi)^3}\frac{k_1^2+2k_1\cdot q+q^2}{k_1^2}\frac{1}{(k_2-k_1)^2+m_{\sigma}^2}\frac{1}{k_2^2}\frac{1}{((k_2+q)^2+m_{\sigma}^2)^2}.\\
\end{split}
\label{eq:ap.41}
\end{equation}

All the other integrals in Eq.~(\ref{eq:ap.40})  can be calculated to obtain $f_k^{(2)}$. Using this procedure, we find
\begin{equation}
\begin{split}
&f_k^{(2)}(q^2)=\frac{1}{m_{\sigma}^2}f_h^{(2)}(q^2)+\frac{1}{(4\pi)^3m_{\sigma}^4}\left\{\frac{2 \pi m_{\sigma}}{\lambda (t+1)}+\frac{2 \pi}{3 \sqrt{t} (t+1)^3}\left[-\sqrt{t} (t (3 t+5)+6) \log (t+4)\right.\right.\\
&\left.\left.+2  \left(-(t+5) \left(t^2+t+1\right) \cot ^{-1}\left(\frac{t+2}{\sqrt{t}}\right)+(2 t+3) \left(-\sqrt{t}\right) (t+1)\right.\right.\right.\\
&\left.\left.\left.-(3 t (t (t+2)+2)-1) \tan ^{-1}\left(\sqrt{t}\right)+3 (t+1)^3 \cot ^{-1}\left(\frac{2}{\sqrt{t}}\right)\right)\right]\right\},
\end{split}
\label{eq:ap.42}
\end{equation}
where $t\equiv q^2/m_{\sigma}^2$.

Similarly, we find
\begin{equation}
\begin{split}
&f_l^{(2)}(q^2)=\frac{1}{2m_{\sigma}^2}f_i^{(2)}(q^2)+\frac{\pi}{(4\pi)^3m_{\sigma}^4}\left\{\frac{(\log (81)-2 \log (t+4))}{2(1+t)}-\frac{\csc ^{-1}\left(\sqrt{\frac{1}{t}+1}\right)}{\sqrt{t}}\right.\\
&\left.-\frac{ \tan ^{-1}\left(\frac{\sqrt{t}}{2}\right)}{\sqrt{t}}-\frac{\left(\tan ^{-1}\left(\frac{t-1}{2 \sqrt{t}}\right)+\tan ^{-1}\left(\frac{4-t}{4 \sqrt{t}}\right)\right)}{6 \sqrt{t}}\right\},
\end{split}
\label{eq:ap.43}
\end{equation}
\begin{equation}
\begin{split}
&f_m^{(2)}(q^2)=\frac{1}{2m_{\sigma}^2}f_f^{(2)}(q^2)+\frac{1}{(4\pi)^3m_{\sigma}^4}\frac{\pi  \log \left(\frac{1}{t}\right)}{t+1},
\end{split}
\label{eq:ap.44}
\end{equation}
\begin{equation}
\begin{split}
&f_n^{(2)}(q^2)=-\frac{1}{(4\pi)^3m_{\sigma}^4}\frac{\pi}{6 t^2}  \left\{t \left(-\frac{12m_{\sigma}}{\lambda}+\pi  \sqrt{t}+16\right)\right.\\
&\left.+2 t^{3/2} \left(2 \tan ^{-1}\left(\sqrt{t}\right)-3 \tan ^{-1}\left(\frac{t-1}{2 \sqrt{t}}\right)\right)-4 \log (t+1)\right\},
\end{split}
\label{eq:ap.45}
\end{equation}
\begin{equation}
\begin{split}
&f_o^{(2)}(q^2)=\frac{1}{2m_{\sigma}^2}f_j^{(2)}(q^2)+\frac{\pi}{(4\pi)^3m_{\sigma}^4}\left\{\frac{\tan ^{-1}\left(\frac{t+3}{2 \sqrt{t}}\right)}{\sqrt{t}}-\frac{\cot ^{-1}\left(\frac{2}{\sqrt{t}}\right)}{\sqrt{t}}-\frac{\tan ^{-1}\left(\frac{3-t}{4 \sqrt{t}}\right)}{\sqrt{t}}\right.\\
&\left.-\frac{(t+2) \left[2\log \left(\frac{t+9}{9}\right)+\sqrt{t} \left((t+4) \left(2 \tan ^{-1}\left(-\frac{3}{\sqrt{t}}\right)+\pi \right)+i (t+3) \log \left(-1+\frac{6 i}{\sqrt{t}+3 i}\right)\right)\right]}{2t (t+4)}\right.\\
&\left.+\frac{\tan ^{-1}\left(\frac{t+3}{2 \sqrt{t}}\right)+\tan ^{-1}\left(\frac{t-3}{4 \sqrt{t}}\right)-2 \cot ^{-1}\left(\frac{2}{\sqrt{t}}\right)}{6 \sqrt{t}}\right\},
\end{split}
\label{eq:ap.46}
\end{equation}
\begin{equation}
\begin{split}
&f_p^{(2)}(q^2)=\frac{1}{2(4\pi)^3m_{\sigma}^4t^2}\left\{\frac{1}{2} i t^{3/2}\left[-\text{Li}_2\left(i \sqrt{t}+1\right)+\text{Li}_2\left(\frac{1}{i \sqrt{t}+2}\right)+\text{Li}_2\left(1-i \sqrt{t}\right)\right.\right.\\
&\left.\left.-\text{Li}_2\left(\frac{1}{2-i \sqrt{t}}\right)+\text{Li}_2\left(1+\frac{1}{i \sqrt{t}-2}\right)-\text{Li}_2\left(1+\frac{1}{-i \sqrt{t}-2}\right)+\text{Li}_2\left(\frac{2 i}{2 i+\sqrt{t}-\sqrt{t+4}}\right)\right.\right.\\
&\left.\left.-\text{Li}_2\left(\frac{2 \left(i+\sqrt{t}\right)}{2 i+\sqrt{t}-\sqrt{t+4}}\right)-\text{Li}_2\left(\frac{2 i}{2 i-\sqrt{t}+\sqrt{t+4}}\right)+\text{Li}_2\left(\frac{2 i-2 \sqrt{t}}{2 i-\sqrt{t}+\sqrt{t+4}}\right)\right.\right.\\
&\left.\left.+\text{Li}_2\left(\frac{2 \left(-i+\sqrt{t}\right)}{-2 i+\sqrt{t}+\sqrt{t+4}}\right)-\text{Li}_2\left(-\frac{2 i}{-2 i+\sqrt{t}+\sqrt{t+4}}\right)+\text{Li}_2\left(\frac{2 i}{2 i+\sqrt{t}+\sqrt{t+4}}\right)\right.\right.\\
&\left.\left.-\text{Li}_2\left(\frac{2 \left(i+\sqrt{t}\right)}{2 i+\sqrt{t}+\sqrt{t+4}}\right)\right]-\frac{t (t+2) \log (t+1)}{t+4}+\frac{t^{3/2}}{4 (t+4)}\left[-\pi  (t+4) \log \left(\frac{1}{t}+1\right)\right.\right.\\
&\left.\left.+2 i (t+4) \log \left(-\sqrt{t}+i\right)+4 i \log \left(\sqrt{t}+i\right)-2 (t+4) \log (t) \cot ^{-1}\left(\sqrt{t}\right)\right.\right.\\
&\left.\left.+(t+6) \left(\pi -2 i \log \left(1+i \sqrt{t}\right)\right)\right]\right\},
\end{split}
\label{eq:ap.47}
\end{equation}
where $t\equiv q^2/m_{\sigma}^2$.

We also obtain
\begin{equation}
\begin{split}
&f_k^{(2)}(q^2=0)=\frac{1}{2(4\pi)^2m_{\sigma}^4}\left(-2+\frac{m_{\sigma}}{\lambda}-\log (4)\right),\;f_l^{(2)}(q^2=0)=\frac{1}{2(4\pi)^2m_{\sigma}^4}\left(2 \log \left(\frac{3}{2}\right)-\frac{1}{3}\right),\\
&f_m^{(2)}(q^2=0)=\frac{1}{2(4\pi)^2m_{\sigma}^4}\left(-1-2\log(3)-\log\left(\frac{\lambda^2}{m_{\sigma}^2}\right)\right).\\
\end{split}
\label{eq:ap.48}
\end{equation}

\subsubsection{\label{sec:a25}The fourth type of 2-loop diagrams.}

\begin{figure*}
\begin{center}
\includegraphics[scale=0.7]{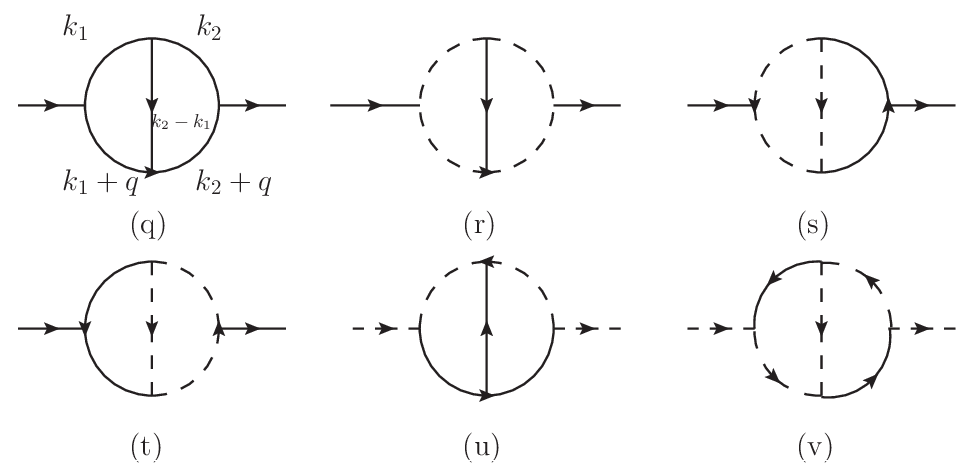}
\caption{The  fourth type of 2-loop diagrams.}
\label{fig:2loopclass4}
\end{center}
\end{figure*}

The second type of 2-loop diagrams are shown in Fig.~\ref{fig:2loopclass4}. The diagrams in (q), (r), (s), (t), (u) and (v) of Fig.~\ref{fig:2loopclass4}  are denoted as $f^{(2)}_{\alpha}(q^2)$, with $\alpha =   q,r,s,t,u,v$,
\begin{equation}
\begin{split}
&f_q^{(2)}(q^2)=\int \frac{d^3k_1}{(2\pi)^3}\int \frac{d^3k_2}{(2\pi)^3}\frac{1}{k_1^2+m_{\sigma}^2}\frac{1}{(k_1+q)^2+m_{\sigma}^2}\frac{1}{(k_1-k_2)^2+m_{\sigma}^2}\\
&\times \frac{1}{k_2^2+m_{\sigma}^2}\frac{1}{(k_2+q)^2+m_{\sigma}^2},\\
&f_r^{(2)}(q^2)=\int \frac{d^3k_1}{(2\pi)^3}\int \frac{d^3k_2}{(2\pi)^3}\frac{1}{k_1^2}\frac{1}{(k_1+q)^2}\frac{1}{(k_1-k_2)^2+m_{\sigma}^2}\frac{1}{k_2^2}\frac{1}{(k_2+q)^2},\\
&f_s^{(2)}(q^2)=f_t^{(2)}(q^2)=\int \frac{d^3k_1}{(2\pi)^3}\int \frac{d^3k_2}{(2\pi)^3}\frac{1}{k_1^2}\frac{1}{(k_1+q)^2}\frac{1}{(k_1-k_2)^2}\frac{1}{k_2^2+m_{\sigma}^2}\frac{1}{(k_2+q)^2+m_{\sigma}^2},\\
&f_u^{(2)}(q^2)=\int \frac{d^3k_1}{(2\pi)^3}\int \frac{d^3k_2}{(2\pi)^3}\frac{1}{k_1^2+m_{\sigma}^2}\frac{1}{(k_1-q)^2}\frac{1}{(k_1-k_2)^2+m_{\sigma}^2}\frac{1}{k_2^2+m_{\sigma}^2}\frac{1}{(k_2-q)^2},\\
&f_v^{(2)}(q^2)=\int \frac{d^3k_1}{(2\pi)^3}\int \frac{d^3k_2}{(2\pi)^3}\frac{1}{k_1^2+m_{\sigma}^2}
\frac{1}{(k_1+q)^2}\frac{1}{(k_1-k_2)^2}\frac{1}{k_2^2}
\frac{1}{(k_2+q)^2+m_{\sigma}^2}, \\
\end{split}
\label{eq:ap.49}
\end{equation}
which  are UV finite.

One can use IBP relation to obtain the differential equations for the diagrams~\cite{IBPwithDE}. Take $f_q^{(2)}$ as the example,
\begin{equation}
\begin{split}
&{\bf 1}^+f_q^{(2)}={\bf 2}^+f_q^{(2)}={\bf 3}^+f_q^{(2)}={\bf 4}^+f_q^{(2)},\\
\end{split}
\label{eq:ap.50}
\end{equation}
thus
\begin{equation}
\begin{split}
&\frac{\partial}{\partial (m^2)}f_q^{(2)}={\bf 5}^+f_q^{(2)}+4{\bf 1}^+f_q^{(2)}.\\
\end{split}
\label{eq:ap.51}
\end{equation}

Using Eqs.~(\ref{eq:ap.36}), (\ref{eq:ap.50}) and (\ref{eq:ap.51}), we find
\begin{equation}
\begin{split}
&\left(2+\frac{1}{3u}+(2u+\frac{2}{3})\frac{\partial }{\partial u}\right)f_q^{(2)}=\left(\frac{2}{3u}({\bf 3}^--{\bf 5}^-){\bf 4}^+-2{\bf 2}^-{\bf 4}^+\right)f_q^{(2)},\\
\end{split}
\label{eq:ap.52}
\end{equation}
where $u\equiv m_{\sigma}^2/q^2$. Defining
\begin{equation}
\begin{split}
&g_q(u)=\frac{1}{\sqrt{u} \sqrt{3 u+1}}f_q^{(2)}(u),
\end{split}
\label{eq:ap.53}
\end{equation}
we find
\begin{equation}
\begin{split}
&\frac{2u+\frac{2}{3}}{\sqrt{u} \sqrt{3 u+1}}g_q'(u)=\left(\frac{2}{3u}({\bf 3}^--{\bf 5}^-){\bf 4}^+-2{\bf 2}^-{\bf 4}^+\right)f_q^{(2)}, \\
\end{split}
\label{eq:ap.54}
\end{equation}
so
\begin{equation}
\begin{split}
&g_q(u)=\int _0^{\frac{m^2}{q^2}} \frac{\sqrt{u} \sqrt{3 u+1}}{2u+\frac{2}{3}}\left(\frac{2}{3u}({\bf 3}^--{\bf 5}^-){\bf 4}^+-2{\bf 2}^-{\bf 4}^+\right)f_q^{(2)}(u) du+C,
\end{split}
\label{eq:ap.55}
\end{equation}
where $C$ is a constant.

The integrals in Eq.~(\ref{eq:ap.55}) only have four denominators, and  are thus easier to evaluate. One can obtain
\begin{equation}
\begin{split}
&f_q^{(2)}(t)=-\frac{1}{(4\pi)^3m_{\sigma}^4}\frac{1}{t \sqrt{t+3}} \int _0^{\frac{1}{t}}du\\
&\times \frac{2 \pi  \left(\sqrt{u} \log \left(\frac{9 u}{9 u+1}\right)+(4 u+2) \cot ^{-1}\left(\frac{6 u+1}{\sqrt{u}}\right)\right)}{u \sqrt{3 u+1} (4 u+1)}
\end{split}
\label{eq:ap.56}
\end{equation}

Using such a procedure, we find
\begin{equation}
\begin{split}
&f_q^{(2)}(t)=\frac{1}{(4\pi)^3m_{\sigma}^4}\frac{1}{t \sqrt{t+3}} \left\{f_{q_1}(\frac{1}{t})+4\pi i\left[\sum _{i=2}^{6}f_{q_i}\left(\sin^{-1}
\left(\sqrt{1+\frac{3}{t}}\right)\right)-C_{q_1}
-C_{q_2}\right]\right\},
\end{split}
\label{eq:ap.57}
\end{equation}
where
\begin{equation}
\begin{split}
&f_{q_1}(u)=4 i \pi  \log \left(\frac{6 u-i \sqrt{u}+1}{6 u+i \sqrt{u}+1}\right) \left(\tanh ^{-1}\left(\sqrt{3 u+1}\right)-\tanh ^{-1}\left(2 \sqrt{3 u+1}\right)\right)\\
&-2 \pi  \left\{\frac{1}{3} i \left[\pi ^2-3 \left(\text{Li}_2\left(\frac{1}{5} \left(2 \sqrt{6}+7\right)\right)+\text{Li}_2\left(\frac{1}{5} \left(7-2 \sqrt{6}\right)\right)\right)\right]\right.\\
&\left.+\pi  \log \left(\frac{9}{5}\right)+2 \log \left(\frac{1}{5} \left(7-2 \sqrt{6}\right)\right) \sin ^{-1}\left(\sqrt{\frac{6}{5}}\right)\right\}+2 \pi  \left\{2 i \text{Li}_2\left(-e^{2 i \tan ^{-1}\left(\sqrt{3+\frac{1}{u}}\right)}\right)\right.\\
&\left.-i \text{Li}_2\left(\frac{1}{5} \left(2 \sqrt{6}-7\right) e^{2 i \tan ^{-1}\left(\sqrt{3+\frac{1}{u}}\right)}\right)-i \text{Li}_2\left(-\frac{1}{5} \left(2 \sqrt{6}+7\right) e^{2 i \tan ^{-1}\left(\sqrt{3+\frac{1}{u}}\right)}\right)\right.\\
&\left.+2 \left[\log \left(\frac{9}{5}\right) \tan ^{-1}\left(\sqrt{\frac{1}{u}+3}\right)+\log \left(\frac{1}{5} \left(7-2 \sqrt{6}\right)\right) \sin ^{-1}\left(\sqrt{\frac{6}{5}}\right)\right]\right\},\\
\end{split}
\label{eq:ap.58}
\end{equation}
\begin{equation}
\begin{split}
&f_{q_2}(a)=-\frac{1}{2} \left(4 \tanh ^{-1}\left(\tan \left(\frac{a}{2}\right)\right)+\log (3)\right) \log \left(\sqrt{3}-\tan \left(\frac{a}{2}\right)-2\right)\\
&-\log \left(\sqrt{3}-\tan \left(\frac{a}{2}\right)+2\right) \log \left(\tan \left(\frac{a}{2}\right)-1\right)\\
&+\tanh ^{-1}\left(\frac{\tan \left(\frac{a}{2}\right)-2}{\sqrt{3}}\right) \left[\log \left(-3-\frac{6}{\sin (a)-1}\right)-2 \log \left(\tan \left(\frac{a}{2}\right)+1\right)\right]\\
&-\tanh ^{-1}(2 \sin (a)) \log \left(\frac{\left(\sqrt{2}+\sqrt{6}-2 \tan \left(\frac{a}{2}\right)\right) \left(\sqrt{3}+\tan \left(\frac{a}{2}\right)-2\right) \left(\sqrt{3}-\tan \left(\frac{a}{2}\right)-2\right)}{\left(-\sqrt{2}+\sqrt{6}-2 \tan \left(\frac{a}{2}\right)\right) \left(\sqrt{3}-\tan \left(\frac{a}{2}\right)+2\right) \left(\sqrt{3}+\tan \left(\frac{a}{2}\right)+2\right)}\right)\\
&+\tanh ^{-1}(\sin (a)) \log \left(\frac{\left(\sqrt{2}+\sqrt{6}-2 \tan \left(\frac{a}{2}\right)\right) \left(\sqrt{2}+\sqrt{6}+2 \tan \left(\frac{a}{2}\right)\right) \left(\sqrt{3}-\tan \left(\frac{a}{2}\right)-2\right)}{\left(-\sqrt{2}+\sqrt{6}-2 \tan \left(\frac{a}{2}\right)\right) \left(-\sqrt{2}+\sqrt{6}+2 \tan \left(\frac{a}{2}\right)\right) \left(\sqrt{3}+\tan \left(\frac{a}{2}\right)+2\right)}\right),\\
\end{split}
\label{eq:ap.59}
\end{equation}
\begin{equation}
\begin{split}
&f_{q_3}(a)=\frac{1}{4} \left[\log ^2\left(\frac{\sqrt{3}-\tan \left(\frac{a}{2}\right)-2}{\sqrt{3}-\tan \left(\frac{a}{2}\right)+2}\right)-\log ^2\left(\frac{\sqrt{3}+\tan \left(\frac{a}{2}\right)-2}{\sqrt{3}+\tan \left(\frac{a}{2}\right)+2}\right)\right.\\
&\left.-2 \log \left(\sqrt{3}+\tan \left(\frac{a}{2}\right)-2\right) \log \left(\sqrt{3}-\tan \left(\frac{a}{2}\right)+2\right)\right.\\
&\left.-2 \log \left(\sqrt{3}-\tan \left(\frac{a}{2}\right)+2\right) \log \left(-\sqrt{3}+\tan \left(\frac{a}{2}\right)+2\right)\right.\\
&\left.+2 \log \left(-\sqrt{3}-\tan \left(\frac{a}{2}\right)+2\right) \log \left(\sqrt{3}+\tan \left(\frac{a}{2}\right)+2\right)\right.\\
&\left.+2 \log \left(\sqrt{3}-\tan \left(\frac{a}{2}\right)-2\right) \log \left(\sqrt{3}+\tan \left(\frac{a}{2}\right)+2\right)\right]+\log \left(1-\tan \left(\frac{a}{2}\right)\right) \log \left(\sqrt{3}+\tan \left(\frac{a}{2}\right)-2\right)\\
&+\frac{1}{2} \left[\log (3)+2 \log \left(\frac{-\tan \left(\frac{a}{2}\right)-1}{1-\tan \left(\frac{a}{2}\right)}\right)\right] \log \left(\sqrt{3}+\tan \left(\frac{a}{2}\right)+2\right)\\
&+\log \left(\frac{\left(2-2 \tan \left(\frac{a}{2}\right)\right) \left(-\sqrt{2}-\sqrt{6}+2\right)}{\left(\sqrt{2}+\sqrt{6}+2\right) \left(2 \tan \left(\frac{a}{2}\right)+2\right)}\right) \log \left(\sqrt{2}+\sqrt{6}+2 \tan \left(\frac{a}{2}\right)\right)\\
&-\frac{1}{2} \left\{\left[\log \left(\frac{10 \sin (a)+5}{\left(11-4 \sqrt{6}\right) (2 \sin (a)-1)}\right)+\log \left(\frac{1-2 \sin (a)}{2 \sin (a)+1}\right)\right] \log \left(-\sqrt{2}+\sqrt{6}+2 \tan \left(\frac{a}{2}\right)\right)\right.\\
&\left.-\left[\log \left(\frac{1-2 \sin (a)}{2 \sin (a)+1}\right)+\log \left(\frac{\left(4 \sqrt{6}+11\right) (2 \sin (a)+1)}{10 \sin (a)-5}\right)\right] \log \left(\sqrt{2}+\sqrt{6}+2 \tan \left(\frac{a}{2}\right)\right)\right.\\
&\left.+\log \left(\frac{\left(4 \sqrt{6}+11\right) (2 \sin (a)-1)}{10 \sin (a)+5}\right) \left(-\log \left(-\sqrt{2}+\sqrt{6}-2 \tan \left(\frac{a}{2}\right)\right)\right)\right.\\
&\left.+\log \left(\frac{\left(4 \sqrt{6}+11\right) (2 \sin (a)-1)}{10 \sin (a)+5}\right) \log \left(\sqrt{2}+\sqrt{6}-2 \tan \left(\frac{a}{2}\right)\right)\right\}\\
&+\log \left(-\sqrt{2}+\sqrt{6}+2 \tan \left(\frac{a}{2}\right)\right) \log \left(\frac{\left(2 \left(\tan \left(\frac{a}{2}\right)+1\right)\right) \left(-\sqrt{2}+\sqrt{6}+2\right)}{\left(\sqrt{2}-\sqrt{6}+2\right) \left(2-2 \tan \left(\frac{a}{2}\right)\right)}\right)\\
&+\log \left(\frac{\left(2 \left(\tan \left(\frac{a}{2}\right)+1\right)\right) \left(\sqrt{2}-\sqrt{6}+2\right)}{\left(-\sqrt{2}+\sqrt{6}+2\right) \left(2-2 \tan \left(\frac{a}{2}\right)\right)}\right) \log \left(-\sqrt{2}+\sqrt{6}-2 \tan \left(\frac{a}{2}\right)\right)\\
&+\log \left(\frac{\left(2 \left(\tan \left(\frac{a}{2}\right)-1\right)\right) \left(\sqrt{2}+\sqrt{6}+2\right)}{\left(\sqrt{2}+\sqrt{6}-2\right) \left(2 \left(\tan \left(\frac{a}{2}\right)+1\right)\right)}\right) \log \left(\sqrt{2}+\sqrt{6}-2 \tan \left(\frac{a}{2}\right)\right)\\
&-\frac{1}{2} \left\{-\log \left(\sqrt{3}-\tan \left(\frac{a}{2}\right)+2\right) \log \left(\frac{1}{3} \sqrt{48} \left(\left(\sqrt{3}+2\right) \tan \left(\frac{a}{2}\right)-1\right)\right)\right.\\
&\left.-\log \left(\sqrt{3}+\tan \left(\frac{a}{2}\right)-2\right) \log \left(\frac{1}{4} \sqrt{3} \left(\left(\sqrt{3}+2\right) \tan \left(\frac{a}{2}\right)+1\right)\right)\right.\\
&\left.+\log \left(\sqrt{3}-\tan \left(\frac{a}{2}\right)-2\right) \log \left(\frac{1}{4} \sqrt{3} \left(1-\left(\sqrt{3}+2\right) \tan \left(\frac{a}{2}\right)\right)\right)\right.\\
&\left.+\log \left(\sqrt{3}+\tan \left(\frac{a}{2}\right)+2\right) \log \left(\frac{4}{3} \left(-\sqrt{3}-\left(2 \sqrt{3}+3\right) \tan \left(\frac{a}{2}\right)\right)\right)\right\},\\
\end{split}
\label{eq:ap.60}
\end{equation}
\begin{equation}
\begin{split}
&f_{q_4}(a)=\text{Li}_2\left(\frac{\sqrt{3}-\tan \left(\frac{a}{2}\right)-2}{\sqrt{3}-3}\right)-\text{Li}_2\left(\frac{\sqrt{3}+\tan \left(\frac{a}{2}\right)-2}{\sqrt{3}-3}\right)\\
&-\text{Li}_2\left(\frac{\sqrt{3}-\tan \left(\frac{a}{2}\right)-2}{\sqrt{3}-1}\right)+\text{Li}_2\left(\frac{\sqrt{3}+\tan \left(\frac{a}{2}\right)-2}{\sqrt{3}-1}\right)-\text{Li}_2\left(\frac{\sqrt{3}-\tan \left(\frac{a}{2}\right)+2}{\sqrt{3}+1}\right)\\
&+\text{Li}_2\left(\frac{\sqrt{3}+\tan \left(\frac{a}{2}\right)+2}{\sqrt{3}+1}\right)+\text{Li}_2\left(\frac{\sqrt{3}-\tan \left(\frac{a}{2}\right)+2}{\sqrt{3}+3}\right)-\text{Li}_2\left(\frac{\sqrt{3}+\tan \left(\frac{a}{2}\right)+2}{\sqrt{3}+3}\right)\\
&-\text{Li}_2\left(\frac{\sqrt{2}-\sqrt{6}+2 \tan \left(\frac{a}{2}\right)}{\sqrt{2}-\sqrt{6}+2}\right)+\text{Li}_2\left(\frac{\sqrt{2}-\sqrt{6}-2 \tan \left(\frac{a}{2}\right)}{\sqrt{2}-\sqrt{6}+2}\right)+\text{Li}_2\left(\frac{-\sqrt{2}+\sqrt{6}-2 \tan \left(\frac{a}{2}\right)}{-\sqrt{2}+\sqrt{6}+2}\right)\\
&+\text{Li}_2\left(\frac{\sqrt{2}+\sqrt{6}-2 \tan \left(\frac{a}{2}\right)}{\sqrt{2}+\sqrt{6}-2}\right)-\text{Li}_2\left(\frac{\sqrt{2}+\sqrt{6}-2 \tan \left(\frac{a}{2}\right)}{\sqrt{2}+\sqrt{6}+2}\right)-\text{Li}_2\left(\frac{-\sqrt{2}+\sqrt{6}+2 \tan \left(\frac{a}{2}\right)}{-\sqrt{2}+\sqrt{6}+2}\right)\\
&-\text{Li}_2\left(\frac{\sqrt{2}+\sqrt{6}+2 \tan \left(\frac{a}{2}\right)}{\sqrt{2}+\sqrt{6}-2}\right)+\text{Li}_2\left(\frac{\sqrt{2}+\sqrt{6}+2 \tan \left(\frac{a}{2}\right)}{\sqrt{2}+\sqrt{6}+2}\right),
\end{split}
\end{equation}
\label{eq:ap.61}
\begin{equation}
\begin{split}
&f_{q_5}(a)=\frac{1}{2}\left\{\text{Li}_2\left(\frac{\sqrt{2}-\sqrt{6}+2 \tan \left(\frac{a}{2}\right)}{\sqrt{2}+2 \sqrt{3}-\sqrt{6}+4}\right)+\text{Li}_2\left(\frac{\sqrt{2}-\sqrt{6}+2 \tan \left(\frac{a}{2}\right)}{\sqrt{2}-2 \sqrt{3}-\sqrt{6}+4}\right)\right.\\
&\left.-\text{Li}_2\left(\frac{-\sqrt{2}+\sqrt{6}-2 \tan \left(\frac{a}{2}\right)}{-\sqrt{2}+2 \sqrt{3}+\sqrt{6}+4}\right)-\text{Li}_2\left(\frac{\sqrt{2}+\sqrt{6}-2 \tan \left(\frac{a}{2}\right)}{\sqrt{2}+2 \sqrt{3}+\sqrt{6}-4}\right)+\text{Li}_2\left(\frac{\sqrt{2}+\sqrt{6}-2 \tan \left(\frac{a}{2}\right)}{\sqrt{2}+2 \sqrt{3}+\sqrt{6}+4}\right)\right.\\
&\left.-\text{Li}_2\left(\frac{-\sqrt{2}+\sqrt{6}-2 \tan \left(\frac{a}{2}\right)}{-\sqrt{2}-2 \sqrt{3}+\sqrt{6}+4}\right)-\text{Li}_2\left(\frac{\sqrt{2}+\sqrt{6}-2 \tan \left(\frac{a}{2}\right)}{\sqrt{2}-2 \sqrt{3}+\sqrt{6}-4}\right)+\text{Li}_2\left(\frac{\sqrt{2}+\sqrt{6}-2 \tan \left(\frac{a}{2}\right)}{\sqrt{2}-2 \sqrt{3}+\sqrt{6}+4}\right)\right.\\
&\left.-\text{Li}_2\left(\frac{\sqrt{2}-\sqrt{6}-2 \tan \left(\frac{a}{2}\right)}{\sqrt{2}+2 \sqrt{3}-\sqrt{6}+4}\right)-\text{Li}_2\left(\frac{-\sqrt{2}+\sqrt{6}+2 \tan \left(\frac{a}{2}\right)}{-\sqrt{2}+2 \sqrt{3}+\sqrt{6}-4}\right)+\text{Li}_2\left(\frac{-\sqrt{2}+\sqrt{6}+2 \tan \left(\frac{a}{2}\right)}{-\sqrt{2}+2 \sqrt{3}+\sqrt{6}+4}\right)\right.\\
&\left.+\text{Li}_2\left(\frac{\sqrt{2}+\sqrt{6}+2 \tan \left(\frac{a}{2}\right)}{\sqrt{2}+2 \sqrt{3}+\sqrt{6}-4}\right)-\text{Li}_2\left(\frac{\sqrt{2}+\sqrt{6}+2 \tan \left(\frac{a}{2}\right)}{\sqrt{2}+2 \sqrt{3}+\sqrt{6}+4}\right)+\text{Li}_2\left(\frac{-\sqrt{2}+\sqrt{6}+2 \tan \left(\frac{a}{2}\right)}{-\sqrt{2}-2 \sqrt{3}+\sqrt{6}+4}\right)\right.\\
&\left.+\text{Li}_2\left(\frac{\sqrt{2}+\sqrt{6}+2 \tan \left(\frac{a}{2}\right)}{\sqrt{2}-2 \sqrt{3}+\sqrt{6}-4}\right)-\text{Li}_2\left(\frac{\sqrt{2}+\sqrt{6}+2 \tan \left(\frac{a}{2}\right)}{\sqrt{2}-2 \sqrt{3}+\sqrt{6}+4}\right)\right\},
\end{split}
\end{equation}
\label{eq:ap.62}
\begin{equation}
\begin{split}
&f_{q_6}(a)=\frac{1}{2}\left\{\text{Li}_2\left(\frac{\sqrt{3}-\tan \left(\frac{a}{2}\right)-2}{2 \sqrt{3}}\right)+\text{Li}_2\left(\frac{\sqrt{3}-\tan \left(\frac{a}{2}\right)+2}{2 \sqrt{3}}\right)-\text{Li}_2\left(\frac{\sqrt{3}+\tan \left(\frac{a}{2}\right)-2}{2 \sqrt{3}}\right)\right.\\
&\left.-\text{Li}_2\left(\frac{\sqrt{3}+\tan \left(\frac{a}{2}\right)+2}{2 \sqrt{3}}\right)+\text{Li}_2\left(\frac{\sqrt{3}+\tan \left(\frac{a}{2}\right)-2}{2 \sqrt{3}-4}\right)-\text{Li}_2\left(\frac{\sqrt{3}-\tan \left(\frac{a}{2}\right)+2}{2 \sqrt{3}+4}\right)\right.\\
&\left.+\text{Li}_2\left(\frac{\sqrt{3}+\tan \left(\frac{a}{2}\right)+2}{2 \sqrt{3}+4}\right)+\text{Li}_2\left(\frac{1}{4} \left(-\sqrt{3}-\tan \left(\frac{a}{2}\right)+2\right)\right)-\text{Li}_2\left(\frac{1}{4} \left(\sqrt{3}-\tan \left(\frac{a}{2}\right)+2\right)\right)\right.\\
&\left.-\text{Li}_2\left(\frac{1}{4} \left(-\sqrt{3}+\tan \left(\frac{a}{2}\right)+2\right)\right)+\text{Li}_2\left(\frac{1}{4} \left(\sqrt{3}+\tan \left(\frac{a}{2}\right)+2\right)\right)\right.\\
&\left.-\text{Li}_2\left(\frac{1}{2} \left(\left(\sqrt{3}+2\right) \tan \left(\frac{a}{2}\right)+1\right)\right)\right\},
\end{split}
\label{eq:ap.63}
\end{equation}
\begin{equation}
\begin{split}
&C_{q_1}=\frac{1}{4} \left\{3 \pi ^2-2 \log \left(3-\sqrt{3}\right) \log \left(\sqrt{3}+1\right)+2 \log \left(\sqrt{3}-1\right) \log \left(\sqrt{3}+3\right)\right.\\
&\left.-i \pi  \left[\log \left(\frac{648}{25}\right)+\log \left(3255 \sqrt{3}-88 \sqrt{2} \left(26 \sqrt{3}+45\right)+5642\right)+2 \log \left(-6 \sqrt{2}-5 \sqrt{3}+4 \sqrt{6}+10\right)\right]\right\}\\
&+\frac{1}{2} \text{Li}_2\left(\frac{1}{2}-\frac{\sqrt{3}}{2}\right)+\frac{1}{2} \text{Li}_2\left(\frac{3}{2}-\frac{\sqrt{3}}{2}\right)+\frac{1}{2} \text{Li}_2\left(-\frac{\sqrt{3}}{2}-\frac{1}{2}\right)+\frac{1}{2} \text{Li}_2\left(\frac{1}{4} \left(1-\sqrt{3}\right)\right)\\
&-\frac{1}{2} \text{Li}_2\left(\frac{1}{4} \left(3-\sqrt{3}\right)\right)-\frac{1}{2} \text{Li}_2\left(\frac{1}{6} \left(3-\sqrt{3}\right)\right)-\frac{1}{2} \text{Li}_2\left(\frac{1}{2} \left(\sqrt{3}-1\right)\right)-\frac{1}{2} \text{Li}_2\left(\frac{1}{2} \left(\sqrt{3}+1\right)\right)\\
&-\frac{1}{2} \text{Li}_2\left(\frac{1}{4} \left(\sqrt{3}+1\right)\right)-\frac{1}{2} \text{Li}_2\left(\frac{1}{2} \left(\sqrt{3}+3\right)\right)+\frac{1}{2} \text{Li}_2\left(\frac{1}{4} \left(\sqrt{3}+3\right)\right)+\frac{1}{2} \text{Li}_2\left(\frac{1}{6} \left(\sqrt{3}+3\right)\right)\\
&-\frac{1}{2} \text{Li}_2\left(\frac{\sqrt{2}-\sqrt{6}-2}{\sqrt{2}+2 \sqrt{3}-\sqrt{6}+4}\right)+\frac{1}{2} \text{Li}_2\left(\frac{\sqrt{2}-\sqrt{6}+2}{\sqrt{2}+2 \sqrt{3}-\sqrt{6}+4}\right)+\frac{1}{2} \text{Li}_2\left(\frac{\sqrt{2}-\sqrt{6}+2}{\sqrt{2}-2 \sqrt{3}-\sqrt{6}+4}\right)\\
&+\text{Li}_2\left(\frac{-\sqrt{2}+\sqrt{6}+2}{-\sqrt{2}+\sqrt{6}-2}\right)+\text{Li}_2\left(\frac{-\sqrt{2}+\sqrt{6}-2}{-\sqrt{2}+\sqrt{6}+2}\right)-\text{Li}_2\left(\frac{\sqrt{2}+\sqrt{6}+2}{\sqrt{2}+\sqrt{6}-2}\right)-\text{Li}_2\left(\frac{\sqrt{2}+\sqrt{6}-2}{\sqrt{2}+\sqrt{6}+2}\right)\\
&-\frac{1}{2} \text{Li}_2\left(\frac{-\sqrt{2}+\sqrt{6}+2}{-\sqrt{2}+2 \sqrt{3}+\sqrt{6}-4}\right)-\frac{1}{2} \text{Li}_2\left(\frac{-\sqrt{2}+\sqrt{6}-2}{-\sqrt{2}+2 \sqrt{3}+\sqrt{6}+4}\right)+\frac{1}{2} \text{Li}_2\left(\frac{-\sqrt{2}+\sqrt{6}+2}{-\sqrt{2}+2 \sqrt{3}+\sqrt{6}+4}\right),\\
\end{split}
\label{eq:ap.64}
\end{equation}
\begin{equation}
\begin{split}
&C_{q_2}=-\frac{1}{2} \text{Li}_2\left(\frac{\sqrt{2}+\sqrt{6}-2}{\sqrt{2}+2 \sqrt{3}+\sqrt{6}-4}\right)+\frac{1}{2} \text{Li}_2\left(\frac{\sqrt{2}+\sqrt{6}+2}{\sqrt{2}+2 \sqrt{3}+\sqrt{6}-4}\right)\\
&+\frac{1}{2} \text{Li}_2\left(\frac{\sqrt{2}+\sqrt{6}-2}{\sqrt{2}+2 \sqrt{3}+\sqrt{6}+4}\right)-\frac{1}{2} \text{Li}_2\left(\frac{\sqrt{2}+\sqrt{6}+2}{\sqrt{2}+2 \sqrt{3}+\sqrt{6}+4}\right)-\frac{1}{2} \text{Li}_2\left(\frac{-\sqrt{2}+\sqrt{6}-2}{-\sqrt{2}-2 \sqrt{3}+\sqrt{6}+4}\right)\\
&+\frac{1}{2} \text{Li}_2\left(\frac{-\sqrt{2}+\sqrt{6}+2}{-\sqrt{2}-2 \sqrt{3}+\sqrt{6}+4}\right)-\frac{1}{2} \text{Li}_2\left(\frac{\sqrt{2}+\sqrt{6}-2}{\sqrt{2}-2 \sqrt{3}+\sqrt{6}-4}\right)+\frac{1}{2} \text{Li}_2\left(\frac{\sqrt{2}+\sqrt{6}+2}{\sqrt{2}-2 \sqrt{3}+\sqrt{6}-4}\right)\\
&+\frac{1}{2} \text{Li}_2\left(\frac{\sqrt{2}+\sqrt{6}-2}{\sqrt{2}-2 \sqrt{3}+\sqrt{6}+4}\right)-\frac{1}{2} \text{Li}_2\left(\frac{\sqrt{2}+\sqrt{6}+2}{\sqrt{2}-2 \sqrt{3}+\sqrt{6}+4}\right),
\end{split}
\label{eq:ap.65}
\end{equation}
and
\begin{equation}
\begin{split}
&f_r=\frac{1}{2(4\pi)^2m_{\sigma}^4}\times \frac{1}{t \sqrt{1-t}}\left\{2 i \text{Li}_2\left(\frac{2 \left(\sqrt{\frac{1}{1-t}}+1\right)}{\sqrt{2}+2}\right)+2 i \text{Li}_2\left(\frac{i}{\sqrt{\frac{1}{t}-1}}\right)\right.\\
&\left.-2 i \text{Li}_2\left(\frac{i \sqrt{2}}{\sqrt{\frac{1}{t}-1}}\right)+2 i \text{Li}_2\left(-\frac{i}{\sqrt{\frac{1}{t}-1}}\right)-2 i \text{Li}_2\left(-\frac{i \sqrt{2}}{\sqrt{\frac{1}{t}-1}}\right)-4 i \text{Li}_2\left(\sqrt{\frac{1}{1-t}}+1\right)\right.\\
&\left.+2 i \text{Li}_2\left[\left(\sqrt{2}+2\right) \left(\sqrt{\frac{1}{1-t}}+1\right)\right]+2 i \text{Li}_2\left[-i \left(\sqrt{2}-1\right) \tan \left(\frac{1}{2} \sin ^{-1}\left(\sqrt{\frac{1}{t}}\right)\right)\right]\right.\\
&\left.-2 i \text{Li}_2\left[i \left(\sqrt{2}-1\right) \tan \left(\frac{1}{2} \sin ^{-1}\left(\sqrt{\frac{1}{t}}\right)\right)\right]-2 i \text{Li}_2\left[-i \left(\sqrt{2}+1\right) \tan \left(\frac{1}{2} \sin ^{-1}\left(\sqrt{\frac{1}{t}}\right)\right)\right]\right.\\
&\left.+2 i \text{Li}_2\left[i \left(\sqrt{2}+1\right) \tan \left(\frac{1}{2} \sin ^{-1}\left(\sqrt{\frac{1}{t}}\right)\right)\right]+4 \pi  \log \left(\sqrt{2}+2\right)\right.\\
&\left.+\frac{1}{2} i \left[2 \text{Li}_2(2)-4 \left(\text{Li}_2\left(2-\sqrt{2}\right)+\text{Li}_2\left(\sqrt{2}+2\right)\right)+\pi ^2\right]\right.\\
&\left.+2 \log \left(\tan \left(\frac{\csc ^{-1}(\sqrt{t})}{2}\right)\right)\left[i \log \left(1+\frac{2 i}{\sqrt{t}-i}\right)-2 \tan ^{-1}\left((\sqrt{2}-1) \tan \left(\frac{1}{2} \csc ^{-1}(\sqrt{t})\right)\right)\right.\right.\\
&\left.\left.+2 \tan ^{-1}\left(\left(\sqrt{2}+1\right) \tan \left(\frac{1}{2} \csc ^{-1}\left(\sqrt{t}\right)\right)\right)\right] \right\}, \\
&f_r^{(2)}(q^2)=\begin{cases}f_r,\;\;t\geq 1,\\
                             f_r-\frac{1}{2(4\pi)^2m_{\sigma}^4}\times \frac{4\pi \log (2)}{t \sqrt{1-t}},\;\;0<t<1,
                             \end{cases}\\
\end{split}
\label{eq:ap.66}
\end{equation}
\begin{equation}
\begin{split}
&f_s^{(2)}(q^2)=\frac{1}{(4\pi)^3m_{\sigma}^4}\frac{\pi}{3t}\left\{-6 \text{Li}_2\left(-1+\frac{2 i}{\sqrt{t}}\right)+6 \text{Li}_2\left(1+\frac{2 i}{\sqrt{t}}\right)-12 \tan ^{-1}\left(\frac{2}{\sqrt{t}}\right) \cot ^{-1}\left(\sqrt{t}\right)\right.\\
&\left.-6 \text{Li}_2\left(-1-\frac{2 i}{\sqrt{t}}\right)+6 \text{Li}_2\left(1-\frac{2 i}{\sqrt{t}}\right)+\log (27) \left[\log \left(\frac{1}{\sqrt{t}}+i\right)+\log \left(-\frac{1}{\sqrt{t}}+i\right)\right]\right.\\
&\left.-3 \left[\log \left(1+\frac{2 i}{\sqrt{t}}\right) \log \left(2+\frac{2 i}{\sqrt{t}}\right)+\log \left(1-\frac{2 i}{\sqrt{t}}\right) \log \left(2-\frac{2 i}{\sqrt{t}}\right)+\pi ^2+i \pi  \log (9)+\log (3) \log \left(\frac{9}{4}\right)\right.\right.\\
&\left.\left.+\log \left(-\frac{2}{\sqrt{t}}+i\right) \left(\log \left(2+\frac{2 i}{\sqrt{t}}\right)+\log \left(\frac{1}{-6-\frac{6 i}{\sqrt{t}}}\right)\right)-\log \left(\frac{2}{3}-\frac{2 i}{3 \sqrt{t}}\right) \log \left(\frac{1}{3}+\frac{2 i}{3 \sqrt{t}}\right)\right.\right.\\
&\left.\left.+\log \left(\frac{2}{\sqrt{t}}+i\right) \left(-\log \left(-6 \sqrt{t}+6 i\right)+\log \left(-2 \sqrt{t}+2 i\right)-i \pi \right)-\log \left(\frac{1}{3}-\frac{2 i}{3 \sqrt{t}}\right) \log \left(\frac{2}{3}+\frac{2 i}{3 \sqrt{t}}\right)\right]\right\}.
\end{split}
\label{eq:ap.67}
\end{equation}

We cannot find the result for $f_u^{(2)}$ and $f_v^{(2)}$. But  they can be expressed   as
\begin{equation}
\begin{split}
&f_u^{(2)}(q^2)=\frac{1}{(4\pi)^3m_{\sigma}^4}\frac{1}{\sqrt{t \left(t^2+t+1\right)}}\int _0^{\frac{1}{t}}du\frac{\pi  \left((u-1) \left(4 \sqrt{u} \log \left(\frac{3 u}{4 u+1}\right)\right)-8 \tan ^{-1}\left(\frac{\sqrt{u}}{2 u+1}\right)\right)}{2 (u+1) u \sqrt{u^2+u+1}},\\
&f_v^{(2)}(q^2)=\frac{1}{(4\pi)^3m_{\sigma}^4}\frac{1}{\sqrt{t (t+2)}}\int _0^{\frac{1}{t}}du\frac{\pi  \left(-8 \sqrt{u} \log (u)+4 \sqrt{u} \log (4 u+1)-8 \tan ^{-1}\left(\frac{\sqrt{u}}{2 u+1}\right)\right)}{2 (u+1) \sqrt{u (2 u+1)}},\\
\end{split}
\label{eq:ap.68}
\end{equation}
for which numerical integrations can be used.

We also note that when $t$ is real, we should use $f_q^{(2)}(t+0^+i)$ instead of $f_q^{(2)}(t)$, and that when analytically continuing $|q|$ to $\sqrt{-(\omega +i\epsilon)^2}$, for $f_r^{(2)}$, we should use $f_r^{(2)}=f_r$ and $\left(f_r^{(2)}((q^2)^*)\right)^*$, for $f_q^{(2)}$ and $f_s^{(2)}$, we should use $\left(f_q^{(2)}((q^2)^*)\right)^*$ and $\left(f_s^{(2)}((q^2)^*)\right)^*$.

We also find
\begin{equation}
\begin{split}
&f_u^{(2)}(q^2=0)=\frac{1}{2(4\pi)^2m_{\sigma}^4}2\log (\frac{4}{3}),\;\;f_v^{(2)}(q^2=0)
=-\frac{1}{2(4\pi)^3m_{\sigma}^4}
\log\left(\frac{36\lambda^2}{m_{\sigma}^2}\right).\\
\end{split}
\label{eq:ap.69}
\end{equation}

\subsection{\label{sec:a3}Higher order contributions}

\subsubsection{\label{sec:a3.1}RPA-like contributions to \texorpdfstring{$\Pi _{\sigma}$}{Pi sigma}}

\begin{figure*}
\begin{center}
\includegraphics[scale=0.8]{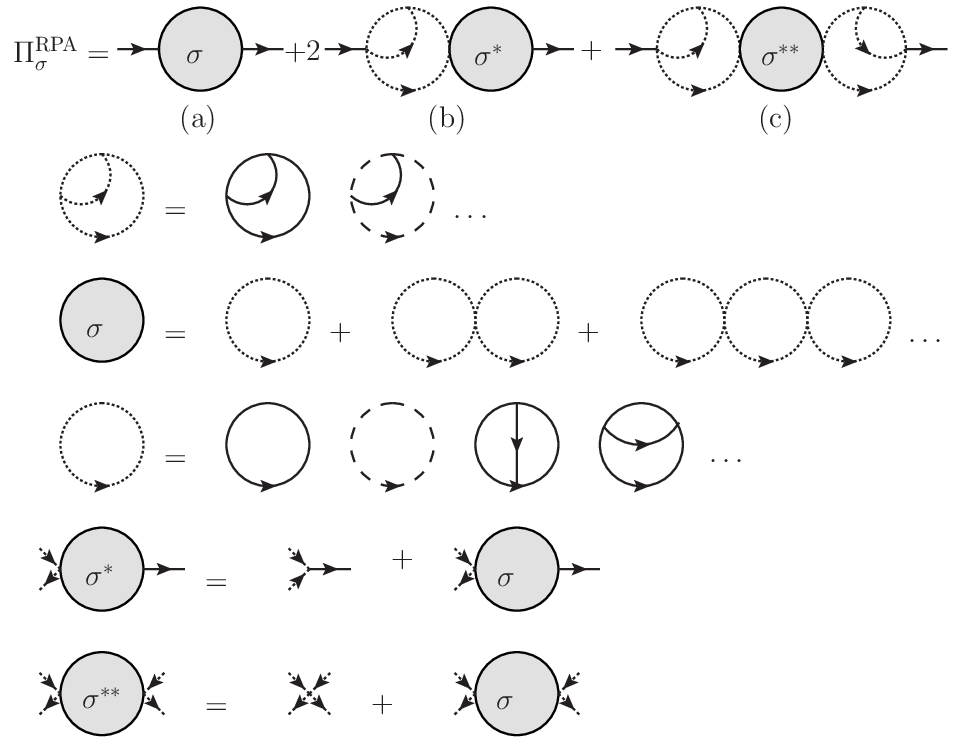}
\caption{The RPA-like contributions to $\Pi _{\sigma}$. The `$\ldots$' on the second row denotes all possible 2-loop diagrams with the external leg connecting to a four-leg vertex, the `$\ldots$' on the third row denotes all diagrams that can be drawn as a string of dotted circles. `$\ldots$' on the fourth row denotes all diagrams that can be drawn as a circle with a vertex connecting two legs on the left and a vertex connecting two legs on the right, up to 2-loop orders.}
\label{fig:sigma}
\end{center}
\end{figure*}
\begin{figure*}
\begin{center}
\includegraphics[scale=0.7]{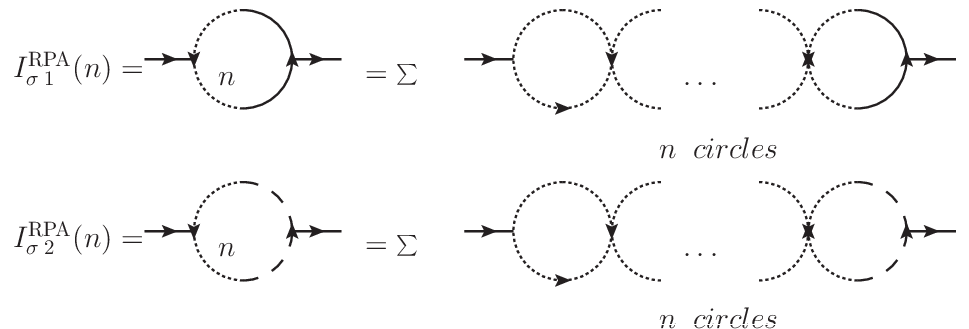}
\caption{Definitions of $I _{\sigma\;1}^{\rm RPA}(n)$ and $I _{\sigma\;2}^{\rm RPA}(n)$. The summation means the sum of all possible kinds of $n$ circles, with each circle up to 2-loop orders.}
\label{fig:rpadefine1}
\end{center}
\end{figure*}
\begin{figure*}
\begin{center}
\includegraphics[scale=0.8]{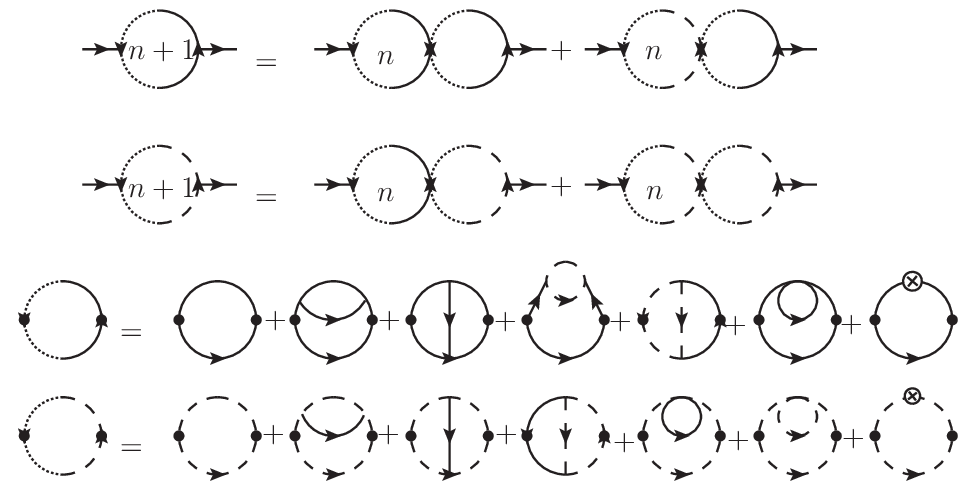}
\caption{The recursive relation used  to calculate $I _{\sigma\;1}^{\rm RPA}(n)$ and $I _{\sigma\;2}^{\rm RPA}(n)$.}
\label{fig:rparecursive1}
\end{center}
\end{figure*}

The RPA-like contribution to $\Pi _{\sigma}$ is  denoted as $\Pi _{\sigma}^{\rm RPA}$ and shown in Fig.~\ref{fig:sigma}. The contributions from  (a) (b) and (c) of Fig.~\ref{fig:sigma}  are denoted as $\Pi _{\sigma}^{\rm RPA\;a}$, $\Pi _{\sigma}^{\rm RPA\;b}$ and $\Pi _{\sigma}^{\rm RPA\;c}$ respectively.  To calculate $\Pi _{\sigma}^{\rm RPA\;a}$, we can define $I _{\sigma\;1}^{\rm RPA}(n)$ and $I _{\sigma\;2}^{\rm RPA}(n)$    as shown in Fig.~\ref{fig:rpadefine1}. They can be calculated by using recursive relation shown in Fig.~\ref{fig:rparecursive1}. The relation can be written as
\begin{equation}
\begin{split}
&I _{\sigma\;1}^{\rm RPA}(1)=x_0,\;\;\;I _{\sigma\;2}^{\rm RPA}(1)=y_0,\\
&I _{\sigma\;1}^{\rm RPA}(n+1)=x_1I _{\sigma\;1}^{\rm RPA}(n)+x_2I _{\sigma\;2}^{\rm RPA}(n),\\
&I _{\sigma\;2}^{\rm RPA}(n+1)=y_2I _{\sigma\;1}^{\rm RPA}(n)+y_1I _{\sigma\;2}^{\rm RPA}(n),\\
\end{split}
\label{eq:ap.70}
\end{equation}
where
\begin{equation}
\begin{split}
&x_0=18U^2v^2f_c^{(1)}(q^2)+36U^2v^2\delta_m^{(1)}f_f^{(1)}(q^2)-108U^3v^2f_a^{(1)}f_f^{(1)}(q^2)+648U^4v^4f_o^{(2)}(q^2)\\
&+72U^4v^4f_p^{(2)}(q^2)+648U^4v^4f_q^{(2)}(q^2)+24U^4v^4f_s^{(2)}(q^2),\\
&x_1=-\frac{1}{6Uv^2}x_0,\;\;\;x_2=-\frac{1}{18Uv^2}\left(x_0-24U^4v^4f_s^{(2)}(q^2)\right)-12U^3v^2f_s^{(2)}(q^2),\\
&y_0=2U^2v^2f_d^{(1)}(q^2)+4U^2v^2\delta _m^{(1)}f_g^{(1)}(q^2)-4U^3v^2f_a^{(1)}f_g^{(1)}(q^2)-12U^3v^2f_b^{(1)}f_g^{(1)}(q^2)\\
&+16U^4v^4f_n^{(2)}(q^2)+8U^4v^4f_r^{(2)}(q^2)+24U^4v^4f_s^{(2)}(q^2),\\
&y_2=-\frac{1}{2Uv^2}y_0,\;\;y_1=-\frac{3}{2Uv^2}\left(y_0-24U^4v^4f_s^{(2)}(q^2)\right)-4U^3v^2f_s^{(2)}(q^2).\\
\end{split}
\label{eq:ap.71}
\end{equation}

By solving the recursive relation, we find
\begin{equation}
\begin{split}
&I _{\sigma\;1}^{\rm RPA}(n)=\frac{2^{-n-1}}{Q (x_1 y_1-x_2 y_2)}\times\\
&\left[\left(Q (x_0y_1-x_2y_0)-\left(x_0x_1y_1+x_1x_2y_0+x_2y_0y_1-2x_0x_2y_2-x_0y_1^2\right)\right)\left(x_1+y_1-Q\right)^n\right.\\
&\left.\left(Q (x_0y_1-x_2y_0)+\left(x_0x_1y_1+x_1x_2y_0+x_2y_0y_1-2x_0x_2y_2-x_0y_1^2\right)\right)\left(x_1+y_1+Q\right)^n\right],\\
&I _{\sigma\;2}^{\rm RPA}(n)=\frac{2^{-n-1}}{Q (x_1 y_1-x_2 y_2)}\times\\
&\left[\left(Q (y_0x_1-y_2x_0)-\left(y_0y_1x_1+y_1y_2x_0+y_2x_0x_1-2y_0y_2x_2-y_0x_1^2\right)\right)\left(x_1+y_1-Q\right)^n\right.\\
&\left.\left(Q (y_0x_1-y_2x_0)+\left(y_0y_1x_1+y_1y_2x_0+y_2x_0x_1-2y_0y_2x_2-y_0x_1^2\right)\right)\left(x_1+y_1+Q\right)^n\right],\\
\end{split}
\label{eq:ap.72}
\end{equation}
where
\begin{equation}
\begin{split}
&Q=\sqrt{(x_1-y_1)^2+4x_2y_2},\\
\end{split}
\label{eq:ap.73}
\end{equation}
Finally, we find
\begin{equation}
\begin{split}
&\Pi _{\sigma}^{\rm RPA\;a}=\sum _{n=1}^{\infty}\left(I _{\sigma\;1}^{\rm RPA}(n)+I _{\sigma\;2}^{\rm RPA}(n)\right)=\frac{x_0 (1-y_1+y_2)+y_0(1-x_1+x_2)}{(x_1-1) (y_1-1)-x_2y_2}.\\
\end{split}
\label{eq:ap.74}
\end{equation}

\begin{figure*}
\begin{center}
\includegraphics[scale=0.7]{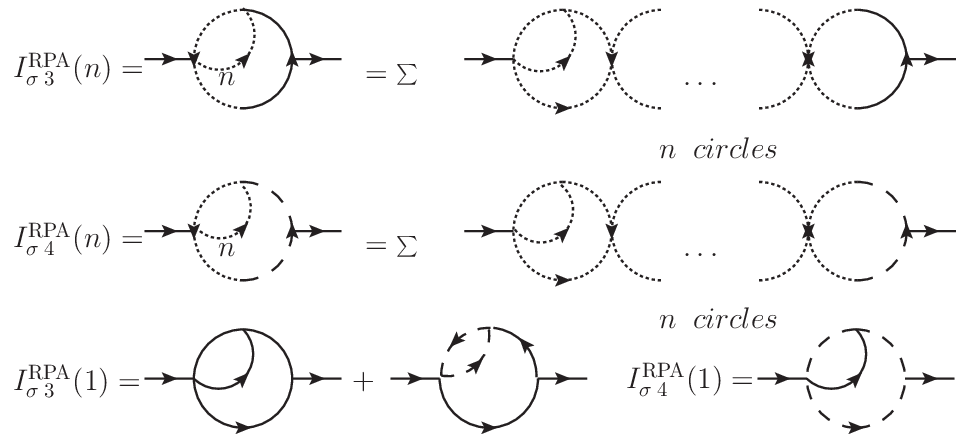}
\caption{Definitions of $I _{\sigma\;3}^{\rm RPA}(n)$ and $I _{\sigma\;4}^{\rm RPA}(n)$. The summation means the  sum of all possible kinds of   $n$ circles.}
\label{fig:rpadefine2}
\end{center}
\end{figure*}

\begin{figure*}
\begin{center}
\includegraphics[scale=0.7]{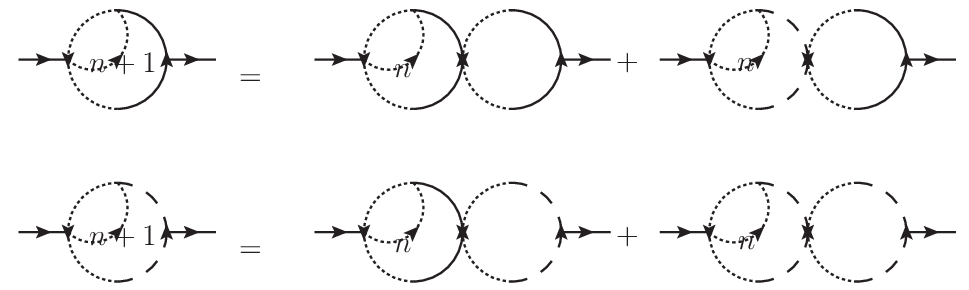}
\caption{The recursive relation used to calculate $I _{\sigma\;3}^{\rm RPA}(n)$ and $I _{\sigma\;4}^{\rm RPA}(n)$.}
\label{fig:rparecursive2}
\end{center}
\end{figure*}

Similarly, we define $I _{\sigma\;3}^{\rm RPA}(n)$ and $I _{\sigma\;4}^{\rm RPA}(n)$  as shown in Fig.~\ref{fig:rpadefine2}. The recursive relation for $I _{\sigma\;3}^{\rm RPA}(n)$ and $I _{\sigma\;4}^{\rm RPA}(n)$ are shown in Fig.~\ref{fig:rparecursive2} and can be written as
\begin{equation}
\begin{split}
&I _{\sigma\;3}^{\rm RPA}(1)=x_4=-108U^3v^2f_j^{(2)}(q^2)-12U^3v^2f_g^{(2)}(q^2),\\
&I _{\sigma\;4}^{\rm RPA}(1)=y_4=-8U^3v^2f_e^{(2)}(q^2),\\
&I _{\sigma\;3}^{\rm RPA}(n+1)=x_1I _{\sigma\;3}^{\rm RPA}(n)+x_2I _{\sigma\;4}^{\rm RPA}(n),\\
&I _{\sigma\;4}^{\rm RPA}(n+1)=y_2I _{\sigma\;3}^{\rm RPA}(n)+y_1I _{\sigma\;4}^{\rm RPA}(n).\\
\end{split}
\label{eq:ap.75}
\end{equation}
We find
\begin{equation}
\begin{split}
&\Pi _{\sigma}^{\rm RPA\;b}=\sum _{n=1}^{\infty}\left(I _{\sigma\;3}^{\rm RPA}(n)+I _{\sigma\;4}^{\rm RPA}(n)\right)=\frac{x_4 (1-y_1+y_2)+y_4(1-x_1+x_2)}{(x_1-1) (y_1-1)-x_2y_2}.\\
\end{split}
\label{eq:ap.76}
\end{equation}

$\Pi _{\sigma}^{\rm RPA\;c}$ can also be obtained with the help of $I _{\sigma\;3}^{\rm RPA}(n)$ and $I _{\sigma\;4}^{\rm RPA}(n)$,
\begin{equation}
\begin{split}
&\Pi _{\sigma}^{\rm RPA\;c}=\sum _{n=1}^{\infty}\left(\frac{1}{-6Uv}I _{\sigma\;3}^{\rm RPA}(n)\left(-108U^3vf_j^{(2)}(q^2)-12U^3vf_g^{(2)}(q^2)-8U^3vf_e^{(2)}(q^2)\right)\right.\\
&\left.+\frac{1}{-2Uv}I _{\sigma\;4}^{\rm RPA}(n)\left(-36U^3vf_j^{(2)}(q^2)-4U^3vf_g^{(2)}(q^2)-24U^3vf_e^{(2)}(q^2)\right)\right)\\
&=\frac{y_4 (y_5-x_1y_5+x_2x_5)+x_4(x_5-y_1x_5+y_2y_5)}{(x_1-1) (y_1-1)-x_2y_2},\\
\end{split}
\label{eq:ap.77}
\end{equation}
where
\begin{equation}
\begin{split}
&x_5=\frac{1}{6Uv}\left(108U^3vf_j^{(2)}(q^2)+12U^3vf_g^{(2)}(q^2)+8U^3vf_e^{(2)}(q^2)\right),\\
&y_5=\frac{1}{2Uv}\left(36U^3vf_j^{(2)}(q^2)+4U^3vf_g^{(2)}(q^2)+24U^3vf_e^{(2)}(q^2)\right).\\
\end{split}
\label{eq:ap.78}
\end{equation}

Finally,   $\Pi _{\sigma}^{\rm RPA}$ can be obtained as shown in Fig.~\ref{fig:sigma}, and can be written as
\begin{equation}
\begin{split}
&\Pi _{\sigma}^{\rm RPA}(q^2)=\Pi _{\sigma}^{\rm RPA\;a}+2\Pi _{\sigma}^{\rm RPA\;b}+\Pi _{\sigma}^{\rm RPA\;c}.
\end{split}
\label{eq:ap.79}
\end{equation}

\subsubsection{\label{sec:a3.2}RPA-like contributions to Cross-Susceptibilities}

\begin{figure*}
\begin{center}
\includegraphics[scale=0.7]{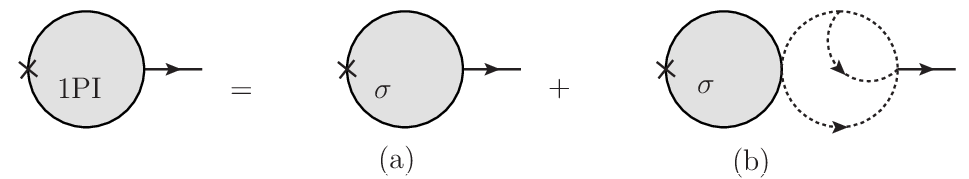}
\caption{The RPA-like contributions to $\Pi _{\rm A^2\sigma}$.}
\label{fig:rpacs}
\end{center}
\end{figure*}

\begin{figure*}
\begin{center}
\includegraphics[scale=0.7]{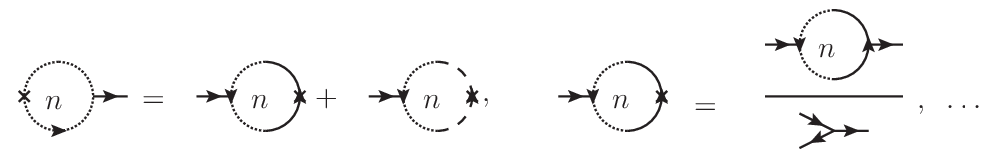}
\caption{The diagrams used  to calculate $\Pi^{\rm RPA\; a} _{A^2\sigma}$. The horizontal line means division, and the diagrams are divided by the coupling constant as shown in Eq.~(\ref{eq:ap.80}).}
\label{fig:rpacsdefine1}
\end{center}
\end{figure*}

The RPA-like contributions to $\Pi _{A^2\sigma}$ are shown in Fig.~\ref{fig:rpacs}. The diagram shown in (a) and (b)  of Fig.~\ref{fig:rpacs} are denoted as $\Pi _{A^2\sigma}^a$ and $\Pi _{A^2\sigma}^b$, which can be  calculated with the help of the diagram  defined in Fig.~\ref{fig:rpacsdefine1} which  is denoted as $I_{A^2\sigma}^{\rm RPA\;a}(n)$ and can be  written as
\begin{equation}
\begin{split}
&I_{A^2\sigma}^{\rm RPA\;a}(n)=\frac{2}{-6Uv}I^{\rm RPA}_{\sigma\;1}(n)+\frac{2}{-2Uv}I^{\rm RPA}_{\sigma\;2}(n).\\
\end{split}
\label{eq:ap.80}
\end{equation}

When the propagators connected to each initial state are the same, as   in (4) of Fig.~\ref{fig:cross2loop1}, a factor $2$ is added by removing  the symmetry factor. When the propagators are different,  as in the diagram in (1) of Fig.~\ref{fig:cross2loop1}, a factor $2$ is added by the  exchange of the initial states.

As a result,
\begin{equation}
\begin{split}
&\Pi _{A^2\sigma}^a=\sum _{n=1}^{\infty}I_{A^2\sigma}^{\rm RPA\;a}(n)=-\frac{2x_0 (1-y_1+3y_2)+2y_0 (3-3x_1+x_2)}{6Uv \left((x_1-1) (y_1-1)-x_2y_2\right)}.
\end{split}
\label{eq:ap.81}
\end{equation}
Similarly
\begin{equation}
\begin{split}
&\Pi _{A^2\sigma}^b=\sum _{n=1}^{\infty}\left(\frac{2}{-6Uv}I^{\rm RPA}_{\sigma\;3}(n)+\frac{2}{-2Uv}I^{\rm RPA}_{\sigma\;4}(n)\right)\\
&=-\frac{2x_4 (1-y_1+3y_2)+2y_4 (3-3x_1+x_2)}{6Uv \left((x_1-1) (y_1-1)-x_2y_2\right)},
\end{split}
\label{eq:ap.82}
\end{equation}
and
\begin{equation}
\begin{split}
&\Pi _{A^2\sigma}=\Pi _{A^2\sigma}^a+\Pi _{A^2\sigma}^b.
\end{split}
\label{eq:ap.83}
\end{equation}

\begin{figure*}
\begin{center}
\includegraphics[scale=0.7]{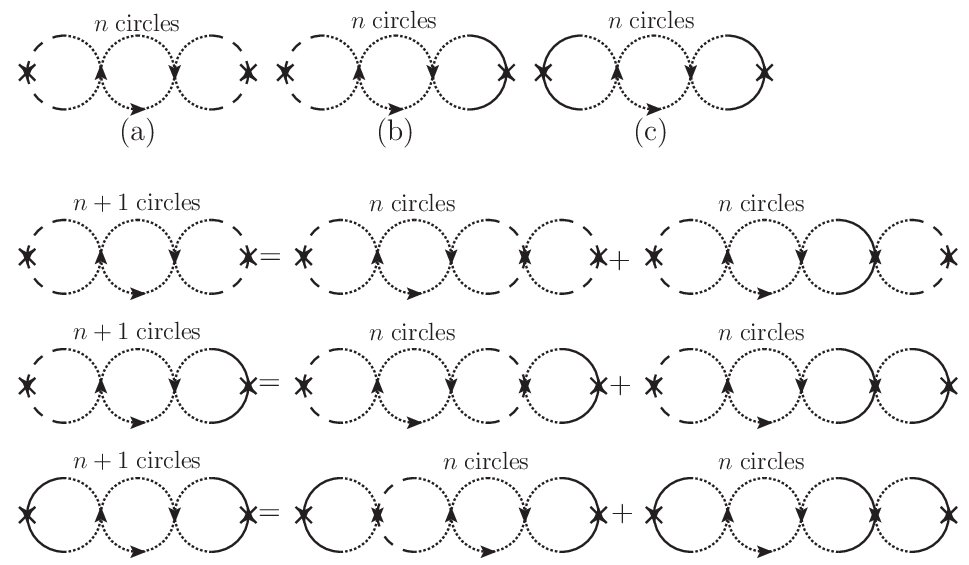}
\caption{The diagrams and recursive relation used to calculate $\Pi _{A^2B^2}$.}
\label{fig:rpacsrecursive}
\end{center}
\end{figure*}

RPA-like contributions $\Pi _{A^2B^2}$ can be calculated with the help of diagrams shown in Fig.~\ref{fig:rpacsrecursive}. The diagrams in (a) (b) and (c) of  Fig.~\ref{fig:rpacsrecursive}  are defined as $I_{\pi^2\pi^2}(n)$, $I_{\sigma^2\pi^2}(n)$ and $I_{\sigma^2\sigma^2}(n)$. As shown in Fig.~\ref{fig:rpacsrecursive},  we find
\begin{equation}
\begin{split}
&I_{\pi^2\pi^2}(1)=x_6,\;I_{\pi^2\sigma^2}(1)=y_6,\;I_{\sigma^2\sigma^2}(1)=z_6,\\
&I_{\pi^2\pi^2}(n+1)=y_1I_{\pi^2\pi^2}(n)+y_2I_{\pi^2\sigma^2}(n),\\
&I_{\pi^2\sigma^2}(n+1)=x_2I_{\pi^2\pi^2}(n)+x_1I_{\pi^2\sigma^2}(n),\\
&I_{\sigma^2\sigma^2}(n+1)=x_2I_{\pi^2\sigma^2}(n)+x_1I_{\sigma^2\sigma^2}(n),\\
\end{split}
\label{eq:ap.84}
\end{equation}
with
\begin{equation}
\begin{split}
&x_6=2f_d^{(1)}(q^2)+4\delta _m^{(1)}f_g^{(1)}(q^2)-4Uf_a^{(1)}f_g^{(1)}(q^2)-12Uf_b^{(1)}f_g^{(1)}(q^2)\\
&+16U^2v^2f_n^{(2)}(q^2)+8U^2v^2f_r^{(2)}(q^2),\\
&y_6=8U^2v^2f_s^{(2)}(q^2),\\
&z_6=2f_c^{(1)}(q^2)+4\delta _m^{(1)}f_f^{(1)}(q^2)-12Uf_a^{(1)}f_f^{(1)}(q^2)+72U^2v^2f_o^{(2)}(q^2)\\
&+8U^2v^2f_p^{(2)}(q^2)+72U^2v^2f_q^{(2)}(q^2).\\
\end{split}
\label{eq:ap.85}
\end{equation}
$\Pi _{A^2B^2}$ can be written as
\begin{equation}
\begin{split}
&\Pi _{A^2B^2}=\sum _{n=1}^{\infty} \left(I_{\pi^2\pi^2}(n)+I_{\sigma^2\sigma^2}(n)+2I_{\pi^2\sigma^2}(n)\right)\\
&=\frac{4}{(1-x_1) \left(Q^2-(x_1+y_1-2)^2\right)}\times \left\{y_6 ((1-y_1) (2 x_1-x_2-2)+x_1 y_2-y_2)\right.\\
&\left.+z_6 (x_2y_2-(1-x_1)(1-y_1))-x_6(x_2-x_1+1)^2\right\}.\\
\end{split}
\label{eq:ap.86}
\end{equation}

\subsubsection{\label{sec:a3.3}1PI summation of \texorpdfstring{$\chi _{A^2B^2}$}{chi A2B2}}

\begin{figure}
\includegraphics[scale=0.7]{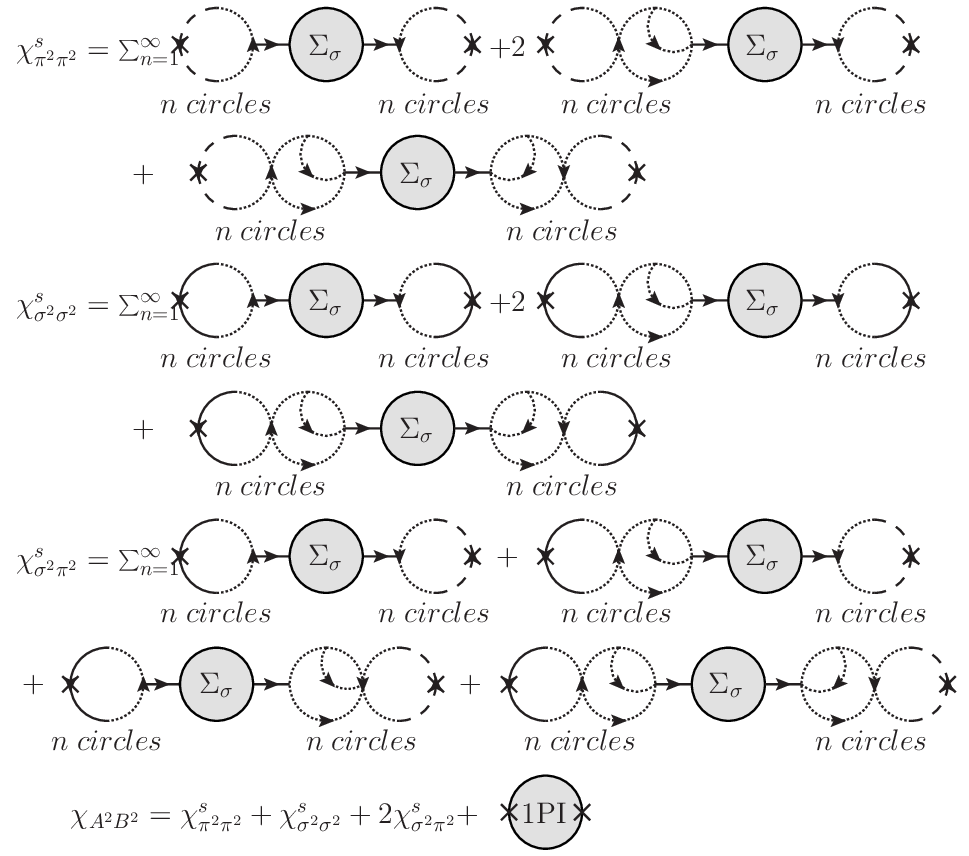}
\caption{1PI summation of $\chi _{A^2B^2}$.}
\label{fig:1pisumcs}
\end{figure}

The 1PI sum of $\chi _{A^2B^2}$ are shown in Fig.~\ref{fig:1pisumcs}, and can be written as
\begin{equation}
\begin{split}
&\chi _{\sigma^2\sigma^2}^s(q^2)=\frac{4\Sigma _{\sigma}(q^2)}{36U^2v^2}\sum _{n=1}^{\infty}\left(I_{\sigma\;1}^{\rm RPA}(n)+I_{\sigma\;3}^{\rm RPA}(n)\right)^2,\\
&\chi _{\pi^2\pi^2}^s(q^2)=\frac{4\Sigma _{\sigma}(q^2)}{4U^2v^2}\sum _{n=1}^{\infty}\left(I_{\sigma\;2}^{\rm RPA}(n)+I_{\sigma\;4}^{\rm RPA}(n)\right)^2,\\
&\chi _{\sigma^2\pi^2}^s(q^2)=\frac{4\Sigma _{\sigma}(q^2)}{12U^2v^2}\sum _{n=1}^{\infty}\left(I_{\sigma\;1}^{\rm RPA}(n)+I_{\sigma\;3}^{\rm RPA}(n)\right)\left(I_{\sigma\;2}^{\rm RPA}(n)+I_{\sigma\;4}^{\rm RPA}(n)\right),\\
&\chi _{A^2B^2}^s(q^2)=\chi _{\sigma^2\sigma^2}^s(q^2)+\chi _{\pi^2\pi^2}^s(q^2)+2\chi _{\sigma^2\pi^2}^s(q^2)\\
&=4\Sigma _{\sigma}(q^2)\sum _{n=1}^{\infty}\left(\frac{I_{\sigma\;1}^{\rm RPA}(n)+I_{\sigma\;3}^{\rm RPA}(n)}{6Uv}+\frac{I_{\sigma\;2}^{\rm RPA}(n)+I_{\sigma\;4}^{\rm RPA}(n)}{2Uv}\right)^2\\
&=\frac{4\Sigma _{\sigma}(q^2)}{9U^2v^2(Q^2-(x_1+y_1-2)^2)((1+x_1) (1+y_1)-x_2y_2)(x_1y_1-x_2y_2-1)}\\
&\times \left\{(x_0+x_4)^2\left[x_1(y_1(y_1-3y_2-1)(y_1-3y_2+1)+6y_2)\right.\right.\\
&\left.\left.-(y_1-3y_2)(x_2y_1y_2-3y_2(x_2y_2-1)+y_1)-x_2y_2+1\right]\right.\\
&\left.+(y_0+y_4)^2\left[9 x_1^3 y_1-6 x_1^2x_2y_1-9 x_1^2+x_1x_2^2y_1-x_2y_2\left((x_2-3x_1)^2+9\right)-9 x_1y_1+x_2^2+6x_2y_1+9\right]\right.\\
&\left.+2(x_0+x_4)(y_0+y_4)\left[y_1y_2 \left(9-9x_1^2+x_2^2\right)+\left(y_1^2-1\right) \left(3 x_1^2-x_1x_2-3\right)-3 x_2y_2^2 (x_2-3x_1)\right]\right\}\\
\end{split}
\label{eq:ap.87}
\end{equation}
The 1PI summation of $\chi _{A^2B^2}$ can be written as
\begin{equation}
\begin{split}
&\chi _{A^2B^2}(q^2)=\Pi _{A^2B^2}(q^2)+\chi _{A^2B^2}^s(q^2).\\
\end{split}
\label{eq:ap.88}
\end{equation}

This work is supported by National Natural Science Foundation of China (Grant No. 12075059).


\begin{thebibliography}{99}
\bibliographystyle{unsrt}

  \bibitem {higgs}
  P. W. Higgs, Phys. Rev. Lett. 13, 508 - 509, (1964);

  F. Englert and R. Brout, Phys. Rev. Lett. 13, 321 - 323, (1964).

  \bibitem {littlewood}
  P. B. Littlewood and C. M. Varma, Phys. Rev. Lett. {\bf 47}, 811 (1981);

  P. B. Littlewood and C. M. Varma, Phys. Rev. B {\bf 26}, 4883 (1982).

  \bibitem {3DOpticalLattice}
  U. Bissbort, et al. Phys. Rev. Lett. {\bf 106}, 205303 (2011), arXiv:1010.2205.

  \bibitem {2DOpticalLattice}
  M. Endres, et al. Nature {\bf 487}, 454-458 (2012), arXiv:1204.5183.


  \bibitem {OtherExperiments}
  C. R\"{u}egg, et al. Phys. Rev. Lett. {\bf 100}, 205701, (2008), arXiv:0803.3720;

  R. Matsunaga, et al. Phys. Rev. Lett. {\bf 111} 057002, (2011), arXiv:1305.0381;

  R. Matsunaga, et al. Science, {\bf 345}, 1145, (2014);

  D. Sherman, et al. Nature Physics {\bf 11}, 188 - 192, (2015).

\bibitem {Cavity}
  Y.-X. Yu, J. Ye and W. Liu, Scientific Reports 3, Article number: 3476, (2013), arXiv:1312.3404.

  \bibitem {Varma1}
  D. Pekker and C. M. Varma, Annual Reviews of Condensed Matter Physics Volume {\bf 6}, (2015), arXiv:1406.2968.

  \bibitem {Varma2}
  C. M. Varma, arXiv:cond-mat/0109409.

  \bibitem {BECReview}
  A. J. Leggett, Rev. Mod. Phys. {\bf 73}, 307, (2001);

  I. Bloch, J. Dalibard, and W. Zwerger, Rev. Mod. Phys. {\bf 80}, 885, (2008).

  \bibitem {O2Model1}
  E. Altman and A. Auerbach, Phys. Rev. Lett. {\bf 89}, 250404, (2002).

  \bibitem {O2Model2}
  M. P. A. Fisher, et al. Phys. Rev. B {\bf 40}, 546, (1989);

  S. D. Huber, et al. Phys. Rev. B {\bf 75}, 085106 (2007), arXiv:cond-mat/0610773.

  \bibitem {3DO2}
  K. Nagao and I. Danshita, Progress of Theoretical and Experimental Physics, 063I01, (2016), arXiv:1603.02395

  K. Nagao, Y. Takahashi, I. Danshita, arXiv:1710.00547

  \bibitem {Sachdev}
  S. Sachdev, Phys. Rev. B {\bf 59}, 14054 (1999).

  \bibitem {SachdevBook}
  S. Sachdev, \textit{Quantum Phase Transitions} (Cambridge University Press, Cambridge, 2000).

  \bibitem {Podolsky1}
  D. Podolsky, A. Auerbach and D. P. Arovas, Phys. Rev. B {\bf 84}, 174522, (2011), arXiv:1108.5207.

  \bibitem {Simulationstudy}
  L. Pollet and N. Prokof'ev, Phys. Rev. Lett. {\bf 109}, 010401, (2012), arXiv:1204.5190.

  \bibitem {Podolsky2}
  D. Podolsky and S. Sachdev, Phys. Rev. B {\bf 86}, 054508, (2012), arXiv:1205.2700.

  \bibitem {FermionSuperfluid}
  B. Liu, H. Zhai and S. Zhang, Phys. Rev. A {\bf 93}, 033641, (2016), arXiv:1502.00431.

  \bibitem {TsvelikBook}
  A. M. Tsvelik, \textit{Quantum Field Theory in Condensed Matter Physics}, (Cambridge University Press, Cambridge, 1995).

  \bibitem {PeskinBook}
  M. E. Peskin and D. V. Schroeder, \textit{An Introduction to Quantum Field Theory}, (Westview Press, Boulder, 1995).

  \bibitem {DR}
  G. 't Hooft and M.Veltman, Nucl. Phys. B {\bf 44}, 189-213 (1972).

  \bibitem {CondensedMatterBook}
  A. Altland, B. D. Simons, \textit{Condensed Matter Field Theory}, (Cambridge University Press, Cambridge, 2010).

  \bibitem {GrozinMultiLoop}
  A. G. Grozin, Int. J. Mod. Phys. A {\bf 19}, 473-520, (2004), arXiv:hep-ph/0307297.

  \bibitem {MellinBarnes}
  S. Weinzierl, arXiv:hep-ph/0604068;

  M. Czakon, J. Gluza and T. Riemann, Nucl. Phys. B {\bf 751} 1 - 17, (2006), arXiv:hep-ph/0604101.

  \bibitem{HypExp}
  T. Huber and D. Maitre, Comput. Phys. Commun. {\bf 175}, 122 - 144, (2006), arXiv:hep-ph/0507094;

  T. Huber and D. Maitre, Comput. Phys. Commun. {\bf 178} 755 - 776, (2008), arXiv:0708.2443.

  \bibitem{MB}
  M. Czakon, Comput. Phys. Commun. {\bf 175} 559 - 571, (2006), arXiv:hep-ph/0511200.

  \bibitem{MBSum}
  M. Ochman and T. Riemann, Acta Phys. Polon. B {\bf 46} no.11, 2117, (2015), arXiv:1511.01323.

  \bibitem{AMBRE}
  J. Gluza, K. Kajda and T. Riemann, Comput. Phys. Commun. {\bf 177} 879-893, (2007), arXiv:0704.2423.

  \bibitem{C0}
  G.'t Hooft and M. Veltman, Nucl. Phys. B {\bf 153}, 365-401 (1979).

  \bibitem{IBP}
  K. G. Chetyrkin and F. V. Tkachov, Nucl. Phys. B {\bf 192}, 159 (1981);

  A. G. Grozin, Int. J. Mod. Phys. A {\bf 26}, 2807-2854 (2011), arXiv:1104.3993.

  \bibitem{IBPwithDE}
  E. Remiddi, Nuovo Cim. A {\bf 110}, 1435-1452 (1997), arXiv:hep-th/9711188;

  T. Gehrmann and E. Remiddi, Nucl. Phys. B {\bf 580}, 485-518 (2000), arXiv:hep-ph/9912329.

\end{thebibliography}
\end{document}